\documentclass{aa}
\usepackage{graphicx}
\usepackage{natbib}
\usepackage{url}
\usepackage{multirow,bigdelim}
\usepackage{rotating}
\usepackage{placeins}
\usepackage{color}
\usepackage{soul}
\usepackage{lscape}
\begin{document}
%USER DEFINED ShORTCUTS
\newcommand{\zabs}{\ensuremath{z_{\rm abs}}}
\newcommand{\zem}{\ensuremath{z_{\rm em}}}
\newcommand{\zqso}{\ensuremath{z_{\rm QSO}}}
\newcommand{\zgal}{\ensuremath{z_{\rm gal}}}
\newcommand{\HH}{\mbox{H$_2$}}
\newcommand{\HD}{\mbox{HD}}
\newcommand{\DD}{\mbox{D$_2$}}
\newcommand{\CO}{\mbox{CO}}
\newcommand{\dla}{damped Lyman\,$\alpha$}
\newcommand{\Dla}{damped Lyman\,$\alpha$}
\newcommand{\lya}{\ensuremath{{\rm Ly}\,\alpha}}
\newcommand{\Lya}{\ensuremath{{\rm Ly}\,\alpha}}
\newcommand{\lyb}{Ly\,$\beta$}
\newcommand{\Ha}{H\,$\alpha$}
\newcommand{\Hb}{H\,$\beta$}
\newcommand{\lyg}{Ly\,$\gamma$}
\newcommand{\lyd}{Ly\,$\delta$}

%ions A\&A style - can be used both in math or normal mode
\newcommand{\ArI}{\ion{Ar}{i}}
\newcommand{\CaII}{\ion{Ca}{ii}}
\newcommand{\CI}{\ion{C}{i}}
\newcommand{\CII}{\ion{C}{ii}}
\newcommand{\CIV}{\ion{C}{iv}}
\newcommand{\ClI}{\ion{Cl}{i}}
\newcommand{\ClII}{\ion{Cl}{ii}}
\newcommand{\CoII}{\ion{Co}{ii}}
\newcommand{\CrII}{\ion{Cr}{ii}}
\newcommand{\CuII}{\ion{Cu}{ii}}
\newcommand{\DI}{\ion{D}{i}}
\newcommand{\FeI}{\ion{Fe}{i}}
\newcommand{\FeII}{\ion{Fe}{ii}}
\newcommand{\GeII}{\ion{Ge}{ii}}
\newcommand{\HI}{\ion{H}{i}}
\newcommand{\MgI}{\ion{Mg}{i}}
\newcommand{\MgII}{\ion{Mg}{ii}}
\newcommand{\MnII}{\ion{Mn}{ii}}
\newcommand{\NaI}{\ion{Na}{i}}
\newcommand{\NI}{\ion{N}{i}}
\newcommand{\NII}{\ion{N}{ii}}
\newcommand{\NV}{\ion{N}{v}}
\newcommand{\NiII}{\ion{Ni}{ii}}
\newcommand{\OI}{\ion{O}{i}}
\newcommand{\OII}{\ion{O}{ii}}
\newcommand{\OIII}{\ion{O}{iii}}
\newcommand{\OVI}{\ion{O}{vi}}
\newcommand{\PII}{\ion{P}{ii}}
\newcommand{\PbII}{\ion{Pb}{ii}}
\newcommand{\SI}{\ion{S}{i}}
\newcommand{\SII}{\ion{S}{ii}}
\newcommand{\SiII}{\ion{Si}{ii}}
\newcommand{\SiIV}{\ion{Si}{iv}}
\newcommand{\TiII}{\ion{Ti}{ii}}
\newcommand{\ZnII}{\ion{Zn}{ii}}
\newcommand{\AlII}{\ion{Al}{ii}}
\newcommand{\AlIII}{\ion{Al}{iii}}
\def\h2{$\rm H_2$}
%Other shortcuts.
\newcommand{\Ho}{\mbox{$H_0$}}
\newcommand{\angstrom}{\mbox{{\rm \AA}}}
\newcommand{\abs}[1]{\left| #1 \right|} % for absolute value
\newcommand{\avg}[1]{\left< #1 \right>} % for average
\newcommand{\kms}{\ensuremath{{\rm km\,s^{-1}}}}
\newcommand{\cmsq}{\ensuremath{{\rm cm}^{-2}}}
\newcommand{\ergs}{\ensuremath{{\rm erg\,s^{-1}}}}
\newcommand{\ergsa}{\ensuremath{{\rm erg\,s^{-1}\,{\AA}^{-1}}}}
\newcommand{\ergscm}{\ensuremath{{\rm erg\,s^{-1}\,cm^{-2}}}}
\newcommand{\ergscma}{\ensuremath{{\rm erg\,s^{-1}\,cm^{-2}\,{\AA}^{-1}}}}
\newcommand{\msyr}{\ensuremath{{\rm M_{\rm \odot}\,yr^{-1}}}}
\newcommand{\nhi}{n_{\rm HI}}
\newcommand{\fhi}{\ensuremath{f_{\rm HI}(N,\chi)}}
\newcommand{\refs}{{\bf (refs!)}}
\newcommand{\PN}{\color{red} PN:~}
\newcommand{\jz}{J0000$+$0048}

%Institutes
\newcommand{\iap}{Institut d'Astrophysique de Paris, CNRS-UPMC, UMR7095, 98bis bd Arago, 75014 Paris, France\label{iap}}
\newcommand{\lam}{Laboratoire d’Astrophysique de Marseille, CNRS/Aix Marseille Université, UMR 7326, 13388, Marseille, France\label{lam}}
\newcommand{\iucaa}{Inter-University Centre for Astronomy and Astrophysics, Post Bag 4, Ganeshkhind, 411\,007, Pune, India\label{iucaa}} 
\newcommand{\eso}{European Southern Observatory, Alonso de C\'ordova 3107, Vitacura, Casilla 19001, Santiago 19, Chile\label{eso}}
\newcommand{\ioffe}{Ioffe Physical-Technical Institute of RAS, {Polyteknicheskaya 26}, 194021 Saint-Petersburg, Russia\label{ioffe}}
\newcommand{\florida}{Department of Astronomy, University of Florida, 211 Bryant Space Science Center, Gainesville, 32611, USA \label{florida}}
\newcommand{\swin}{Centre for Astrophysics and Supercomputing, Swinburne University of Technology, Melbourne, Victoria 3122, Australia \label{swin}}
\newcommand{\laser}{Department of Physics and Astronomy, LaserLaB, Vrije Universiteit, De Boelelaan 1081, 1081 HV Amsterdam, The Netherlands \label{laser}}
\newcommand{\mpie}{Max-Planck-Institut f\"ur extraterrestrische Physik, Giessenbachstra{\ss}e, D-85748 Garching, Germany  \label{mpie}}
\newcommand{\dark}{Dark Cosmology Centre, Niels Bohr Institute, University of Copenhagen, Juliane Maries Vej 30, 2100 Copenhagen {\O}, Denmark \label{dark}}

%===================================== TITLE / AUTHORS / ABSTRACT ============================================
\title{
  Discovery of a Perseus-like cloud in the early Universe  \thanks{Based on observations collected at the European Organisation for Astronomical Research in the Southern Hemisphere under ESO programmes 093.A-0126(A), 096.A-0354(A) and 096.A-0924(B).}}
\subtitle{H\,{\sc i}-to-H$_2$ transition, carbon monoxide and small dust grains at $\zabs \approx 2.53$ towards the quasar \jz}
\titlerunning{A Perseus-like molecular cloud at $z_{\rm abs}=2.5$ towards \jz }

\author{
  P. Noterdaeme \inst{\ref{iap},\thanks{\email{noterdaeme@iap.fr}}}%1
  \and J.-K.~Krogager \inst{\ref{iap},\ref{dark}}                  %2
  \and S.~Balashev \inst{\ref{ioffe}}                              %3
  \and J.~Ge \inst{\ref{florida}}                                  %alpha                        
  \and N.~Gupta \inst{\ref{iucaa}}
  \and T.~Kr\"uhler \inst{\ref{mpie}}                               
  \and C.~Ledoux \inst{\ref{eso}}                 
  \and M.~T.~Murphy \inst{\ref{swin}} 
  \and I.~P\^aris \inst{\ref{lam}}                
  \and P.~Petitjean \inst{\ref{iap}}
  \and H.~Rahmani \inst{\ref{lam}}                
  \and R.~Srianand \inst{\ref{iucaa}}
  \and W.~Ubachs \inst{\ref{laser}}
  }
  \institute{
    \iap
    \and \dark
      \and \ioffe 
  \and \florida
  \and \iucaa
    \and \mpie
  \and  \eso 
  \and  \swin
    \and \lam
  \and \laser
  }

  \abstract{
    We present the discovery of a molecular cloud at $\zabs \approx 2.5255$ along the line of sight
    to the quasar SDSS J\,000015.17$+$004833.3.
    We use a high-resolution spectrum obtained with the Ultraviolet and Visual Echelle Spectrograph
    together with a deep multi-wavelength medium-resolution spectrum obtained with X-Shooter 
    (both on the Very Large Telescope) to perform a 
    detailed analysis of the absorption lines
    from ionic, neutral atomic and molecular species in different excitation levels, as well as the broad-band
    dust extinction. \\
    We find that the absorber classifies as a Damped Lyman-$\alpha$ system (DLA) with $\log N(\HI) (\cmsq)=20.8\pm 0.1$.
    The DLA has super-Solar metallicity ($Z\sim 2.5 Z_{\odot}$, albeit with factor 2-3 uncertainty) with a depletion pattern typical of cold gas
    and an overall molecular fraction $f=2N($H$_2$)/($2N$(H$_2)+N(\HI)) \sim 50$\%. This is the
    highest $f$-value observed to date
    in a high-$z$ intervening system. Most of the molecular hydrogen arises from a clearly identified
    narrow ($b\sim$0.7~\kms), cold component
    in which carbon monoxide molecules are also found, with $\log N($CO$) \approx 15$.
    With the help of the spectral synthesis code Cloudy, we study the chemical and physical
    conditions in the cold gas. We find that the line of sight probes the gas deep after the H\,{\sc i}-to-H$_2$ transition
    in a $\sim$4-5~pc-size cloud with volumic density $n_{\rm H} \sim$~80~cm$^{-3}$ and temperature of only 50~K.
    Our model suggests that the presence of small dust grains (down to about 0.001\,${\rm \mu m}$) and
    high cosmic ray ionisation rate ($\zeta_{\rm H} \sim $ a few times $10^{-15}$\,s$^{-1}$) are needed to explain
    the observed atomic and molecular abundances. The presence of small grains is also in agreement with the
    observed steep extinction curve that also features a 2175~{\AA} bump.\\
    Interestingly, the chemical and physical properties of this cloud are very similar to what is seen in
    diffuse molecular regions of the nearby Perseus complex, despite the former being observed when the Universe
    was only 2.5~Gyr old. 
    The high excitation temperature of CO rotational levels towards \jz\ betrays however the
    higher temperature of the cosmic microwave background. Using the derived physical conditions, we
    correct
    for a small contribution (0.3~K) of collisional excitation and 
    obtain $T_{\rm CMB}(z=2.53) \approx 9.6$~K,
    in perfect agreement with the predicted adiabatic cooling of the Universe. 
  }

   \date{}

\maketitle
%=============================================================================================================

\section{Introduction}
The formation and evolution of galaxies is strongly dependent on the physical properties of the gas in and
around galaxies. Indeed, the gas is the reservoir of baryons from which stars form and at the same time,
it integrates the chemical and physical outputs from star-formation activity.
The gas that is accreted onto galaxies has to cool down and go through different transitional
processes that will determine its properties during its evolution before the final collapse that give birth to stars.
Different phases are indeed identified in the interstellar
medium, depending on the temperature and density and whether the matter is ionised or neutral (atomic or molecular).
In their two-phase model, \citet{Field69} showed that thermal equilibrium leads neutral gas to
segregate into a dense phase, the cold neutral medium (CNM), embedded into a diffuse intercloud phase, the
warm neutral medium (WNM).
Detailed theoretical and numerical works \citep[e.g.][]{Krumholz09a,Sternberg14} show that a transition from
\HI\ to H$_2$ then occurs in the former phase, depending on the balance between H$_2$ formation on the surface
of dust grains \citep[e.g.][]{Jura74b}, and its dissociation by UV photons \citep[e.g.][]{Dalgarno70}, itself
dependent on both self- and dust-shielding. 

Observationally, UV absorption spectroscopy of Galactic clouds towards nearby stars revealed that the molecular
fraction, $f=2$H$_2/(2$H$_2+\HI)$, sharply increases above a \HI\ column density threshold of 5$\times 10^{20}~\cmsq$. A similar
threshold has been found by \citet{Reach94} from far-infrared emission studies of interstellar clouds,
using dust as a tracer for H$_2$. Higher column-density thresholds were observed in the Magellanic Clouds
\citep{Tumlinson07}, which could be the consequence of a higher UV radiation field together with a lower metallicity
in these environments. However, it is also possible that a significant fraction of the observed \HI\ column density
is actually unrelated to the atomic envelopes of the H$_2$-absorbing clouds \citep{Welty12}, since $N(\HI)$ was derived
through unresolved 21-cm emission, while $N($H$_2)$ was measured in absorption.

This highlights the main difficulty in observing the transition regions: because molecular clouds
have sizes of only a few tens to a few hundred parsec \citep[e.g.][]{Fukui10} it is very difficult to compare H$_2$ with
its associated \HI\ in the cloud envelope without also integrating nearby atomic gas. High spatial resolution
(sub-pc) studies exist for nearby molecular clouds such as the Perseus cloud. \citet{Lee12} observed relatively uniform
\HI\ surface density of $\Sigma_{\rm HI}\sim 6-8$~M$_{\odot}$\,pc$^{-2}$ around H$_2$ clouds, in agreement with the theoretical
expectations based on H$_2$ microphysics at Solar metallicity, assuming CNM a priori \citep{Krumholz09a} or not
\citep[][]{Bialy15}.

Ideally, we would also like to study the atomic to molecular transition and the subsequent star formation
over parsec scales in other galaxies. Observations of nearby galaxies have been possible at slightly
sub-kpc resolution, revealing a saturation value around $\Sigma_{\rm HI}\approx 9$~M$_{\odot}$\,pc$^{-2}$ \citep{Bigiel08}.
However, the observational techniques applied in the local Universe are not applicable yet in the distant Universe without a
further strong loss of spatial resolution.
Prescriptions of star-formation over galactic scales, such as the empirical relation between
the molecular to atomic ratio and the hydrostatic pressure \citep[e.g.][]{Blitz06} are nevertheless available and can be used
in evolution models of galaxies \citep[e.g.][]{Lagos11}, although this corresponds to an extrapolation at high redshift
of a phenomenon observed in the local Universe. 
The increase of sensitivity in sub-mm astromomy has also permited tremendous progress
in recent years with detailed studies of the relation between molecular content and star formation at intermediate redshifts
\citep[e.g.][]{Tacconi13}, although still limited to relatively bright and massive galaxies. In addition,
observations of atomic gas through \HI\ 21-cm emission (currently limited to $z<0.4$, e.g. \citealt{Catinella08, Freudling11, Fernandez16})
will have to await future radio facilities such as the Square Kilometre Array.

At high redshift, information about gas in the Universe can be accurately obtained through absorption studies towards bright
background sources. In particular, damped Lyman-$\alpha$ systems (DLAs, see \citealt{Wolfe05} for a review), 
with $N(\HI)\ge 2\times 10^{20}$~\cmsq, trace the neutral gas in a cross-section weighted manner, independently of the
luminosity of the associated object.
DLAs have been conjectured to be originating from gas associated with galaxies, in particular since DLAs contain the bulk
of the neutral gas at high redshift \citep[e.g.][]{Prochaska05,Prochaska09, Noterdaeme09, Noterdaeme12c} and their metallicity
is increasing with decreasing redshift \citep[e.g.][]{Rao06,Rafelski12}.
While the dust production in the bulk of DLAs seems to be very low \citep{Murphy16}, the excitation of atomic and molecular
species indicate some ongoing star-formation activity \citep[e.g.][]{Wolfe04,Srianand05,Neeleman15}.
This is also suggested by numerical simulations \citep[e.g.][]{Cen12, Bird14} or semi-analytical models \citep[e.g.][]{Berry16} but association
with galaxies remain difficult to establish directly, with only a few associations between intervening DLAs and galaxies have been
revealed so far at $z>2$ 
\citep[][]{Moller93, Moller04, Fynbo10, Krogager12, Noterdaeme12, Bouche13, Kashikawa14, Hartoog15, Srianand16}. Indeed, statistical studies show a low level of
in-situ star formation \citep{Rahmani10, Fumagalli15}, although Ly-$\alpha$ emission has been detected through stacking
in sub-samples with the highest \HI\ column densities \citep{Noterdaeme14}, suggesting the latter arise more likely from gas
associated with galaxies at small impact parameters.

\citet{Noterdaeme15b} suggested that H$_2$ is more frequently found in high column density DLAs, but that the measured 
overall molecular fraction remains much lower than what would be expected from single clouds, even at the typically low
metallicities of DLAs. This indicates that most of the observed H\,{\sc i} column density along the line of sight is actually unrelated to the
H$_2$ core and does not participate in its shielding \citep[see also][]{Noterdaeme15}.
This again marks the difficulty of distinguishing the H\,{\sc i} envelope of molecular clouds from unrelated atomic
gas along the same line of sight.
Several methods have been developped to statistically derive the CNM fraction in DLAs. The low detection
rate of 21-cm absorption in DLAs indicates high average spin temperatures and hence that most DLAs are
dominated by WNM \citep[e.g.][]{Kanekar14}. \citet{Neeleman15} recently suggested that the bulk of neutral gas
could be in the CNM for at least 5\% of DLAs, based on the fine-structure excitation of singly ionised carbon and silicon.
This further indicates that such clouds can be as small as a few parsecs. A small size of CNM clouds is also inferred
from the lack of correspondance between 21-cm and H$_2$ absorption seen in DLAs \citep[][]{Srianand12} and by
the partial coverage of the background quasar's broad line region by H$_2$-bearing clouds \citep[e.g.][]{Balashev11}.

Because H$_2$-bearing systems are rare among the overall DLA population \citep[e.g.][]{Ledoux03,Noterdaeme08,Jorgenson14},
directly targeting H$_2$ (instead of blindly targeting H\,{\sc i} gas) could provide a more efficient way
to study the phase transition. Unfortunately, H$_2$ lines are located in the \lya\ forest and difficult to detect at low
spectral resolution \citep[except when the absorption is in the damped regime,][]{Balashev14}. In turn, neutral carbon provides an excellent tracer
of H$_2$ molecules \citep[e.g.][]{Snow06}, since the ionisation energy of \CI\ is close to that of H$_2$ photodissociation.
Furthermore, several transitions are located out of the \lya\ forest, making it possible to
search for strong C\,{\sc i} absorption even at low spectral resolution \citep[see][]{Ledoux15}. Such selection has led
to the first detections of CO molecules in absorption at $z>1.6$, which also opened the exciting possibility to
directly measure the cosmic microwave background (CMB) temperature through the excitation of CO \citep{Noterdaeme11}.
In the two high redshift cases where H$_2$ lines were
also covered, we measured overall molecular fractions of about 25\% \citep[][]{Srianand08,Noterdaeme10b}, i.e. significantly
higher than in other H$_2$-bearing DLAs, that generally have $f\sim 1\%$ or less \citep{Ledoux03}.

In our quest for molecular-rich systems in the Sloan Digital Sky Survey-III
\citep[see][for the corresponding search in the SDSS-II]{Ledoux15}, we found a new case, at $\zabs\sim 2.5$ towards the quasar SDSS\,J000015.17$+$004833.3 (hereafter \jz) 
with strong \CI\ absorption and prominent 2175~{\AA} bump that we followed-up with the Very Large Telescope. The characteristics
of this system in terms of molecular fraction, CO column density, and metallicity supersede all values measured
in DLAs so far.
A cold, molecular-bearing component is clearly identified, allowing us to perform an unprecedented detailed analysis
of the chemical and physical conditions in the molecular cloud and to study the transition from the atomic to the molecular phase. 
We present our observations in Sect.~\ref{s:obs}, the absorption-line
analysis of ionic, atomic and molecular species in Sect.~\ref{s:abs}. We discuss the metallicity and dust abundance in Sect.~\ref{s:metallicity}, the extinction in Sect.~\ref{s:ext} and 
%We discuss
the physical conditions in the cloud in Sect.~\ref{s:phys}. We use CO to measure the
cosmic microwave background temperature at $z=2.53$ in Sect.~\ref{s:cmb}. We search for star-formation activity in Sect.~\ref{s:sfr} and
conclude in Sect.~\ref{s:concl}.

\section{Observations and data reduction \label{s:obs}}

\subsection{UVES}

High-resolution spectroscopic observations of \jz\ ($z_{\rm em} \approx 3.03$) were carried out using the Ultraviolet and Visual Echelle Spectrograph
\citep[UVES;][]{Dekker00} mounted on the unit 2 of the 8.2~m very large telescope (VLT) at Paranal observatory under two distinct
ESO programmes 093.A-0126(A) in 2014 (P93) and 096.A-0354(A) in 2015 (P96). The former observations were all
performed using the standard beam splitter with 390+564 setting at a slit width of $0.9\arcsec$.

The latter (P96) were mostly performed with the same setting but with a narrower slit width of $0.7\arcsec$ in the red arm, and
attached Th-Ar calibration.
We also observed the quasar 2$\times$4200\,s with central wavelength set to 760~nm in the red arm in order to extend the spectral
coverage over the redshifted absorption
position of useful metal species (\ZnII, \FeII). These were taken with a $0.9\arcsec$-wide slit.
We used a CCD readout with 2$\times$2 binning and set the slit position to paralactic angle for all the observations to
minimize the effects of atmospheric dispersion.  A summary of the observations is shown in Table~\ref{t:log}.

The data were reduced using UVES Common Pipeline Library (CPL) data reduction pipeline release
6.5 using an optimal extraction \citep{Horne86}.
We use 4$^{\rm th}$ order polynomials to find the dispersion solution. The individual science exposures
were shifted to the heliocentric-vacuum frame correcting for the observatory’s motion towards the line of sight at the exposure
mid point, using the air-to-vacuum relation from \citet{Ciddor96}.

All exposures taken with the BLUE arm were obtained using a $0.9''$-wide slit and extracted onto a
fixed wavelength grid with a pixel step of 2.5~\kms\ that corresponds to the pixel size on the CCD. 
The spectrum of each echelle order was interpolated
onto this global grid so that no further rebinning was required neither when merging orders of an exposure
nor when combing different exposures. Similarly, exposures taken with the RED arm have higher resolution and
smaller pixel sizes and were extracted on a grid with a pixel step of 2.0~\kms.

Cosmic ray residuals and bad pixels were flagged using a semi-interactive procedure and the data quality was checked
to remove a few failed exposures. Individual 1D extractions were then scaled and combined together into three final 1D
spectra: a ``blue'' spectrum, with
spectral resolution 6.30~\kms; a ``red'' spectrum with resolution ranging from 5.45 to 5.80~\kms\ corresponding to
all exposures taken with $0.9\arcsec$ slit and a higher resolution ``red'' spectrum, with resolution 4.60-4.70~\kms, combining
the $0.7\arcsec$-wide slit exposures. The average S/N per pixel is about 8 at 4000~{\AA} in the blue spectrum. The combined red spectrum has in turn S/N$\sim$20 at 5300~{\AA}.

\subsection{X-Shooter}
Deep, medium-resolution spectroscopic observations of \jz\ over the full wavelength range from $\sim$320\,nm to
$\sim 2.25$\,$\mu$m were carried out at the VLT unit 3 using X-Shooter under ESO programme 096.A-0924(B). We performed all observations
using the Nodding mode and slit widths of 1.3\arcsec, 0.9\arcsec\ and 1.2\arcsec\ for the UVB, VIS and NIR arm, respectively, and a binning
of 1$\times$2.
We used different slit position angles to maximise the spatial coverage around the quasar location. The log of observations is shown in
Table~\ref{t:log}.

Our X-Shooter data reduction heavily relies on the pipeline supplied by ESO in its version
2.5.2 \citep{Modigliani10}. For every position angle, we used the pipeline to apply
a flat-field correction, order tracing and rectification of individual frames in each nodding
position individually. In the UVB and VIS arm, the sky spectrum was subtracted using regions in the
$11\arcsec$-long X-Shooter slit free of signal. In the NIR, the intensity of the sky spectrum is high
so we used the frame taken at the alternate nodding
position closest in time for background subtraction. Wavelength and flux calibration were 
performed using arc lamp lines observed during daytime and the nightly spectrophotometric
standard, respectively.

This process provided us with sky-subtracted, wavelength- and flux-calibrated 2D spectra. After cosmic-ray and bad-pixel detection using our own algorithm based on
Laplacian-filter edge detection, we averaged these frames with variance weighting. This
yielded a single frame per arm and position angle. We then obtained the 1D spectra by optimal extraction, where the appropriate weights along the spatial direction
were derived using a Moffat-function fitted to the data. Finally, the spectra were corrected
for Galactic foreground reddening \citep{Schlafly11} and converted into a vacuum heliocentric reference frame.
We checked the quality of the flux calibration by comparing the spectra against each other and found
agreement within 15\% in the UVB (and better in the other arms). We also found no evidence for significant
chromatic slit losses in the data by comparing with accurate multi-band photometry (see Sect.~\ref{s:ext}).

\begin{figure}
  \centering
  \includegraphics[bb=65 175 490 400, clip=,width=\hsize]{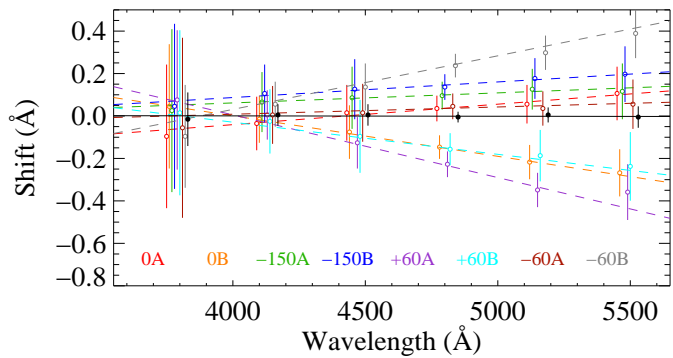}  
  \caption{X-Shooter wavelength distortion in the UVB arm for the different exposures. Points with different
    colours represent averages in 200~{\AA} chunks for different PAs (whose names, given in the bottom, correspond
    to the angle in degrees East of North, and ``A/B'' for first/second observation with same PA). The black filled
    dots correspond to the combined spectrum.} \label{xsdist}
\end{figure}

Because X-Shooter sits at the Cassegrain focus of the VLT, shifts in the wavelength solution by about 0.5~{\AA}
are not uncommon due to flexure \citep[e.g.][]{Bristow11}. This is also expected since the 
observations were performed off the parallactic angle. 
Thanks to UVES observations
of the same object, we indeed noticed and corrected for a wavelength distortion in the UVB arm: We smoothed
the UVES spectrum to X-Shooter's spectral resolution and cross-correlated the resulting spectrum with the X-Shooter data over
200~{\AA} chunks. The wavelength distortion is different for each observation but can be very well approximated by
a linear function of the wavelength \citep[see also][]{Chen14}. We thus corrected for this distortion before combining the individual
1D exposures, improving the wavelength calibration accuracy by a factor of more than 10 compared to the pipeline results
(Fig.~\ref{xsdist}) without losing spectral resolution in the final spectrum due to blurring effect. The combined X-Shooter spectrum has S/N$\sim$ 35, 55 and 25 and resolution around 75, 32 and 35~\kms\ in the UV, visual and near infrared, respectively.

\begin{table*}
  \centering
  \caption{Log of observations \label{t:log}}
  \begin{tabular}{c c c c c}
    \hline \hline
    Program ID              & Setting/Mode  &  Slit widths      &  Observing dates & Exposure time \\
                            &               &  (arcsec)         &                  &  (s)          \\
    \hline
        \multicolumn{5}{c}{UVES}\\
    093.A-0126(A)           & 390$+$564   &  0.9, 0.9    &  Aug 2014         & 5$\times$4800\,s     \\
    096.A-0354(A)           & 390$+$564   &  0.9, 0.7    &  Oct-Nov 2015     & 6$\times$4200\,s     \\
    096.A-0354(A)           & 437$+$760   &  0.9, 0.9    &  Nov 2015         & 2$\times$4200\,s     \\
    \hline
    \multicolumn{5}{c}{X-shooter}\\
    096.A-0924(B)           & Nodding     & 1.3, 0.9, 1.2 & Sep-Dec 2015, Aug 2016     & 8$\times$2$\times$(1400, 1430, $3\times$480)   \\
    \hline
  \end{tabular}
  \tablefoot{
      The different values for slit width correspond to different arms, ordered by increasing wavelength: BLUE, RED for UVES and
      UVB, VIS, NIR for X-shooter. The exposure times for Xshooter data are detailled as (number of OB) $\times$ (nodding positions) $\times$
      (exposure time for a given position), with the NIR being sub-divided in 3 integrations (NDIT).
    }
  \end{table*}

\section{Absorption line analysis \label{s:abs}}

\begin{figure}
  \centering
  \begin{tabular}{c}
    \includegraphics[bb=209 20 400 780, clip=,angle=90, width=\hsize]{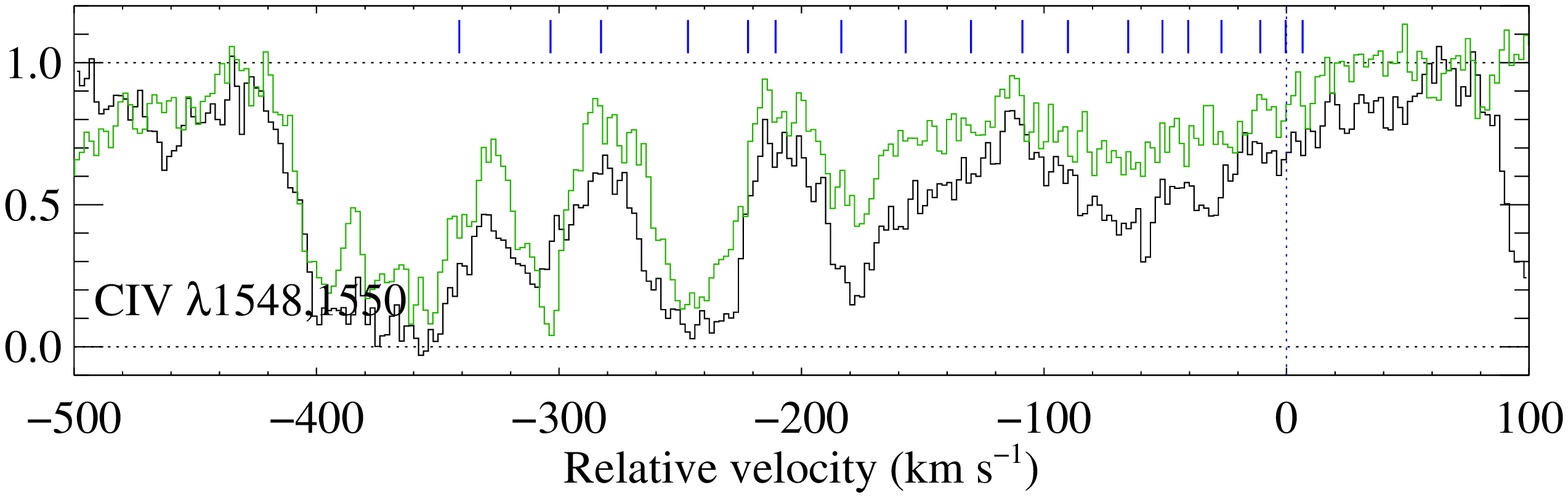}\\
        \includegraphics[bb=150 20 400 780, clip=,angle=90, width=\hsize]{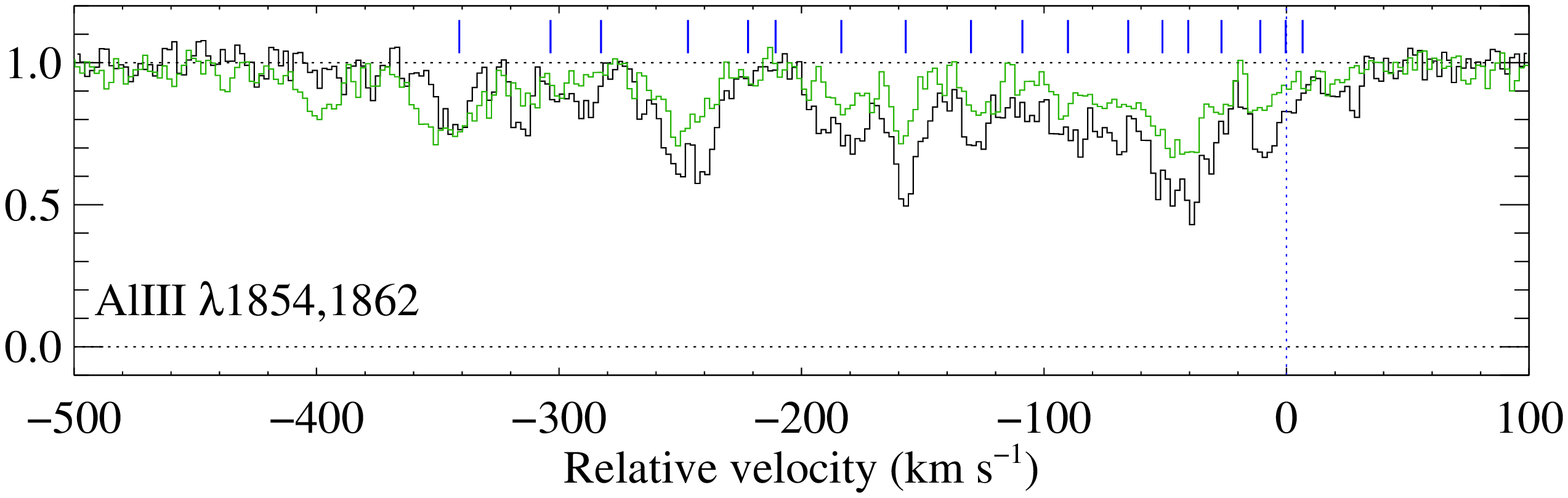}\\
  \end{tabular}
  \caption{High ionisation species detected towards \jz\ in the UVES spectrum. The green profile corresponds the reddest transition for each doublet. The ticks marks correspond to the location of singly ionised metal components,
    and the zero-velocity scale is set at the position of the molecular component. In this, and in all analoguous figures presenting
    absorption lines, the y-axis shows the normalised flux.\label{f:kin}}
\end{figure}

We detect metal absorption lines in the $\zabs=2.525$ absorption system from various ionisation stages from high-ionisation species (e.g. \CIV, \AlIII, see Fig.~\ref{f:kin}) and
singly ionised species (\SiII, \NiII, \ZnII\ and \FeII) spread over about 400~\kms. We detect a narrow component
at the extreme red edge of the profile in which we also detect neutral (\SI, \MgI, \ClI, \CI) and molecular
(H$_2$, CO) species. While the overall kinematic profile is interesting --with a large velocity extent and
absorption strengths varying differently with velocity for different ions, 
suggesting galactic winds-- we mostly focus
on the molecular component in this paper. 
We use {\sc vpfit} \citep{Carswell14} version 10.3 to model the absorption
profiles using multi-component Voigt-profile fitting in order to obtain redshifts, Doppler parameters and column
densities of different species. 

During our analysis, we combine the two red UVES spectra into a single spectrum using an inverse variance weighting.
In principle, the resulting spectral point spread function (SPSF) becomes the
combination of the two Gaussian SPSF as done by \citet{Carswell12}. In practice, because
we are using data from the same instrument with resolutions that differ by
only 20\%, the resulting SPSF can very well be approximated by a single Gaussian 
with resolution ranging from 5 to 5.25~\kms\ over the region covered by both original spectra. We checked
that fitting both red spectra simultaneously or using their combination provides consistent results and therefore provide
here the results using the combined spectrum. For the particular case of CO, we test this more in detail and also provide the
simultaneous fit to the two sets of UVES data.

\subsection{Atomic hydrogen}

\begin{figure}
\centering
\includegraphics[bb=150 220 400 650, clip=,angle=90, width=\hsize]{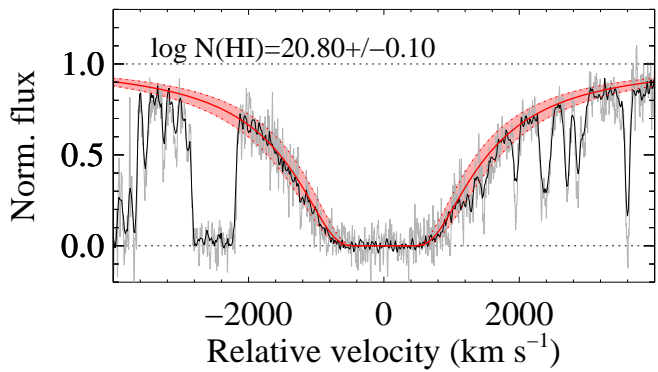} 
\caption{Measurement of the \HI\ column density at $\zabs=2.525$. The X-Shooter data is represented in black, while the UVES data,
  boxcar-smoothed by 5 pixels is represented in grey. The Voigt profile fit is shown in red with the associated uncertainty as a shaded region. \label{f:HI}}
\end{figure}

We determine the \HI\ column density of the system by simultaneously fitting a Voigt profile to the
damped \lya\ absorption line and the continuum of the background quasar. Higher-order Lyman lines
are not usable because of blending with stronger damped H$_2$ lines (see Sect.~\ref{s:H2}).
We use the X-Shooter spectrum since it has a much higher signal-to-noise ratio (S/N) than the UVES spectrum in
this region, and it is flux calibrated, making easier to determine the continuum placement. We obtain $\log N(\HI)=20.8 \pm 0.1$.
As expected, the UVES data is also consitent with this value, see Fig.~\ref{f:HI}.

\subsection{Metals \label{s:met}}

About twenty velocity components with a wide dynamical range of optical depths are distinctly identified in the profiles of
\SiII\ and \FeII\ thanks to several transitions spanning a range of oscillator strengths.
We use these species together with \SI\ (whose 1807~{\AA} transition is blended with \SiII$\lambda$1808) 
to obtain a first solution for the component structure. We then include \NiII\ and \FeII\ and let the column densities vary
freely while the Doppler parameters and redshifts are tied together for singly ionised species. The redshift and Doppler
parameters for neutral species (\SI\ and \MgI) are kept independent.

\begin{figure*}
  \centering
\begin{tabular}{cc}

  \includegraphics[bb=209 20 400 780, clip=,angle=90, width=0.45\hsize]{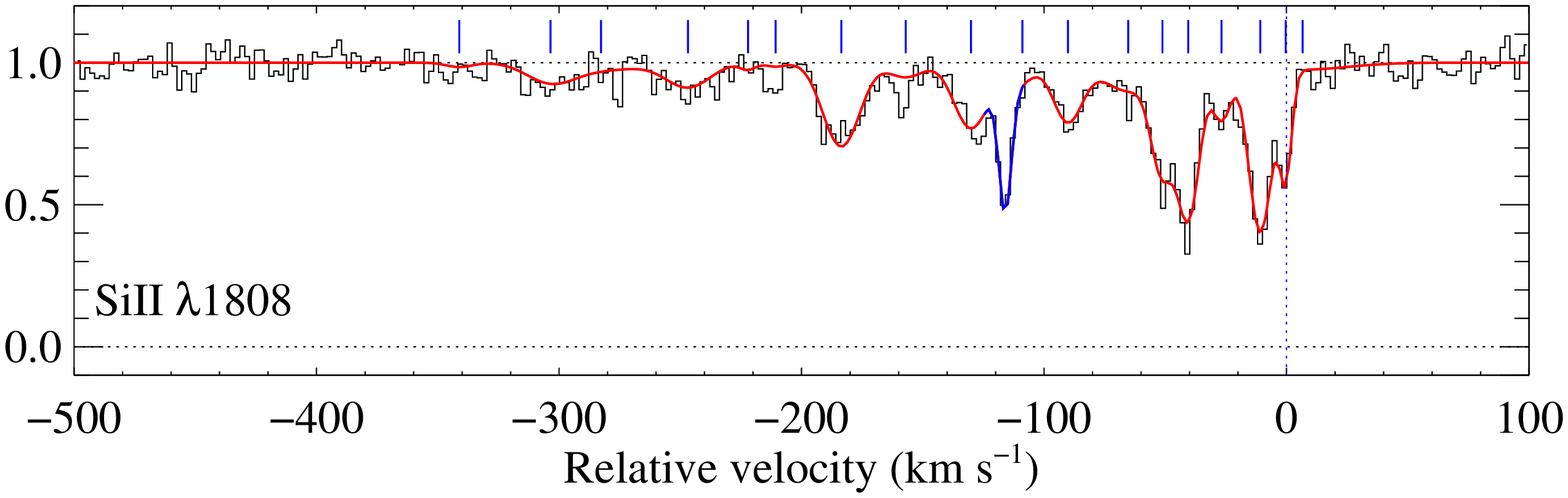}&
  \includegraphics[bb=209 20 400 780, clip=,angle=90, width=0.45\hsize]{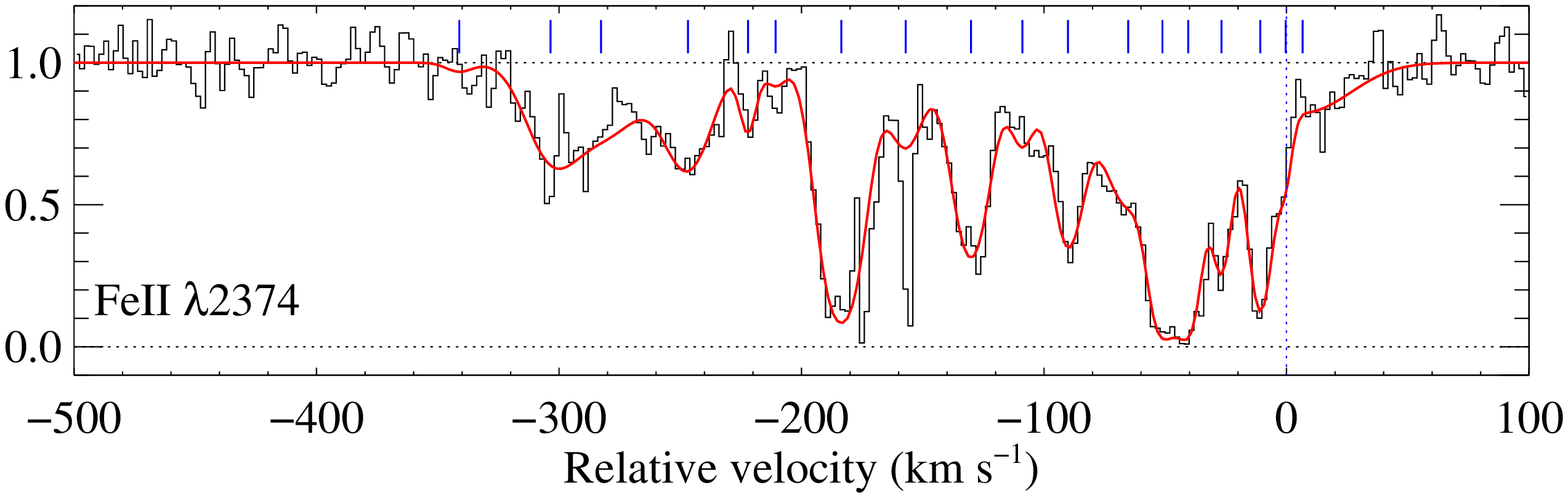}\\[-6pt]

  \includegraphics[bb=209 20 400 780, clip=,angle=90, width=0.45\hsize]{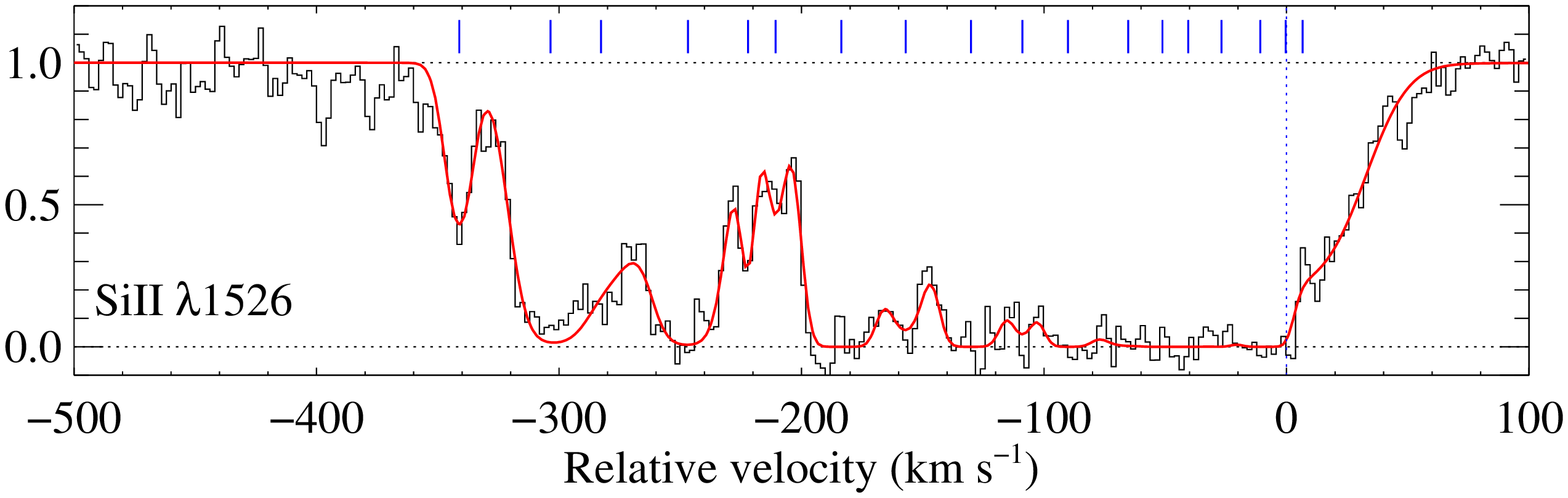}&
  \includegraphics[bb=209 20 400 780, clip=,angle=90, width=0.45\hsize]{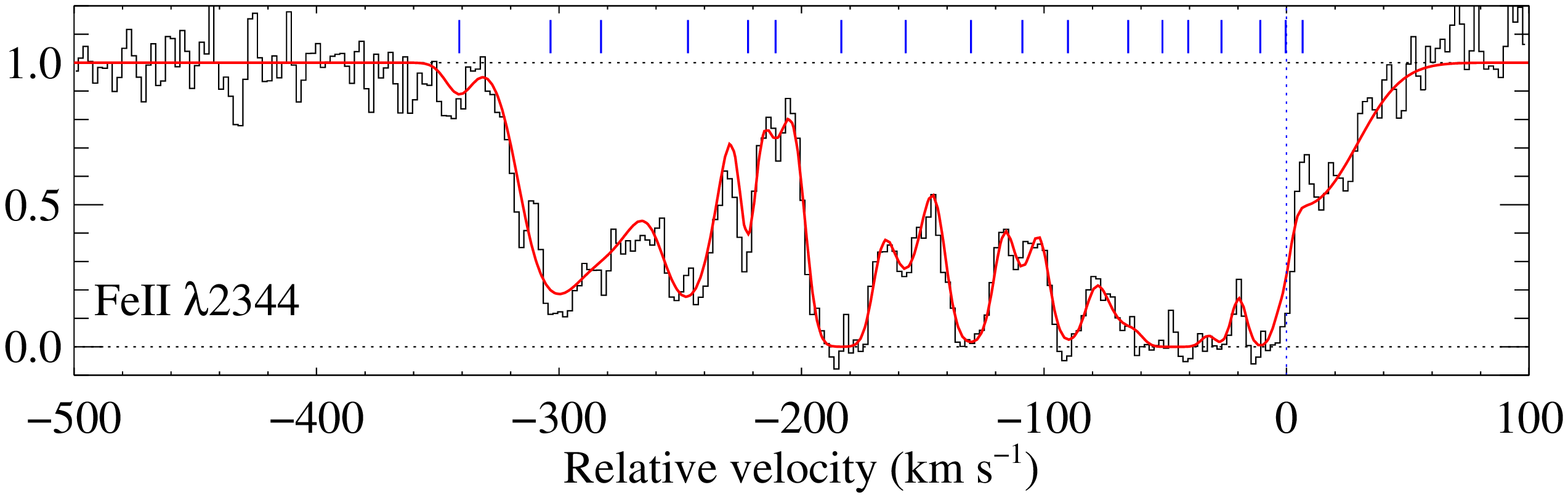}\\[-6pt]

  \includegraphics[bb=209 20 400 780, clip=,angle=90, width=0.45\hsize]{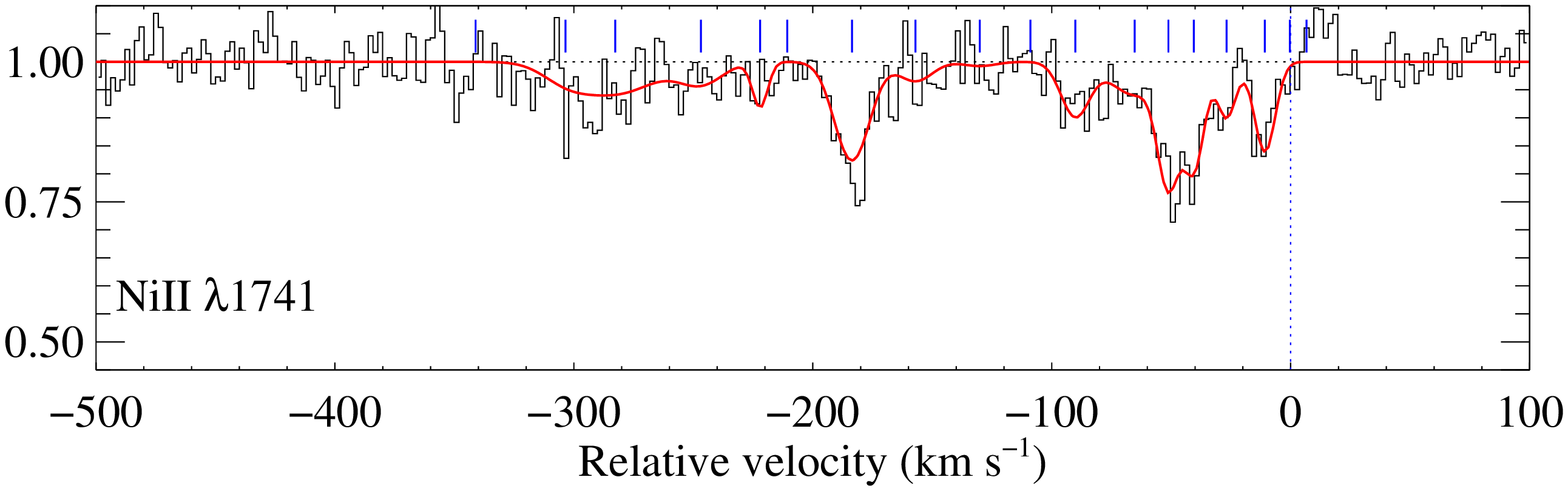}&
  \includegraphics[bb=209 20 400 780, clip=,angle=90, width=0.45\hsize]{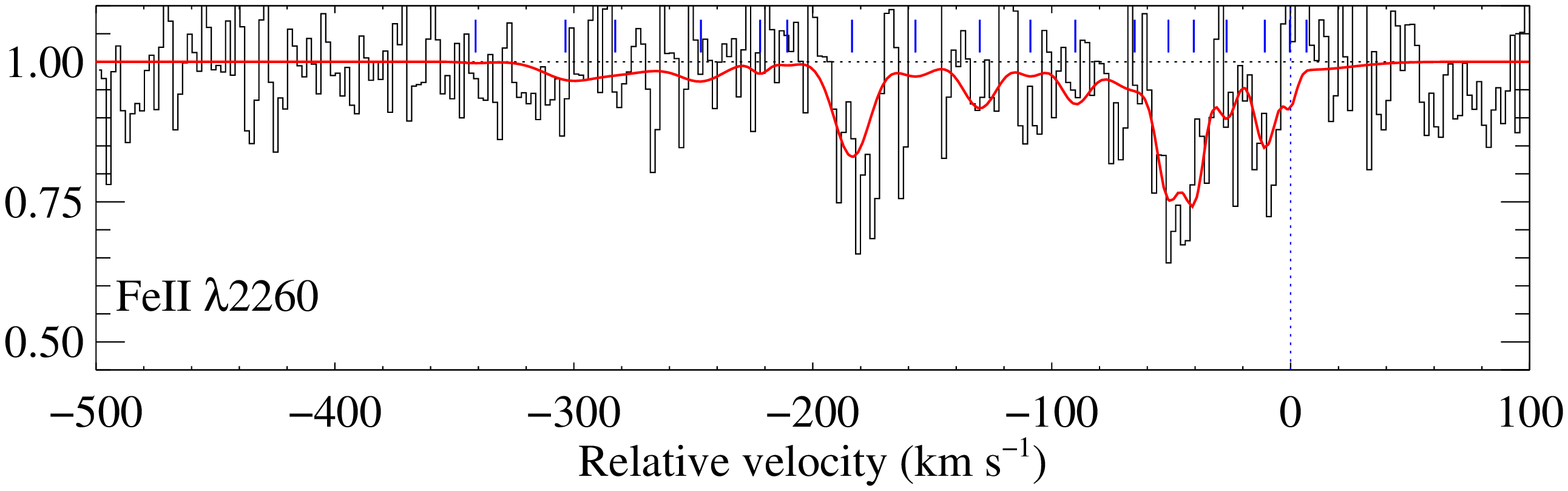}\\[-6pt]

  \includegraphics[bb=209 20 400 780, clip=,angle=90, width=0.45\hsize]{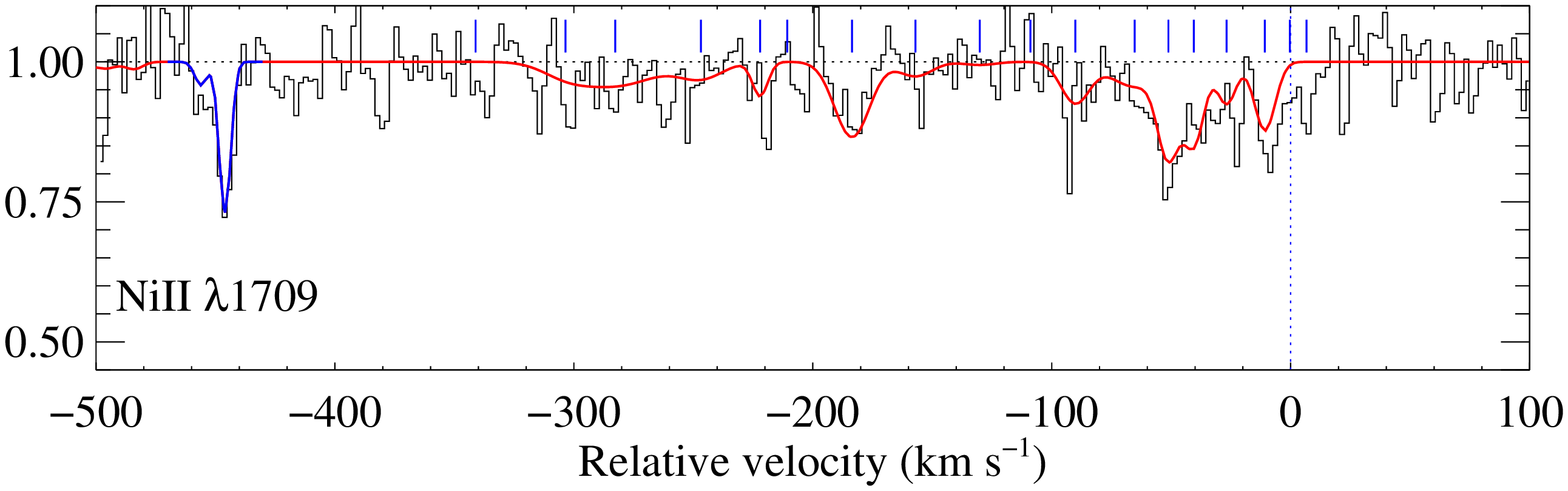}&
  \includegraphics[bb=209 20 400 780, clip=,angle=90, width=0.45\hsize]{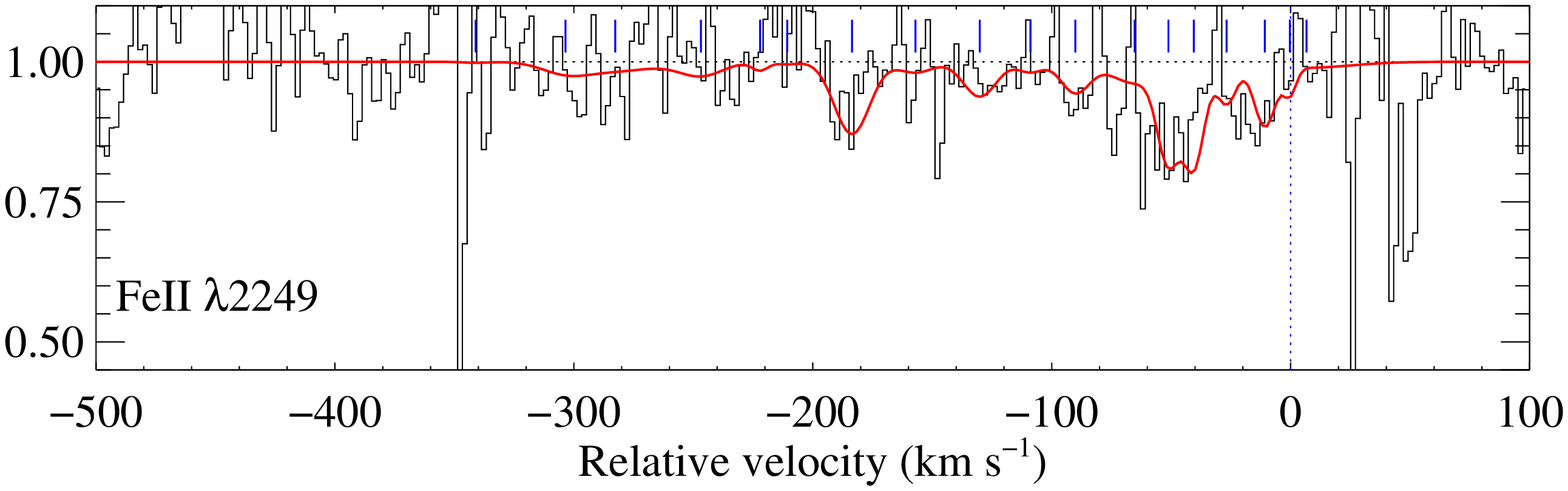}\\[-6pt]

  \includegraphics[bb=209 20 400 780, clip=,angle=90, width=0.45\hsize]{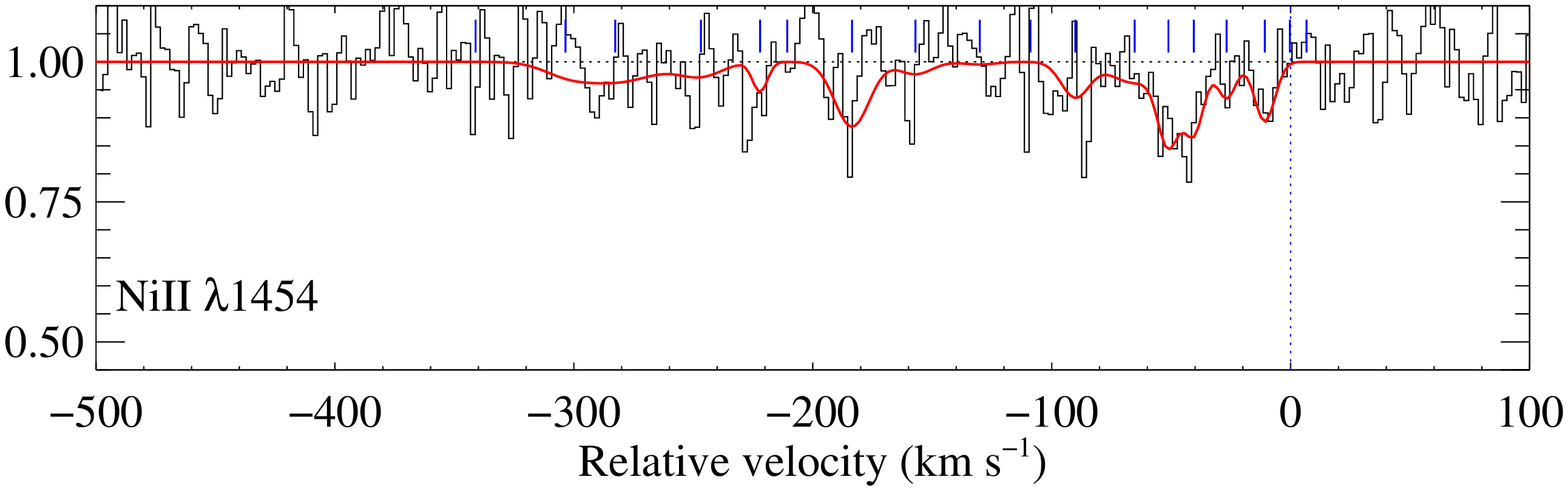}&
  \includegraphics[bb=209 20 400 780, clip=,angle=90, width=0.45\hsize]{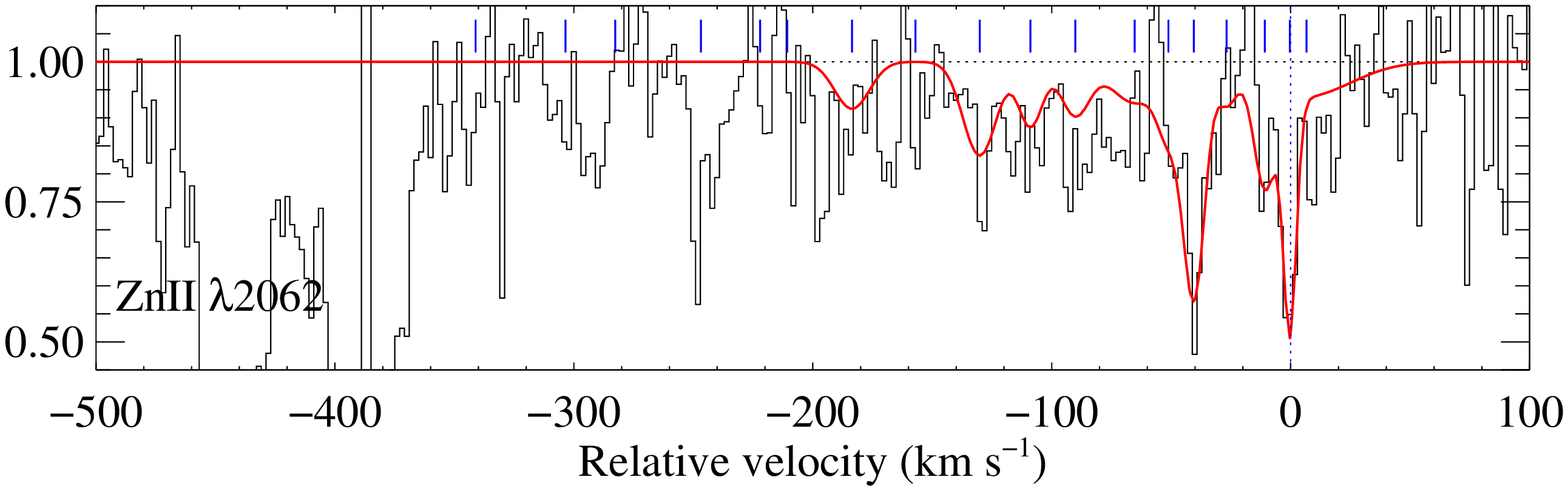}\\[-6pt]

  \includegraphics[bb=150 20 400 780, clip=,angle=90, width=0.45\hsize]{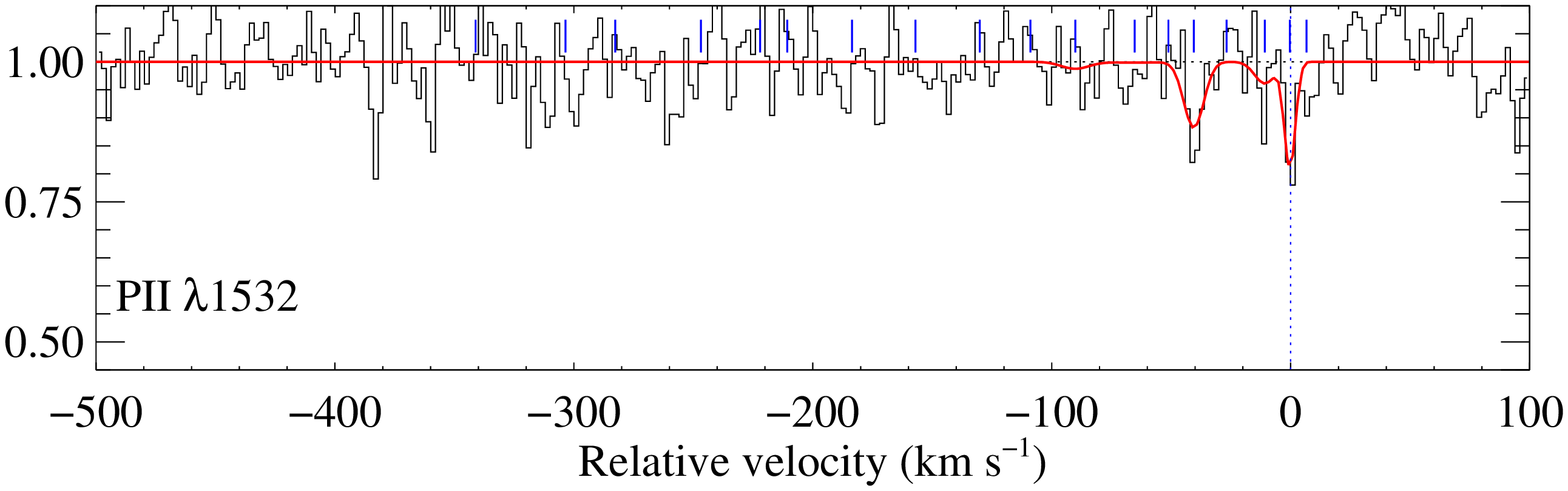}& 
  \includegraphics[bb=150 20 400 780, clip=,angle=90, width=0.45\hsize]{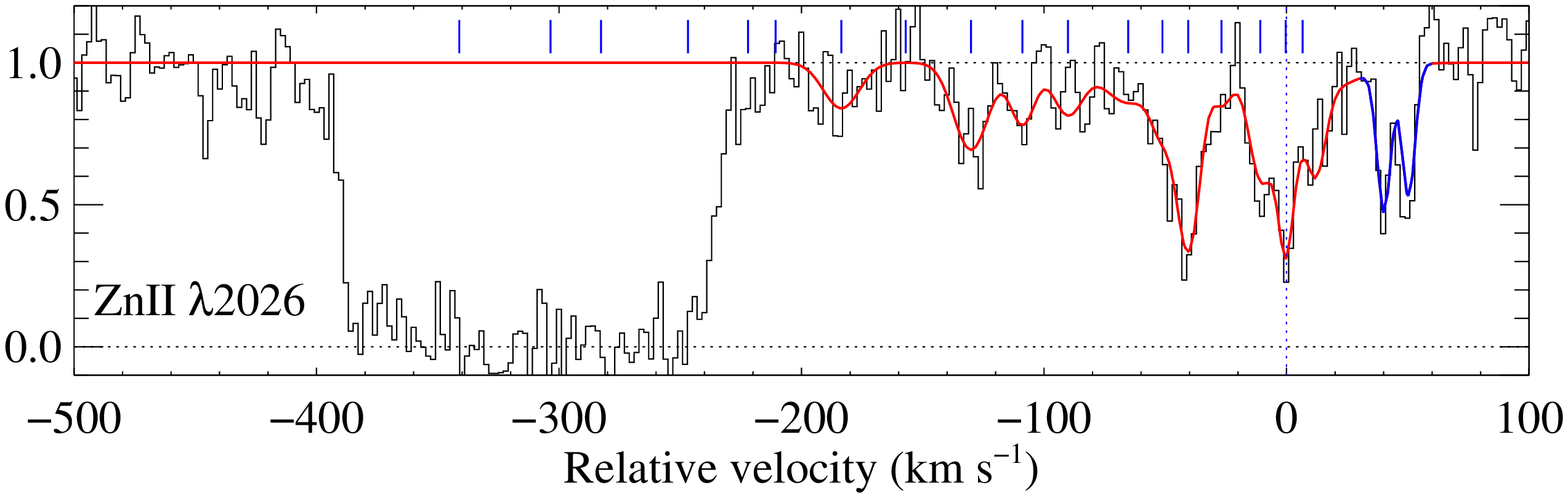}\\[-6pt]

\end{tabular}
\caption{Multi-component Voigt-profile fit to singly ionised species (red), overlayed on top of the normalised UVES data.
  Regions marked in blue correspond to species
  other than labelled: e.g. the blue region in \SiII$\lambda$1808 is due to absorption by \SI$\lambda$1807. \label{f:metalsII}}
\end{figure*}

A very narrow component ($b \approx 0.6~\kms$) corresponding to the neutral and molecular species is clearly seen at the extreme red edge of the profiles of \SiII$\lambda$1808 and \ZnII$\lambda\lambda$2026,2062 while much weaker in \FeII\ lines and not
detected at all in \NiII\ (see the component at $v=0$~\kms in Fig.~\ref{f:metalsII}). This already indicates a high level of dust depletion since the later species are refractory while
zinc is a volatile element \citep[e.g.][]{Pettini97}. We use this narrow component ($z=2.52546$) as the reference for the zero-velocity in all figures and discussions in the paper. We also note that we do not make any assumption on the velocity
structure (e.g. Doppler parameter) of this component and fit the molecular, atomic, and ionic species independently. 
The results from fitting the lines are also shown in Fig.~\ref{f:metalsII} and the corresponding parameters provided in Table~\ref{t:metalsII}.
We measure total column densities of respectively $\log N$(cm$^{-2}$) = 15.93$\pm$0.17 (\SiII), 13.99$\pm$0.04 (\NiII), 14.09$\pm$0.45 (\ZnII)
and 15.14$\pm$0.03 (\FeII). 
  We note that the \ZnII$\lambda$2062 line could be blended with \CrII\ absorption
(\ZnII$\lambda$2026 as well, but the nearby \CrII\ line has very low oscillator strength). However, we do not
detect the unblended \CrII$\lambda\lambda$2056,2066 lines despite strong oscillator strengths. The effect of chromium
on the measurement of $N(\ZnII)$ should therefore be largely negligible.

  We also detect \PII$\lambda$1532 (\PII$\lambda$1301 is unfortunately lost within the saturated \OI$\lambda$1302 profile)
in the two strongest components, although close to the noise level. We therefore fixed the
redshifts and Doppler parameters to the values determined from other metals and obtain $\log N(\PII) \sim 14.5\pm0.1$. 
Finally, \SII\ lines are detected but redshifted in the \lya\ forest. While two of them ($\lambda\lambda1250,1253$) are not severely blended, their oscillator strengths are similar. This, and the low signal-to-noise
ratio achieved in this region of the UVES spectrum prevent us from getting meaningful constraints through line fitting, in particular for the strong narrow component.
However, we checked that the data is consistent with the expected profile assuming a Solar zinc-to-sulphur ratio in the gas phase.

%Automatically generated using results insynth79 on Mon Jun 20 15:38:24 2016
%N(ZnII) *warm*:       13.172322     0.043992910
\begin{table*}
\centering
\caption{Result of Voigt profile fitting to singly ionised metal lines. \label{t:metalsII}}
\begin{tabular}{c c c c c c c}
\hline \hline
{\large \strut}
$z_{\rm abs}$ & $v$ (\kms) & $b$ (km\,s$^{-1}$) & $\log N($SiII)       & $\log N($NiII)       & $\log N($ZnII)       & $\log N($FeII)       \\
\hline         
2.521448 & -341 &   6.64$\pm$ 0.91 &  13.31 $\pm$ 0.04  &                   &                   & 12.34 $\pm$ 0.08 \\
2.521891 & -303 &  13.26$\pm$ 1.61 &  14.21 $\pm$ 0.08  & 12.47 $\pm$ 0.54  &                   & 13.59 $\pm$ 0.19 \\
2.522135 & -283 &  22.62$\pm$ 4.51 &  14.03 $\pm$ 0.15  & 13.08 $\pm$ 0.16  &                   & 13.79 $\pm$ 0.13 \\
2.522557 & -247 &  12.26$\pm$ 0.68 &  14.31 $\pm$ 0.05  & 12.65 $\pm$ 0.20  &                   & 13.72 $\pm$ 0.03 \\
2.522848 & -222 &   3.02$\pm$ 0.44 &  13.28 $\pm$ 0.07  & 12.54 $\pm$ 0.15  &                   & 13.08 $\pm$ 0.04 \\
2.522982 & -211 &   4.49$\pm$ 0.98 &  13.14 $\pm$ 0.06  &                   &                   & 12.64 $\pm$ 0.06 \\
2.523301 & -184 &   9.52$\pm$ 0.28 &  14.80 $\pm$ 0.03  & 13.25 $\pm$ 0.05  & 12.07 $\pm$ 0.15  & 14.36 $\pm$ 0.03 \\
2.523613 & -157 &   9.16$\pm$ 1.08 &  13.96 $\pm$ 0.05  & 12.49 $\pm$ 0.24  &                   & 13.50 $\pm$ 0.04 \\
2.523930 & -130 &   8.89$\pm$ 0.49 &  14.64 $\pm$ 0.04  & 11.81 $\pm$ 1.09  & 12.36 $\pm$ 0.07  & 14.00 $\pm$ 0.02 \\
2.524179 & -109 &   6.08$\pm$ 1.25 &  13.84 $\pm$ 0.10  &                   & 12.06 $\pm$ 0.12  & 13.34 $\pm$ 0.06 \\
2.524400 & -90 &   7.58$\pm$ 0.92 &  14.53 $\pm$ 0.07  & 12.88 $\pm$ 0.10  & 12.04 $\pm$ 0.14  & 13.89 $\pm$ 0.05 \\
2.524692 & -65 &  13.37$\pm$ 6.16 &  14.41 $\pm$ 0.24  & 12.88 $\pm$ 0.25  & 12.14 $\pm$ 0.23  & 13.94 $\pm$ 0.21 \\
2.524858 & -51 &   5.25$\pm$ 1.32 &  14.72 $\pm$ 0.14  & 13.12 $\pm$ 0.14  & 12.01 $\pm$ 0.31  & 14.29 $\pm$ 0.13 \\
2.524983 & -41 &   4.93$\pm$ 0.77 &  14.93 $\pm$ 0.06  & 13.05 $\pm$ 0.10  & 12.65 $\pm$ 0.06  & 14.31 $\pm$ 0.09 \\
2.525145 & -27 &   4.31$\pm$ 0.76 &  14.33 $\pm$ 0.09  & 12.71 $\pm$ 0.12  & 11.70 $\pm$ 0.26  & 13.85 $\pm$ 0.05 \\
2.525333 & -11 &   5.06$\pm$ 0.43 &  14.99 $\pm$ 0.03  & 12.98 $\pm$ 0.07  & 12.25 $\pm$ 0.12  & 14.08 $\pm$ 0.05 \\
2.525456 & 0   &   0.59$\pm$ 0.11 &  15.55 $\pm$ 0.41  &                   & 14.03 $\pm$ 0.52  & 13.51 $\pm$ 0.43 \\
2.525538 & +7  &  26.22$\pm$ 2.94 &  14.06 $\pm$ 0.08  &                   & 12.33 $\pm$ 0.25  & 13.66 $\pm$ 0.07 \\
   Total &  &                  &  15.93 $\pm$ 0.17  & 13.99 $\pm$ 0.04  & 14.09 $\pm$ 0.45  & 15.14 $\pm$ 0.03  \\
\hline
\end{tabular}
\end{table*}

\begin{figure}
  \includegraphics[bb=120 20 450 780, clip=,angle=90, width=\hsize]{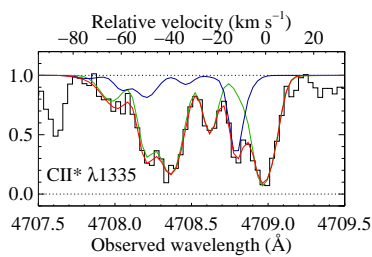}
  \caption{Fit to the \CII$^*\lambda$1335 absorption profile (UVES data). The blue and green curves correspond to the
    contribution from each transition of the doublet. The top axis shows the relative velocity corresponding to the strongest transition (green, with $\lambda=$~1335.7077~{\AA} rest-frame). \label{f:CIIs}}
  \end{figure}

We detect \CII\ lines in the UVES spectrum although these are saturated and do not provide any meaningful constraint on the column
  density of ionised carbon. In turn, the \CII$^*\lambda$1335 fine-structure doublet lines  
are not apparently saturated as can be appreciated from Fig.~\ref{f:CIIs}. $\CII^*\lambda$1037 is unfortunately completely blended with intervening \lya\ forest absorption.
Measuring the corresponding column densities remains therefore hazarduous 
due to the many components overlapping in the doublet and because the velocity decomposition differs from that of other metals.
In addition, the strongest component is on the non-linear part of the curve of
growth. 
We obtain $\log N(\CII^*)\sim 15$ in that component, with $b\sim 0.7$~\kms. These values
should be considered with great caution as the fit is sensitive to the initial guess,
  leading to uncertainties larger than an order of magnitude.

\subsection{Neutral carbon}

\begin{figure}
\centering
\begin{tabular}{c}
\includegraphics[bb=150 20 400 780, clip=,angle=90, width=\hsize]{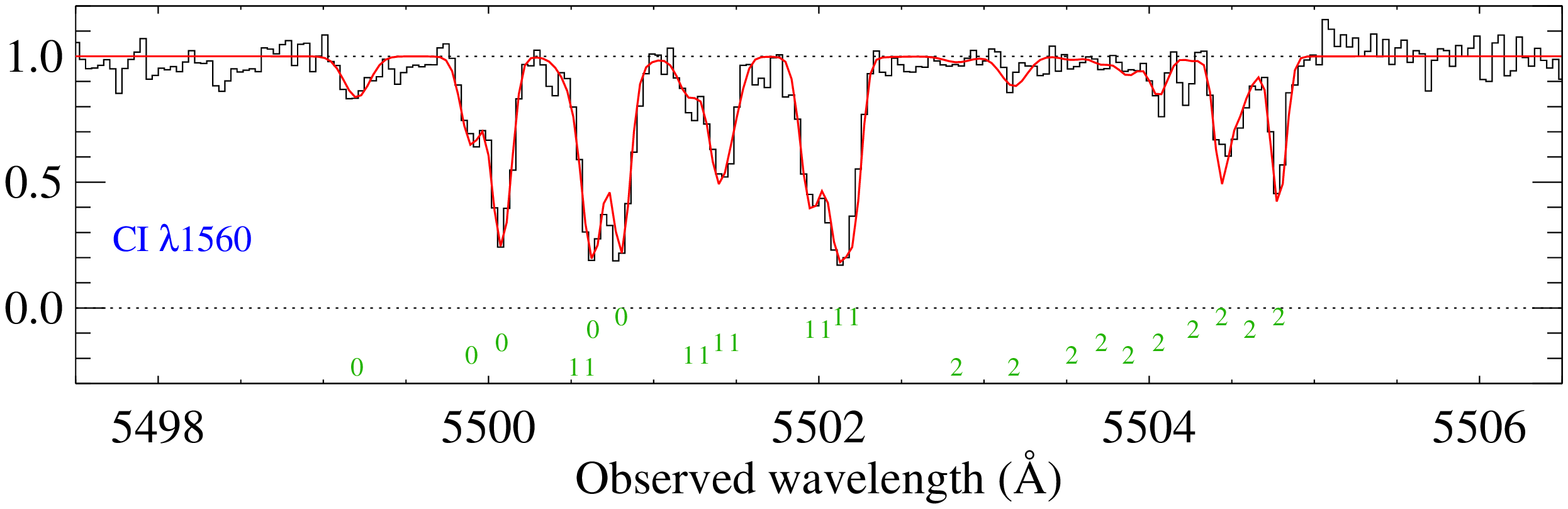}\\
\includegraphics[bb=150 20 400 780, clip=,angle=90, width=\hsize]{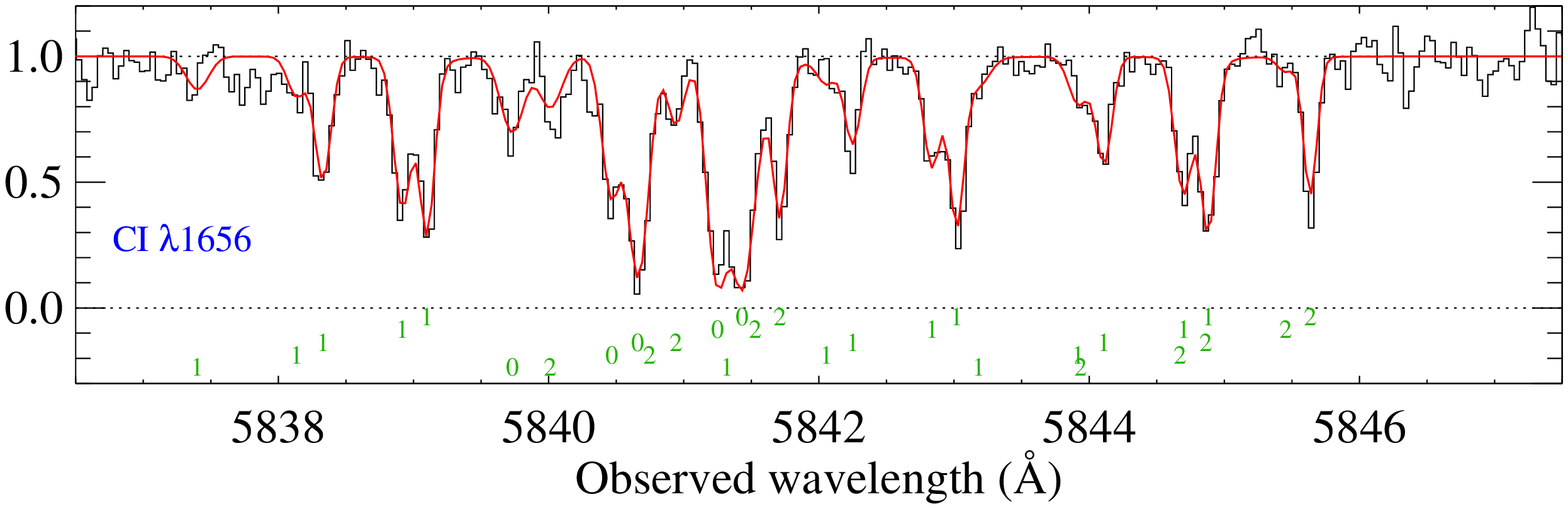}\\
\end{tabular}
\caption{Fit to the neutral carbon lines (UVES data). The green numbers below each plot indicate the fine structure level $J$ for
  each of the five detected \CI\ components. \label{f:CI}}
\end{figure}

The strong \CI\ absorption lines that were used to select the system from the low-resolution SDSS spectrum are resolved in our 
UVES spectrum into different components and different fine-structure levels. 
We detect all three fine-structure levels of neutral carbon's ground state
triplet (2s$^2$2p$^2\,^3$P$^e_{0,1,2}$) in five components, the strongest of which (by a factor of more than a hundred in
column density) is associated with the narrow component seen in both the low-ionisation metal profile
and in molecular absorption (H$_2$ and CO). We simultaneously fit all components
from the fine-structure levels ($J$~=~\{0, 1, 2\}, here denoted \CI, \CI$^*$, \CI$^{**}$, respectively), tying Doppler parameters and redshifts for a given velocity component. We note that while this decreases the number of
free parameters, it is based on the reasonable assumption that the fine-structure levels share the
same physical origin. We use the lines at $\lambda_{\rm rest}\approx 1560$ and 1656~{\AA}, located outside the \lya\ forest, to constrain
the fit (see Fig.~\ref{f:CI}). Including other lines (e.g. \CI$\lambda\lambda$1277,1328) does not improve the constraints due to the
blends, lower S/N and lower spectral resolution. The results are provided in Table~\ref{t:CI}.

 %Automatically generated using results in synthc.res on Mon Jun 13 15:45:51 2016
\begin{table*}
\centering
\caption{Column density of neutral carbon in differemt fine-structure levels \label{t:CI}}
\begin{tabular}{c c c c c}
\hline \hline
{\large \strut}
$z_{\rm abs}$ & $b$ (km\,s$^{-1}$) & $\log N($\CI,J=0)       & $\log N($\CI,J=1)       & $\log N($\CI,J=2)  \\
\hline
2.524432 &  4.96$\pm$ 0.71 &  12.76 $\pm$ 0.05  & 12.74 $\pm$ 0.11  & 12.69 $\pm$ 0.07 \\
2.524876 &  3.41$\pm$ 0.67 &  13.05 $\pm$ 0.04  & 12.73 $\pm$ 0.07  & 12.38 $\pm$ 0.17 \\
2.524993 &  2.57$\pm$ 0.32 &  13.61 $\pm$ 0.07  & 13.31 $\pm$ 0.03  & 12.67 $\pm$ 0.07 \\
2.525347 &  2.84$\pm$ 0.26 &  13.65 $\pm$ 0.06  & 13.46 $\pm$ 0.02  & 12.46 $\pm$ 0.15 \\
2.525458 &  0.81$\pm$ 0.04 &  16.10 $\pm$ 0.08  & 15.54 $\pm$ 0.14  & 14.67 $\pm$ 0.11 \\
\hline
\end{tabular}
\end{table*}

\subsection{Molecular hydrogen \label{s:H2}}

The spectrum of \jz\ is crowded with very strong Lyman (B$^1\Sigma_{\rm u}^+(\nu')$-X$^1\Sigma_{g}^+(0)$)
and Werner (C$^1\Pi_u(\nu')$-X$^1\Sigma_g^+(0)$) lines from molecular hydrogen bluewards
of 4000~{\AA} (see Fig.~\ref{f:H2}). Since the X-Shooter spectrum has a much higher S/N and an extended wavelength coverage
in the blue compared to the UVES spectrum, we use both spectra for the analysis of H$_2$, after normalising the spectra using a spline function. The UVES spectrum
is particularly useful to identify regions blended with intervening \lya\ absorption from the forest, which are subsequently
excluded during the fitting process. 

Because the first ionisation potential of carbon, 11.26~eV, is very close to that of H$_2$ dissociation,
carbon is usually considered as a good tracer of molecular hydrogen \citep[e.g.][]{Srianand05}. While 
there is no one-to-one correspondence, we can expect H$_2$ to be present
in the five components in which \CI\ is detected. Unfortunately, because H$_2$ lines are strongly saturated, it is impossible to distinguish
several close components in their profile. We therefore measure only the total H$_2$ column density by modelling the absorption profile
using a single velocity component. This should be dominated by the reddest narrow component, for which the \CI\ column density is
about two orders of magnitude higher than in the rest of the components. We tie together the redshifts for the different H$_2$ rotational
levels, under the assumption that they arise from the same physical cloud. 
Absorption lines for the low rotational levels ($J\le 2$) are damped, meaning that the column density is well constrained while the profile
does not directly depend on (and therefore does not constrain) the Doppler parameter. 

\begin{figure}
  \centering
\includegraphics[bb=70 170 500 580,clip=,width=0.95\hsize]{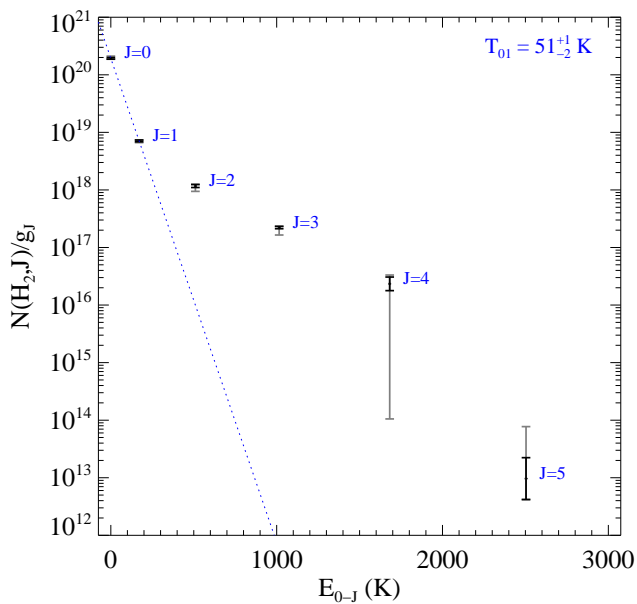}
\caption{H$_2$ excitation diagram. The black points and error bars correspond to the fit with $b=1.0$~\kms.
  Grey error bars correspond to extrema with $b=0.65$ and $5$~km\,s$^{-1}$. \label{f:h2ext}}
\end{figure}

\begin{figure*}
  \centering
\includegraphics[bb=120 45 580 760,angle=90,width=\hsize]{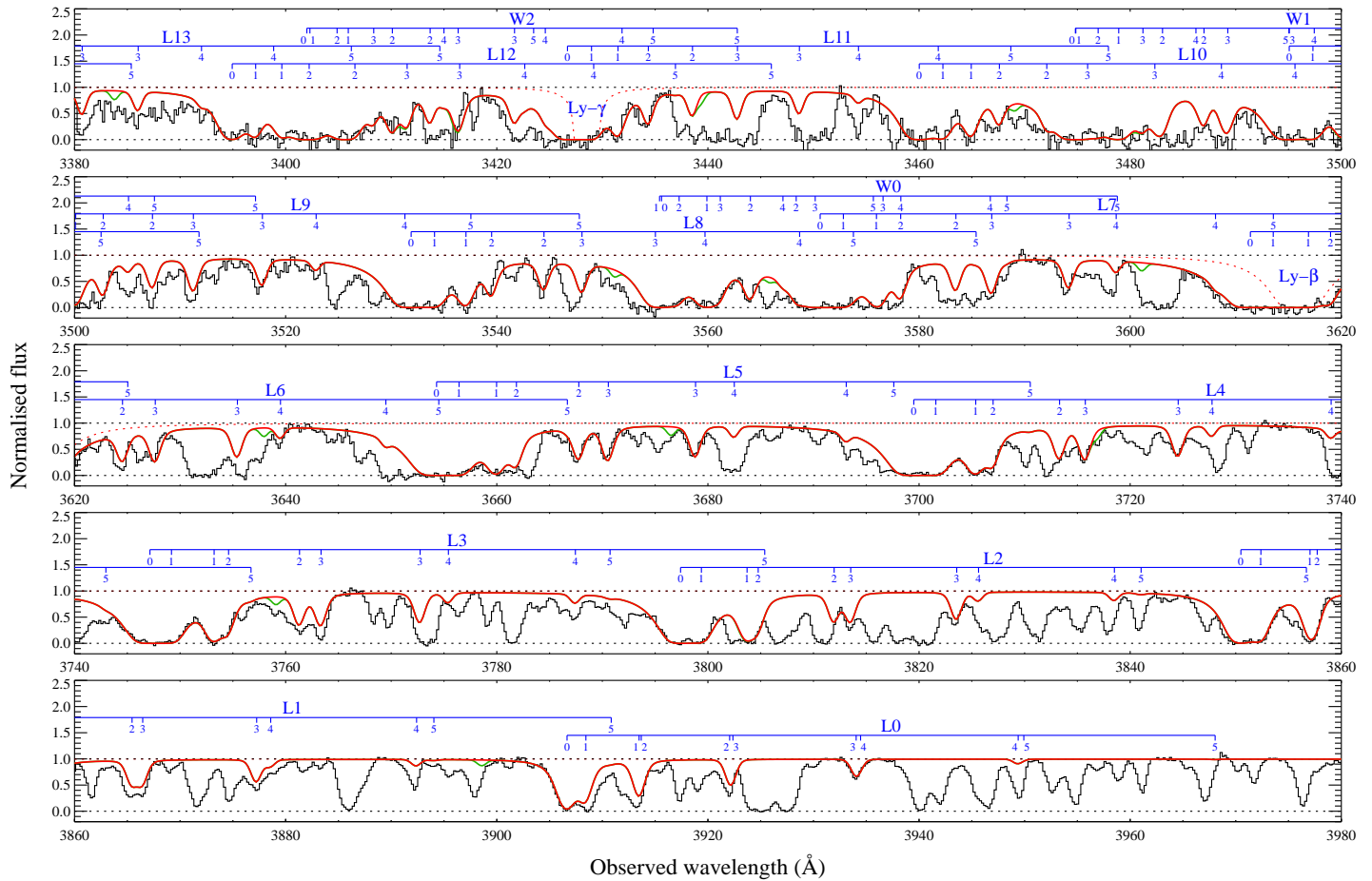}
\caption{Portion of X-Shooter UVB spectrum (black) around the H$_2$ lines, with the 
  best-fit synthetic spectrum for H$_2$ absorption (with $b=1$~\kms) in red. Horizontal blue segments connect rotational levels (short tick marks)
  from a given Lyman (L) or Werner (W) band, as labelled above. \HI\ Ly-$\beta$ and Ly-$\gamma$
  from the DLA are indicated as red dashed profiles. The green profile corresponds to HD lines.
  \label{f:H2}}
\end{figure*}

Figure \ref{f:h2ext} shows the excitation diagram of H$_2$, which presents the population in each rotational level
against the energy of that level:

\begin{equation}
  {{N({\rm H}_2,J')} \over {g({\rm H}_2,J')}} = {{N({\rm H}_2,J)} \over {g({\rm H}_2,J)}} e^{-E_{JJ'}/kT_{JJ'}},
\end{equation}

\noindent where $E_{JJ'}$ is the energy difference between levels $J$ and $J'$, $g({\rm H}_2,J)$, $g({\rm H}_2,J')$ are the respective spin 
statistical weights and $T_{JJ'}$ is the excitation temperature. $T_{01}$ is generally considered as a very good indicator
of the kinetic temperature of the gas at such high column density, where selective self-shielding is no longer
at play and the low rotational levels are easily thermalised thanks to short collisional time-scales \citep{Roy06,LePetit06}.
In turn, the high rotational levels are characterised by a higher excitation temperature. This is expected and
  seen in interstellar clouds because of the very slow infrared relaxation after UV or formation pumping into high-$J$ levels.
  Moreover collisional de-excitation becomes difficult at the high-$J$ levels, where the level spacings become so large
  (several hundred cm$^{-1}$) that these amounts of energy cannot be transferred in collisions, in particular at the
  density and temperatures seen in the ISM. This leads to the observed non-Boltzmann distribution. We also note that the
  observed excitation diagram corresponds to integrated values and that possible additional warmer components with
  lower $N($H$_2$) will mostly contribute to the high-$J$ levels.

We measure $T_{01}=51\pm 2$~K, which is lower than the value typically seen in H$_2$-bearing DLAs \citep[$T\sim 150$~K]{Srianand05},
and closer to what is seen in our Galaxy, with an average of about 77~K \citep{Rachford02}.
The kinetic temperature $T_{\rm kin}\approx T_{01}$ corresponds to a Doppler parameter for H$_2$ of $b_{\rm th} \approx 0.65$~\kms\
if we assume thermal broadening only. However, because turbulent broadening is also likely present, as indicated by similar
Doppler parameters for much heavier species, this value should be considered as a lower limit to the line broadening.
We can get a more realistic estimate of the Doppler parameter by quadratically adding this pure thermal value to the turbulent $b$-value seen for heavier
species (for which $b_{\rm th}$ is negligible, see Fig.~\ref{f:doppler}) and obtain $b=\sqrt{b_{\rm th}^2+b_{\rm turb}^2}\approx$1.0~\kms.

However, it has been observed in several H$_2$-bearing systems that the Doppler parameter can also be an increasing function of
the rotational level \citep[e.g.][]{Lacour05,Noterdaeme07b,AlbornozVasquez14}, possibly due to more turbulent an warmer
external layers where UV pumping of H$_2$ is enhanced \citep{Balashev09}. 
We therefore also perform a fit using
a high $b$-value of 5~\kms. 
The resulting parameters are provided in Table~\ref{t:H2}.
We obtain a total column density of $\log N$(H$_2)=20.43\pm0.02$ which implies an overall molecular fraction, $f=2N($H$_2)/(2N($H$_2)+N(\HI))=0.46 \pm 0.07$,
i.e. the highest value measured to date in a quasar-DLA. This also corresponds to a strict lower limit to the molecular fraction in the cold
component.

\begin{table} 
\centering
\caption{Column density in different rotational levels of H$_2$ for the three values of the Doppler parameter. \label{t:H2}}
\begin{tabular}{c c c c}
\hline \hline
{\large \strut} Rot. level       & \multicolumn{3}{c}{$\log N($H$_2,J)$}     \\
                ($J$)            & $b=1.0$ & $b=0.65$ & $b=5$ \\
                                 & (km\,s$^{-1}$) & (km\,s$^{-1}$) & (km\,s$^{-1}$) \\

\hline
0                 &  20.29 $\pm$ 0.02 & 20.29 $\pm$ 0.02  & 20.31 $\pm$ 0.02  \\           
1                 &  19.81 $\pm$ 0.02 & 19.80 $\pm$ 0.02  & 19.79 $\pm$ 0.02  \\
2                 &  18.77 $\pm$ 0.03 & 18.77 $\pm$ 0.03  & 18.70 $\pm$ 0.03  \\
3                 &  18.67 $\pm$ 0.02 & 18.67 $\pm$ 0.02  & 18.57 $\pm$ 0.03  \\
4                 &  17.32 $\pm$ 0.12 & 17.37 $\pm$ 0.11  & 15.08 $\pm$ 0.10  \\
5                 &  14.50 $\pm$ 0.36 & 14.79 $\pm$ 0.62  & 14.42 $\pm$ 0.16  \\
\hline
Total             & 20.43$\pm$0.02    &  20.43$\pm$0.02   & 20.44$\pm$0.02    \\
\hline
\end{tabular} 
\end{table}

\subsection{Deuterated molecular hydrogen}
\begin{figure}
  \centering
  \includegraphics[width=0.9\hsize]{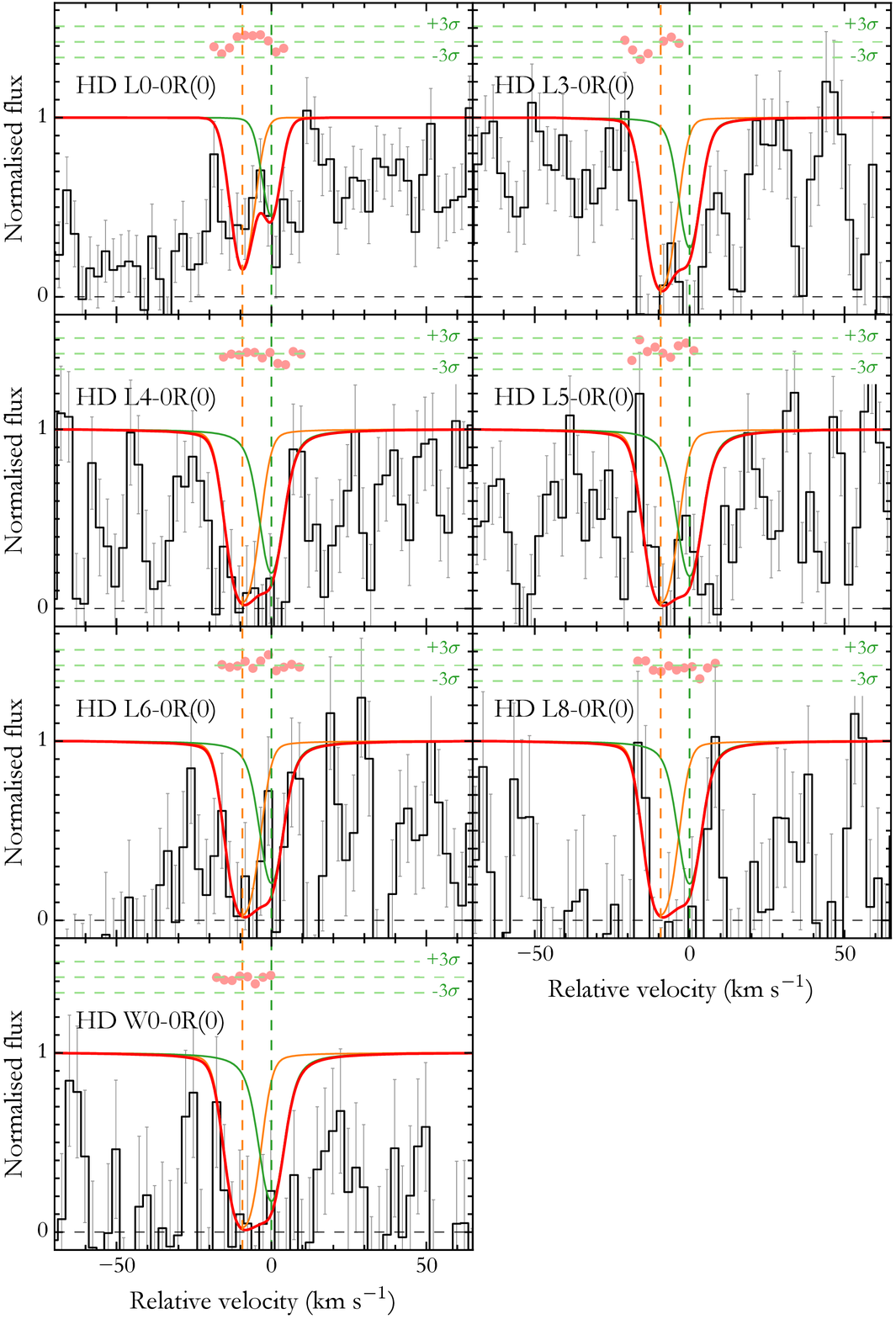}
  \caption{HD synthetic profile overlaid on the UVES spectrum. \label{f:HD}}
  \end{figure}

Several lines of deuterated molecular hydrogen are also detected in both the UVES and X-Shooter spectra.
However, HD lines are often blended with \lya\ forest or H$_2$ absorption and are saturated in the
low S/N UVES spectrum but weak in the medium resolution X-shooter spectrum. We therefore use here
on a different fitting procedure based on Markov Chain Monte Carlo method. We considered
L0R0, L4R0, L5R0, L6R0, L8R0 and W0R0, locally re-normalised, as well as L11R0 and L14R0 (covered only
by X-Shooter). We use two components with fixed redshifts ($z=2.525458$ and $z=2.525348$)
corresponding to the strongest components seen in \CI\ and use \CI\ Doppler parameters as priors. 
The synthetic HD profiles in the UVES spectrum are shown on
Fig.~\ref{f:HD} and the corresponding X-Shooter profile also included in Fig.~\ref{f:H2}.
We only consider the total HD column density as being reasonably trustable, with $\log N_{\rm HD} = 16.64^{+0.16}_{-0.18}$.
This corresponds to ${\rm HD/2H_2} = (8.1^{+3.7}_{-2.8})\times 10^{-5}$, which is significantly higher than the typical ratios observed
in our Galaxy \citep{Snow08} and than the primordial value estimated from \DI/\HI\ in low metallicity high-$z$ DLAs
\citep[(D/H)$_{\rm p}=(2.53\pm0.04)\times 10^{-5}$; ][]{Cooke14}. 
      While a high abundance of deuterium can possibly be explained by strong supply of primordial gas
        \citep[as suggested by][]{Ivanchik10}, the molecular ratio is here more likely explained by chemical
        fractionation and charge exchange processes \citep{Liszt15}. Without entering into
        details of the HD chemistry, we note that the reaction ${\rm D^++H_2} \to {\rm HD + H^+}$ is fast and can lead to an
        increasing of HD compared to H$_2$.
If we call $f_{\rm HD} = {\rm HD/(HD+\DI)}$ the fraction of deuterium in molecular form, then we have

\begin{equation}
  {{\rm HD} \over 2{\rm H_2}} = \left({D \over H}\right) {{f_{\rm HD}} \over {f_{\rm H2}}}
\end{equation}

\noindent Assuming an intrinsic primordial value\footnote{We do not take here
  into account astration of the order of 0.1~dex for the high metallicity and redshift
of our system \citep[see][]{Dvorkin16}.}, the high ${\rm HD}/{\rm H_2}$ ratio can be explained
for $f_{\rm HD}/f_{\rm H2} \simeq 3.2$, which would naturally require that the cloud cannot be fully molecular.
This is indeed what we conclude from modelling the physical conditions in the cloud (Sect.~\ref{s:phys}).
We however caution that a high-resolution spectrum with high S/N ratio is necessary to better take into account
blends with the Ly-$\alpha$ forest and confirm our measurement. 

\subsection{Neutral chlorine}

\begin{figure}
  \centering
  \begin{tabular}{c}
\includegraphics[bb=150 20 400 780, clip=,angle=90, width=\hsize]{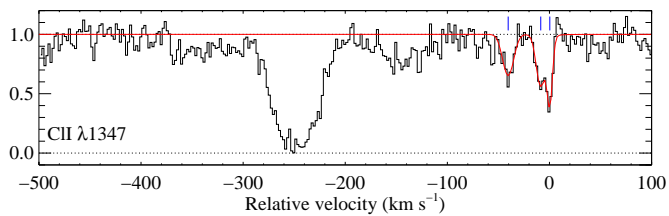}\\
\end{tabular}
\caption{Fit to the absorption profile of neutral chlorine. \label{f:ClI}}
\end{figure}

Chlorine is known to be tightly linked with H$_2$ thanks to rapid chemical reactions \citep[e.g.][]{Jura74}. In our Galaxy,
observations of clouds with $\log N($H$_2)>19$ using the Copernicus satellite have revealed a clear correlation between the
column density of both species \citep{Moomey12}. Recently, \citet{Balashev15} used a sample of known H$_2$-bearing DLAs to
show that this relation stands at high redshifts and down to ten times lower column densities.
Here, only one absorption line of neutral chlorine, \ClI$\lambda$1347 is covered and not blended in our spectrum.
Three components, that match those seen in the neutral carbon profile are detected and used to constrain the column densities,
while those associated to the weakest \CI\ components are below our detection limit.
This again indicates that H$_2$ should actually be present
in more than one component, although too close to be distinguished within the damped profile of the strong, cold
component. Unfortunately, the column density of neutral chlorine in that component is poorly constrained because this line is
in the intermediate regime with a strong dependence on the Doppler parameter. Therefore, the Voigt profile fit keeping all parameters
free leads to
a very high uncertainty in the column density. 
However, we can make the reasonable assumption that
the Doppler parameter should be close to that of other species for this component. Since chlorine is expected to arise from
the H$_2$-bearing gas but with a much higher atomic mass, its thermal broadening should be negligible and its Doppler parameter
close to that of other ``heavy'' species. Assuming $b=0.7$~\kms (see Fig.~\ref{f:doppler}) we obtain a very satisfactory fit with $\log N(\ClI)=14.6 \pm 0.3$
in the narrow component. We also fitted \ClI\ assuming a more relaxed constraint on $b$, that we take in the range 0.6-0.8~\kms, giving
 $\log N(\ClI)=14.43-14.89$ with a similar 0.3~dex fitting uncertainty. This sets the overall uncertainty to about 0.4~dex.
The results are shown in Fig.~\ref{f:ClI} and Table~\ref{t:ClI}, where we also provide the fitting
results leaving the Doppler parameter totally free, for completeness.

\begin{table}
  \caption{Neutral chlorine fitting parameters \label{t:ClI}}
  \begin{tabular}{c c c  c c}
    \hline\hline
    \zabs     &  $b$           & $\log N(\ClI)$  &   $b$        & $\log N(\ClI)$  \\
              & (\kms)         &                 & (\kms)       &                 \\
    \hline
    2.52500   & 6.5$\pm$1.3  & 13.00$\pm$0.06    & 6.6$\pm$1.2  & 13.00$\pm$0.06  \\
    2.52536   & 4.8$\pm$1.5  & 13.03$\pm$0.10    & 4.9$\pm$1.1  & 13.04$\pm$0.06  \\
    2.52546   & 0.8$\pm$0.6  & 14.55$\pm$1.92    & 0.7\tablefootmark{a}   & 14.63$\pm$0.29  \\
    Total     &              & 14.58$\pm$1.81    &                        & 14.65$\pm$0.27 \\
    \hline
  \end{tabular}
  \tablefoot{
    \tablefoottext{a}{Fixed value (see text).}
    }
\end{table}

\subsection{Neutral sulphur \label{s:s1}}

\begin{figure}%[!ht]
\centering
\begin{tabular}{c}
\includegraphics[bb=209 20 400 780, clip=,angle=90, width=\hsize]{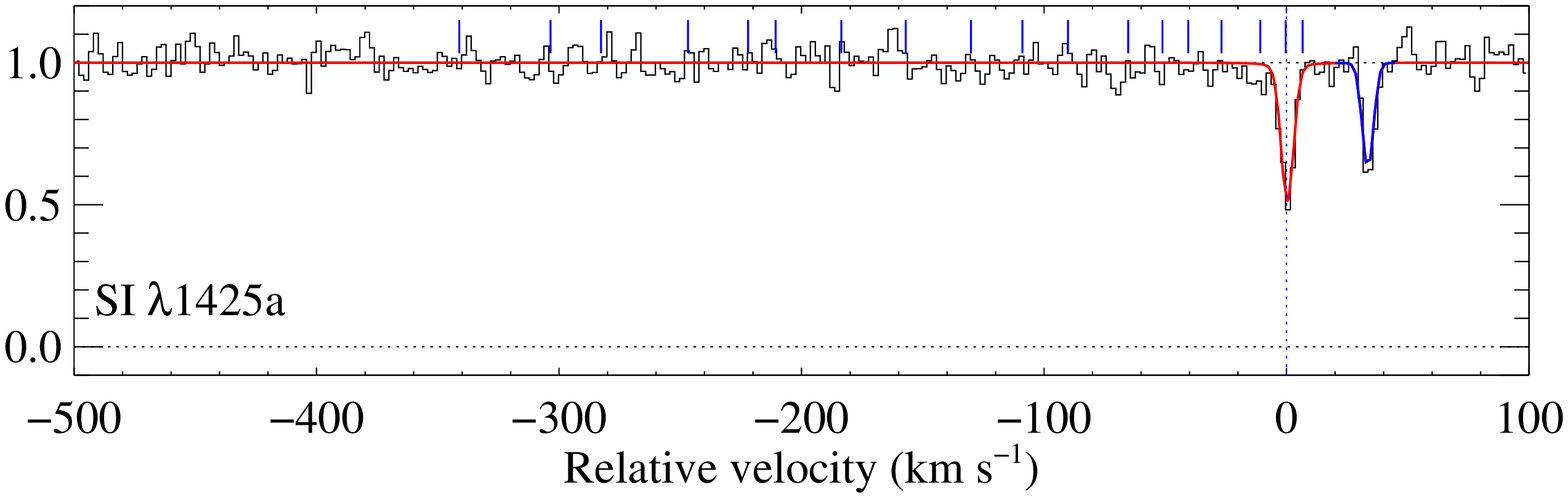}\\[-6pt]
\includegraphics[bb=209 20 400 780, clip=,angle=90, width=\hsize]{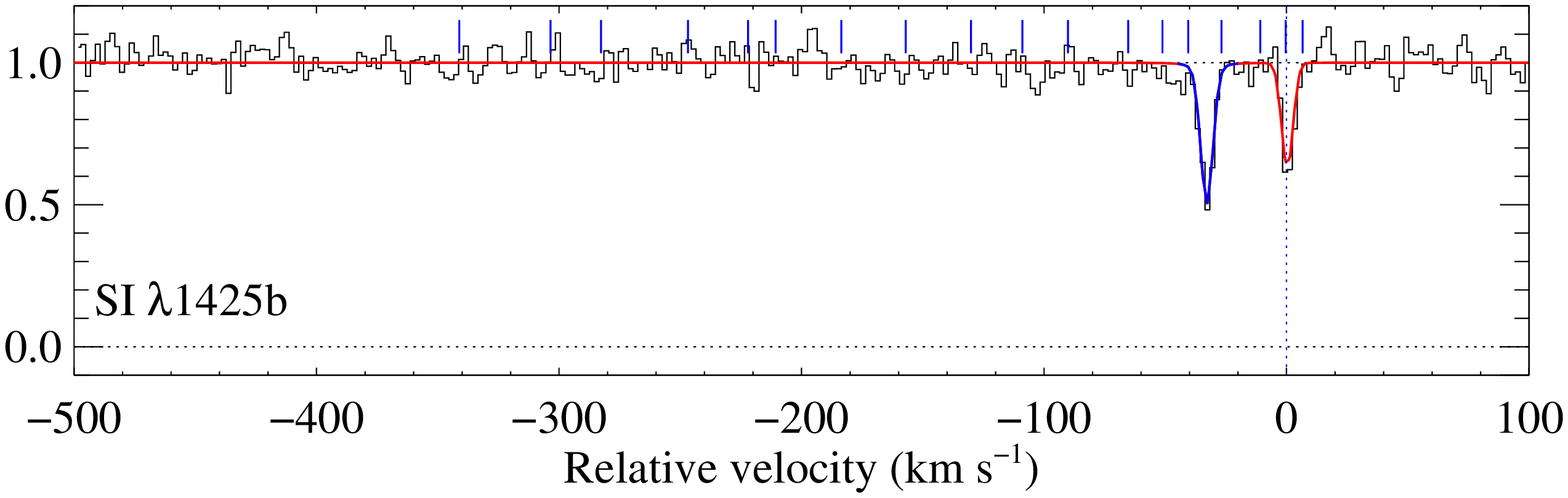}\\[-6pt]
\includegraphics[bb=209 20 400 780, clip=,angle=90, width=\hsize]{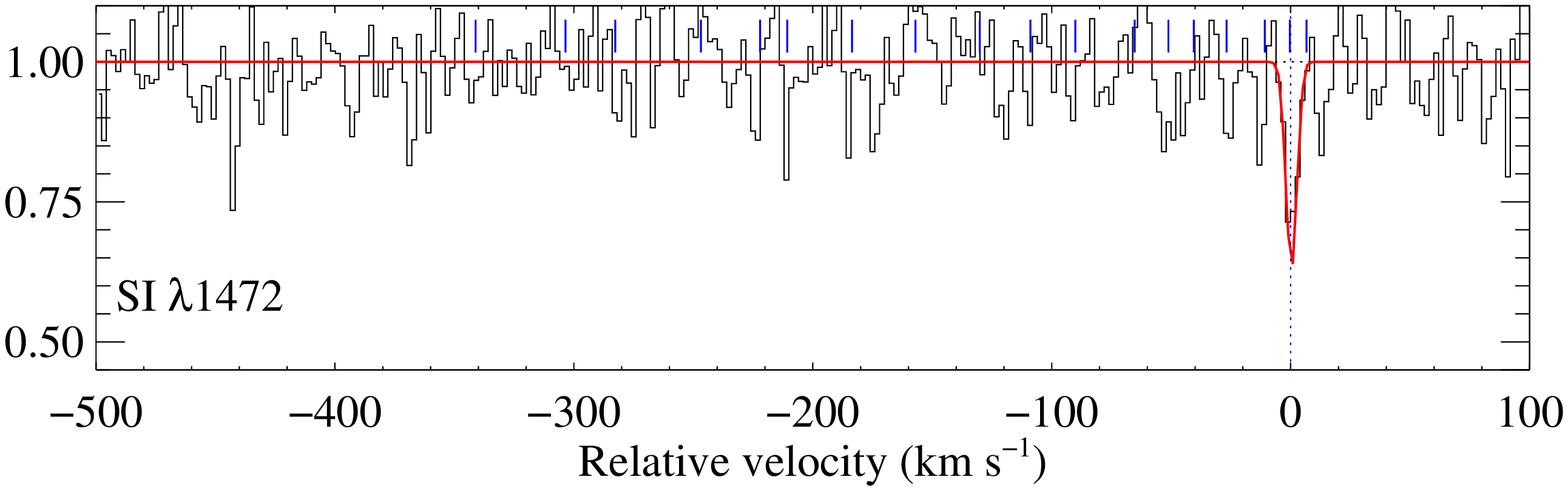}\\[-6pt]
\includegraphics[bb=209 20 400 780, clip=,angle=90, width=\hsize]{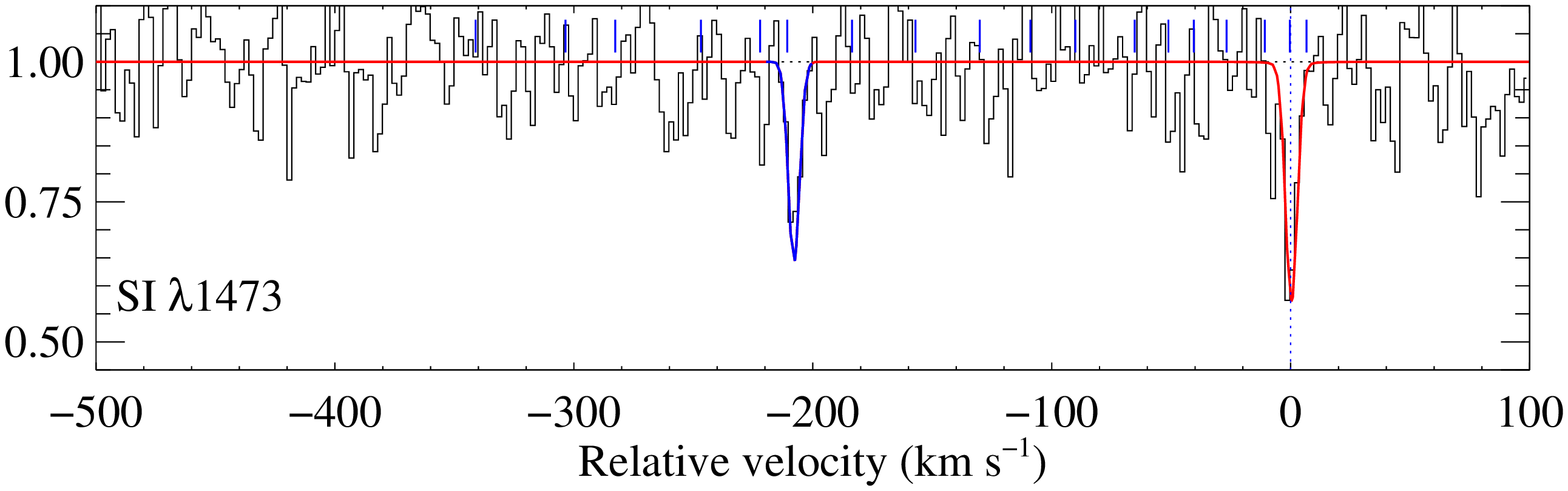}\\[-6pt]
\includegraphics[bb=150 20 400 780, clip=,angle=90, width=\hsize]{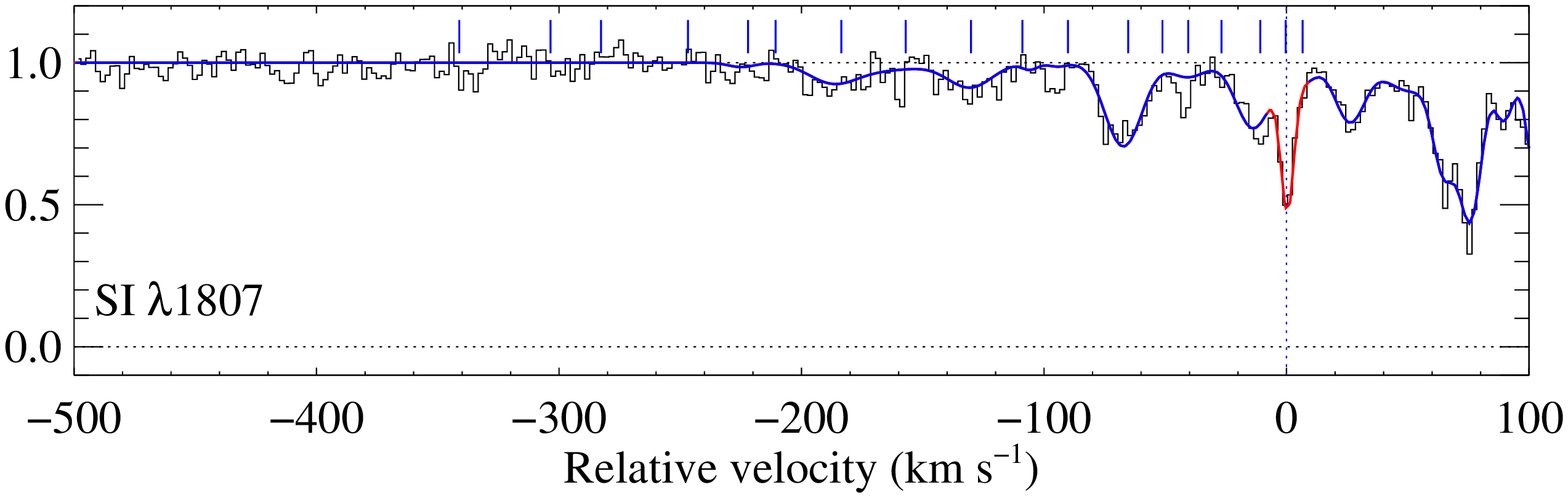}\\[-6pt]
\end{tabular}
\caption{Neutral sulphur absorption lines (UVES data). As for Fig.~\ref{f:metalsII}, the continuous red line correspond to the best fit
  model, with regions corresponding to other absorption lines marked in blue (for example, most of the profile seen in the \SI$\lambda$1807
  actually corresponds to \SiII$\lambda$1808, except around $v=0$~\kms). The short tick marks show the position of the
  singly ionised metal lines. Note that \SI\ is detected only in the component that has CO. \label{f:s1}}
\end{figure}

Because the first ionisation potential of sulphur is 10.36~eV, neutral sulphur is only expected to be
found in very shielded regions. To our knowledge, only a handful detections of \SI\ have been reported so far in DLAs,
  all associated to a molecular absorber featuring CO \citep[][]{Srianand08} and/or strong H$_2$/HD
  \citep{Milutinovic10,Balashev11}.
Here, we detect \SI\ absorption lines with $\log N(\SI)=14.85\pm0.18$ 
in our UVES spectrum from five transitions in a single narrow ($b=0.50 \pm 0.07$~\kms) component, see Fig.~\ref{f:s1}.
This suggests that \SI\ can be used as a tracer for CO \citep{Noterdaeme10b}, just like the presence of \CI\ implies that of H$_2$.
However, because \SI\ lines have similar strengths and are located in the same spectral region as CO lines, this is of little
practical use to identify CO systems. Still, \SI\ can be helpful in determining the velocity structure of multi-component CO absorption
systems \citep[e.g.][]{Srianand08,Noterdaeme09b}. 

\subsection{Neutral magnesium and neutral sodium}

\begin{figure}
\centering
\begin{tabular}{c}
\includegraphics[bb=209 20 400 780, clip=,angle=90, width=\hsize]{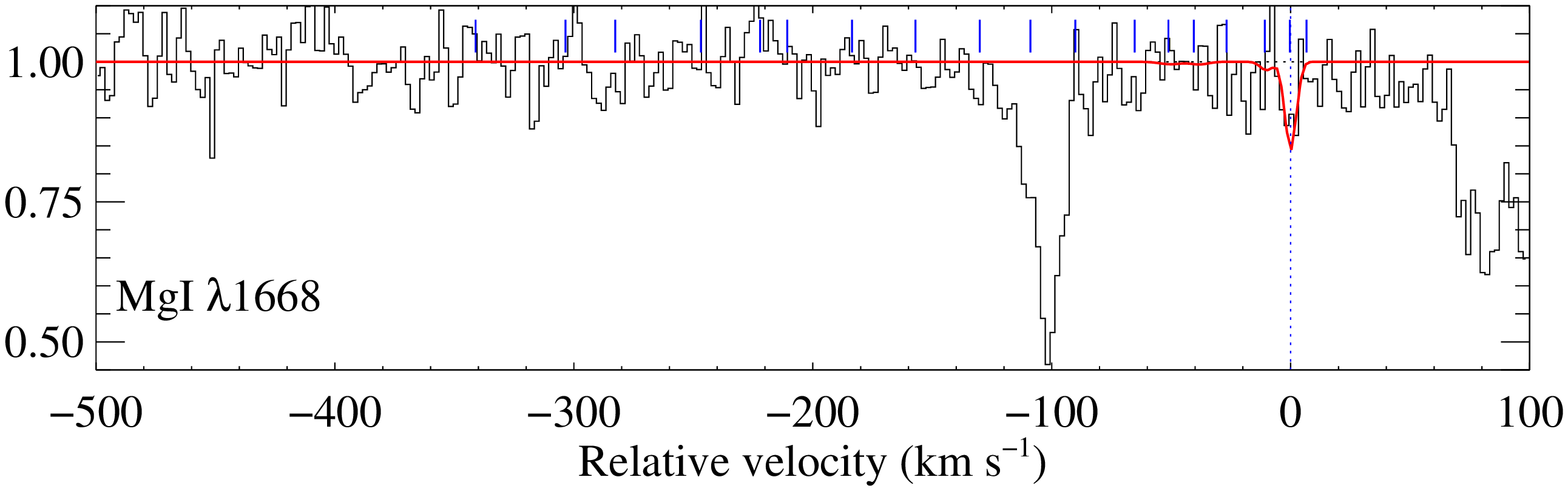}\\[-6pt]
\includegraphics[bb=209 20 400 780, clip=,angle=90, width=\hsize]{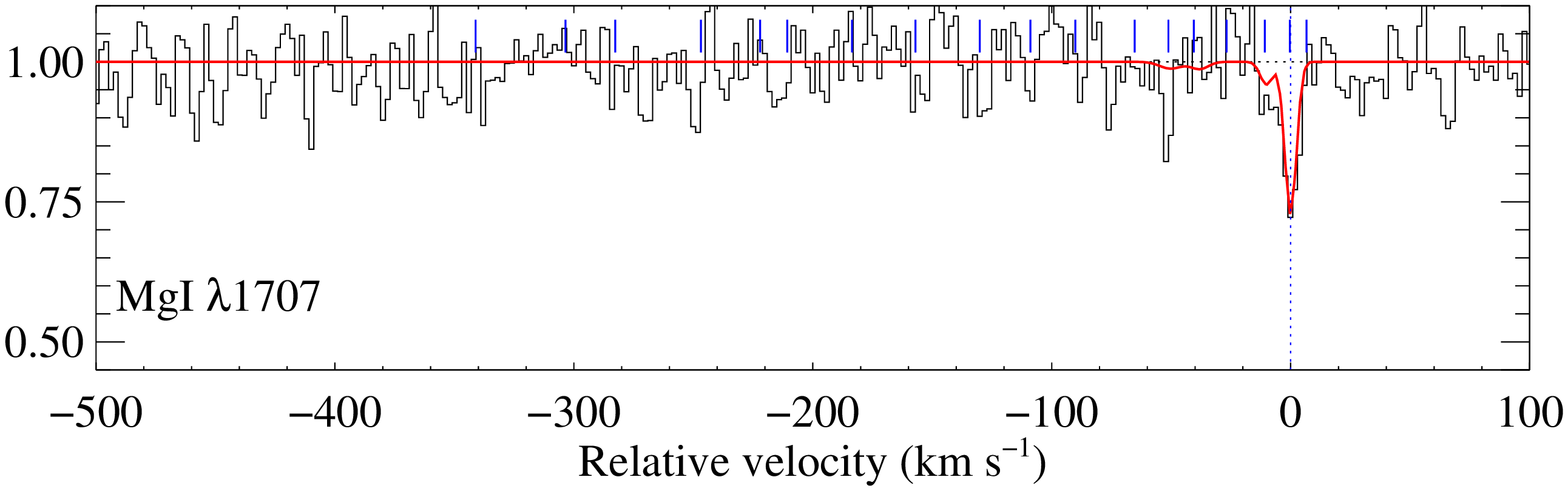}\\[-6pt]
\includegraphics[bb=209 20 400 780, clip=,angle=90, width=\hsize]{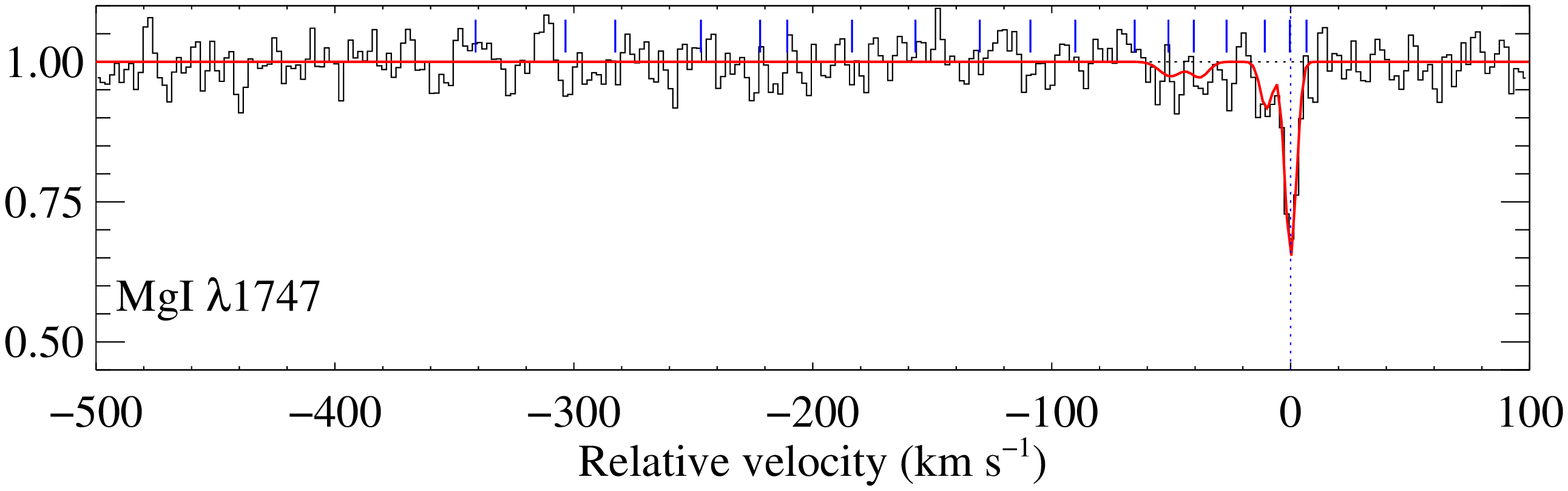}\\[-6pt]
\includegraphics[bb=209 20 400 780, clip=,angle=90, width=\hsize]{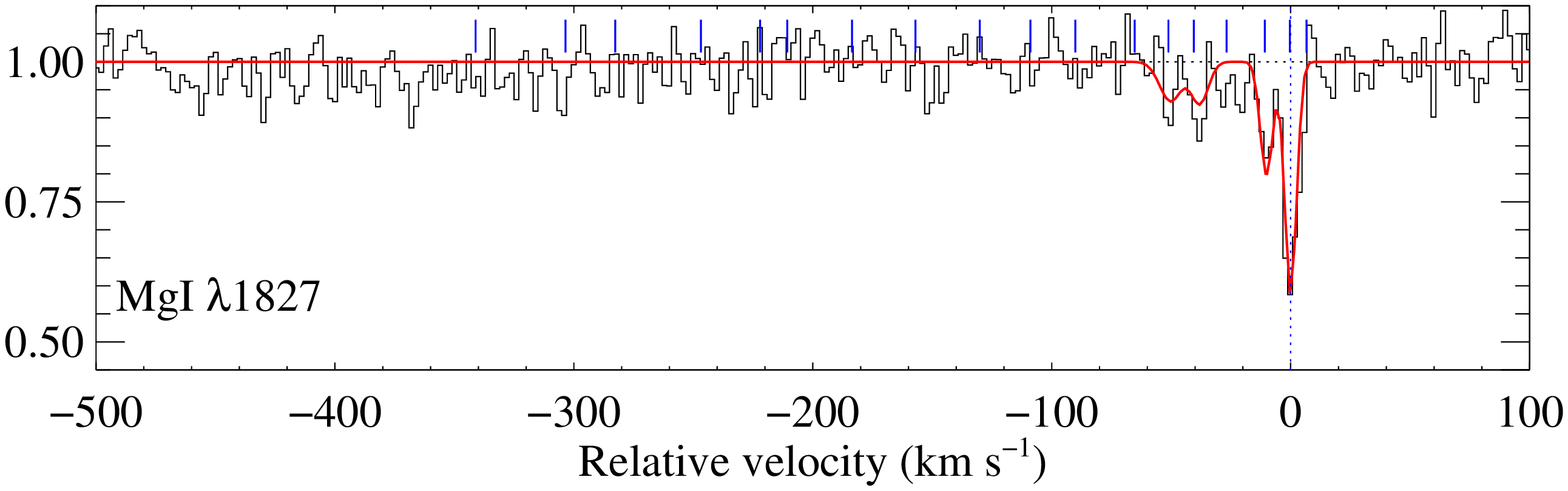}\\[-6pt]
\includegraphics[bb=150 20 400 780, clip=,angle=90, width=\hsize]{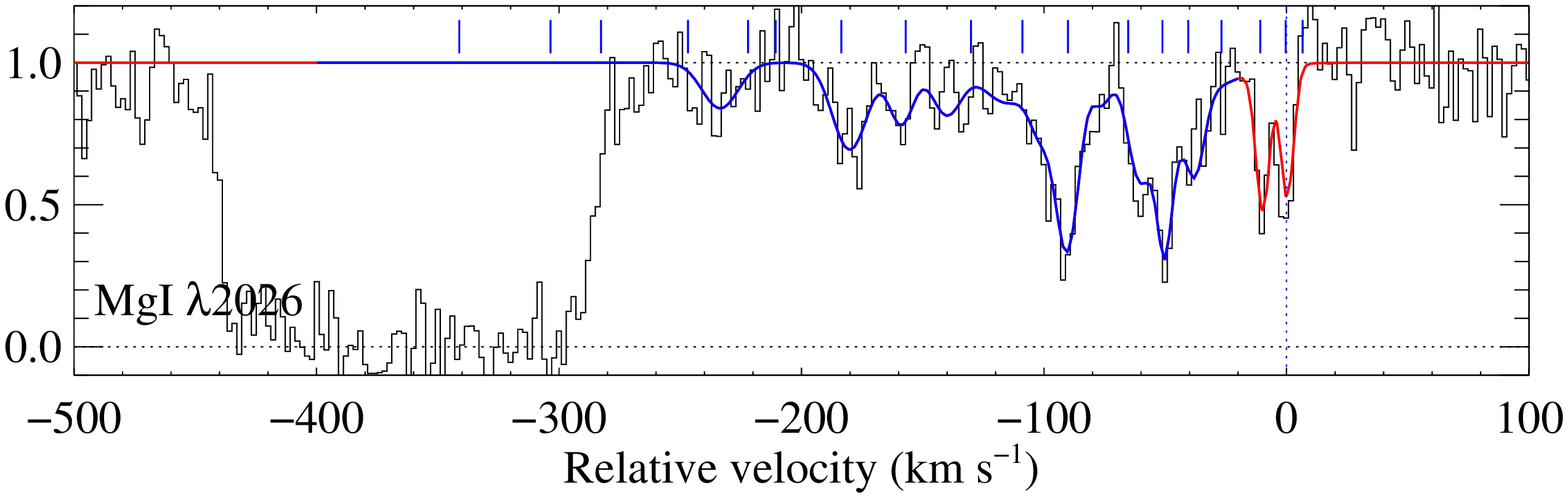}\\[-6pt]
\end{tabular}
\caption{Fit to the neutral magnesium absorption lines (UVES data). The blue region in the bottom panel corresponds to \ZnII$\lambda$2026. \label{f:mg1}}
\end{figure}

Neutral magnesium (\MgI) is detected in five transitions in our UVES spectrum (see Fig.~\ref{f:mg1}).
We clearly detect two components and possibly two additional weak components. The main component, again corresponding to the
molecular one at $v=0$~\kms, contains more than 80\% of the total column density
with $\log N(\MgI)_{\rm c}=14.1 \pm 0.1$ . Interestingly, the hidden saturation of this component with very small Doppler
parameter ($b=0.86\pm0.1$~\kms) is directly evidenced by its relative strengths to the second strongest component: both these
components have similar observed optical depth for $\MgI\lambda2026$, but the former is also seen in transitions with much smaller oscillator
strengths. 

\begin{figure}
\centering
\includegraphics[bb=120 20 400 780, clip=,angle=90, width=\hsize]{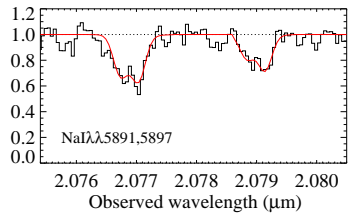}
\caption{Portion of the X-Shooter NIR spectrum around the \NaI\ doublet. \NaI$\lambda$5891 and \NaI$\lambda$5897
    are redshifted at 2.077 and 2.079$\mu$m, respectively. The red curve corresponds to the
  Voigt-profile fit using fixed velocity decomposition from \MgI. \label{f:na1}}
\end{figure}

We also detect the \NaI$\lambda\lambda$5891,5897 doublet in the NIR X-Shooter spectrum. The non-gaussian profile indicates
that several components are present, although blended at the achieved spectral resolution (around $R\sim$8500). We therefore
use the velocity decomposition of \MgI\ obtained at high spectral resolution, i.e., the redshifts and Doppler parameters
of \NaI\ are fixed to the value previously determined for \MgI\ and only the column density is allowed to vary.
This assumption leads to a good fit of the observed \NaI\ absorption features (Fig.~\ref{f:na1}), with a column density in
the main component of $\log(\NaI)_c=15.0\pm0.3$, with a large fitting uncertainty due to the line being much narrower than the resolution
element. We also note that the value is very dependent on the exact normalisation and that 
the first ionisation potential of sodium (5.14~eV) is even lower than that of magnesium (7.65~eV), meaning
that \NaI\ may arise from deeper regions in the cloud. This means that \NaI\ column densities should
be considered with great caution until a high resolution infra-red spectrum is obtained.
{Using the empirical correlation observed in the Milky Way between \NaI\ equivalent width and $E(B-V)$ from
  \citet{Poznanski12}, we expect $A_{\rm V} \sim 0.3$ for their implicitely assumed $R_V=3.1$.
We will see in Sect.~\ref{s:ext} that this is not far from the observed value.}

\subsection{Carbon monoxide \label{s:CO}}

\begin{figure*}
\centering
\begin{tabular}{cc}
    \includegraphics[bb=192 20 435 780, clip=,angle=90, width=0.48\hsize]{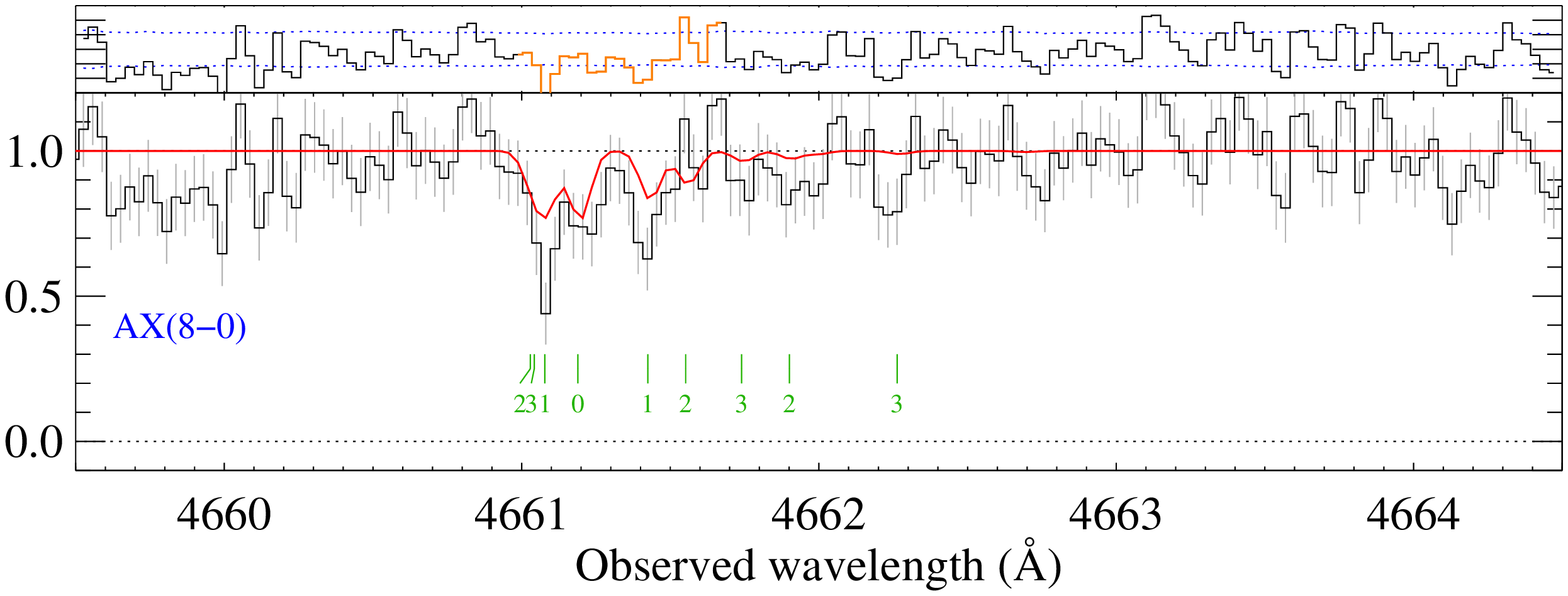}&
    \includegraphics[bb=192 20 435 780, clip=,angle=90, width=0.48\hsize]{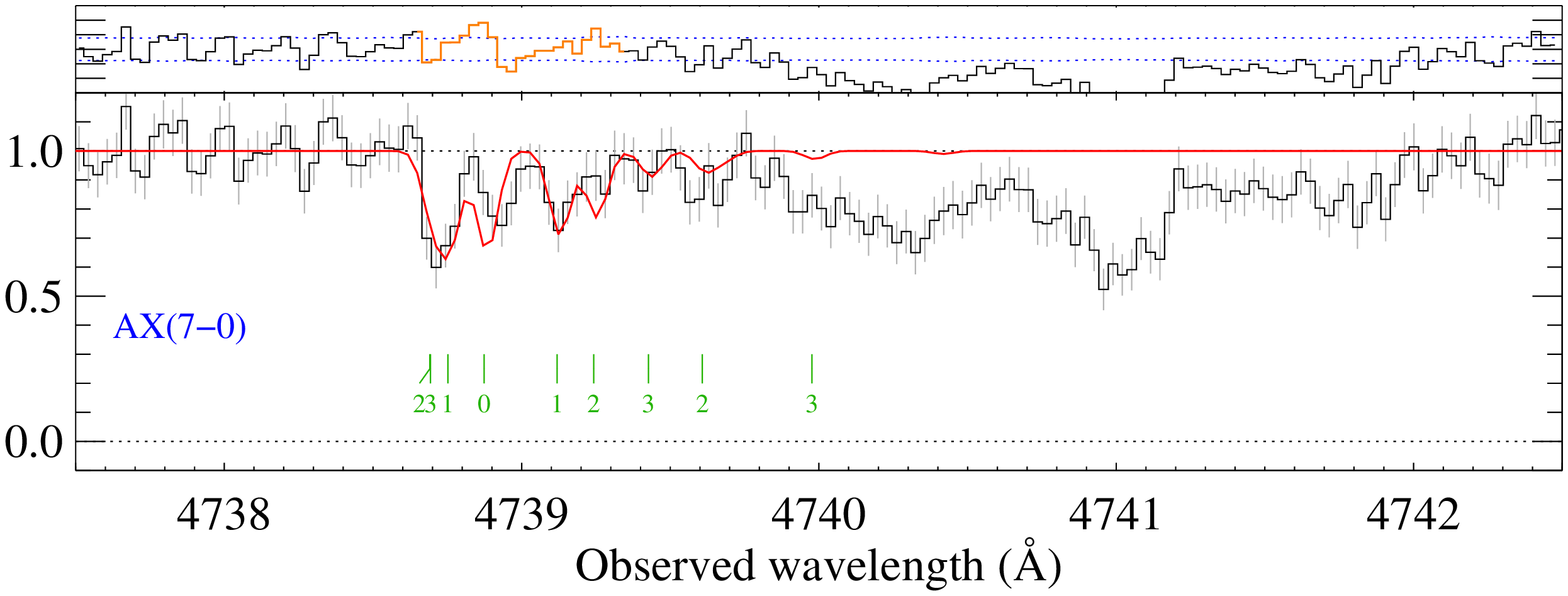}\\
    \includegraphics[bb=192 20 435 780, clip=,angle=90, width=0.48\hsize]{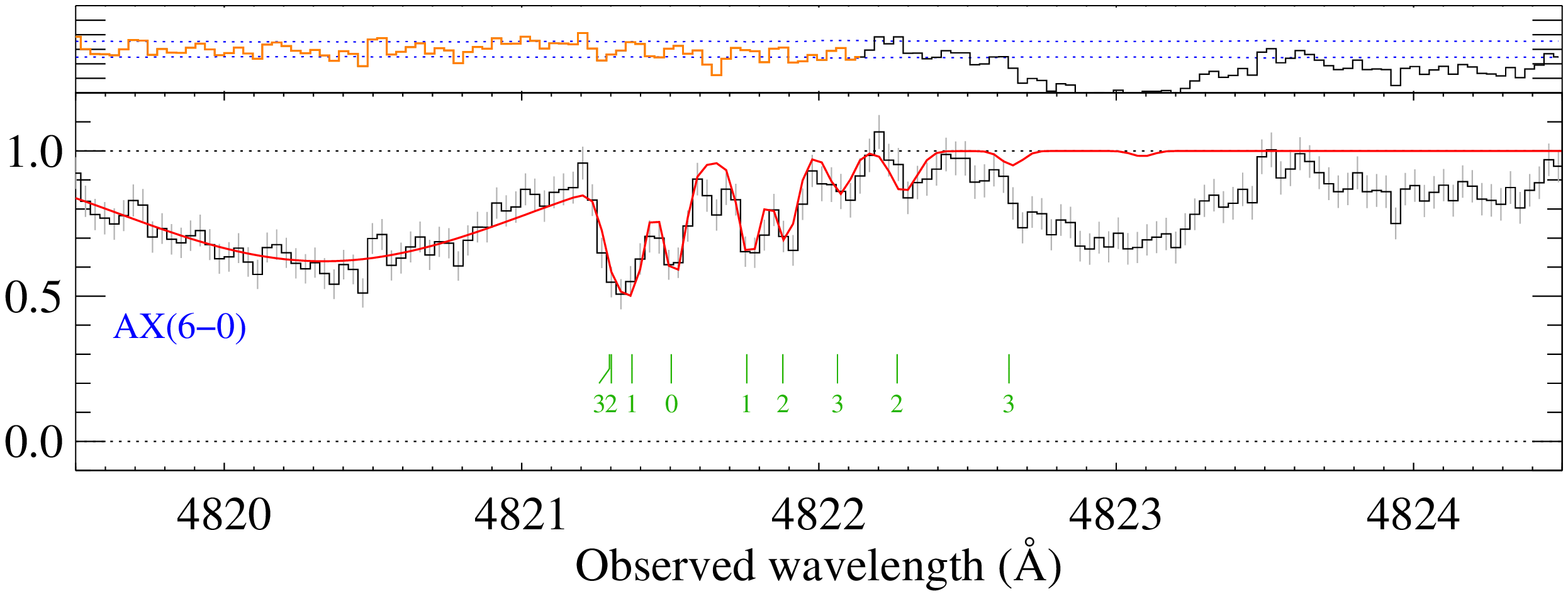}&
    \includegraphics[bb=192 20 435 780, clip=,angle=90, width=0.48\hsize]{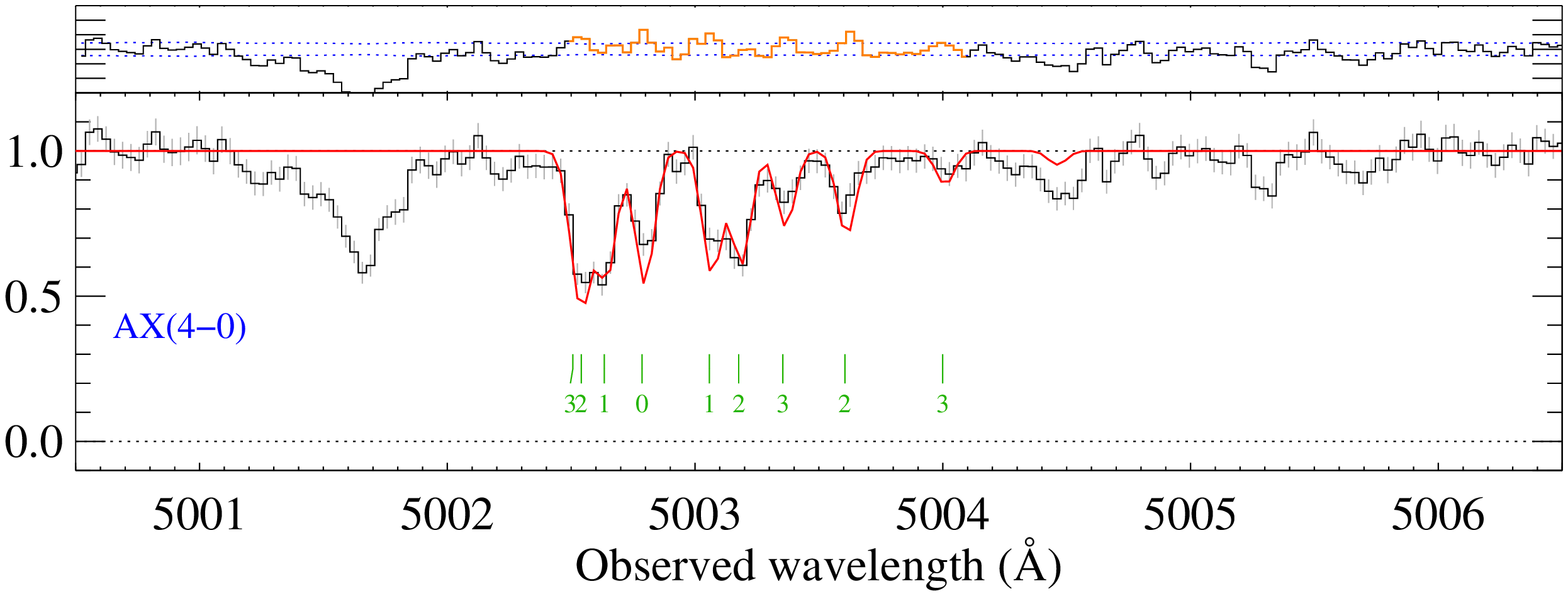}\\
    \includegraphics[bb=192 20 435 780, clip=,angle=90, width=0.48\hsize]{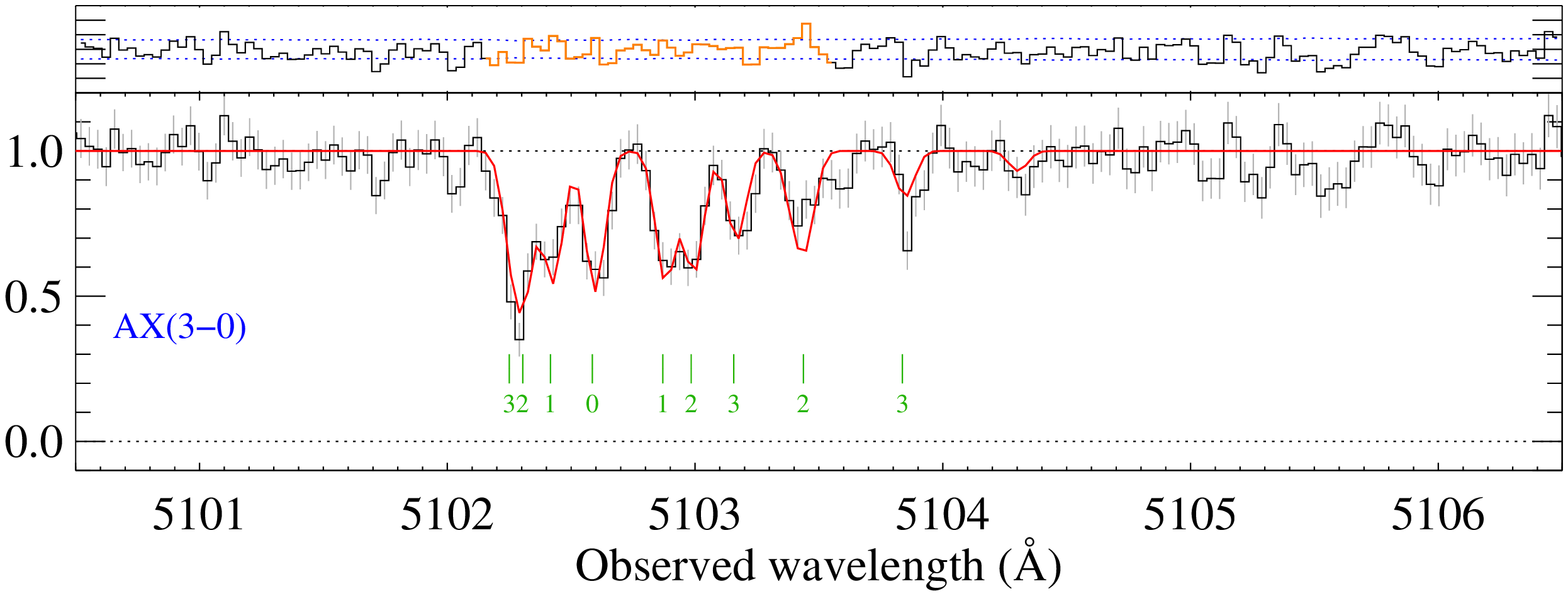}&
    \includegraphics[bb=192 20 435 780, clip=,angle=90, width=0.48\hsize]{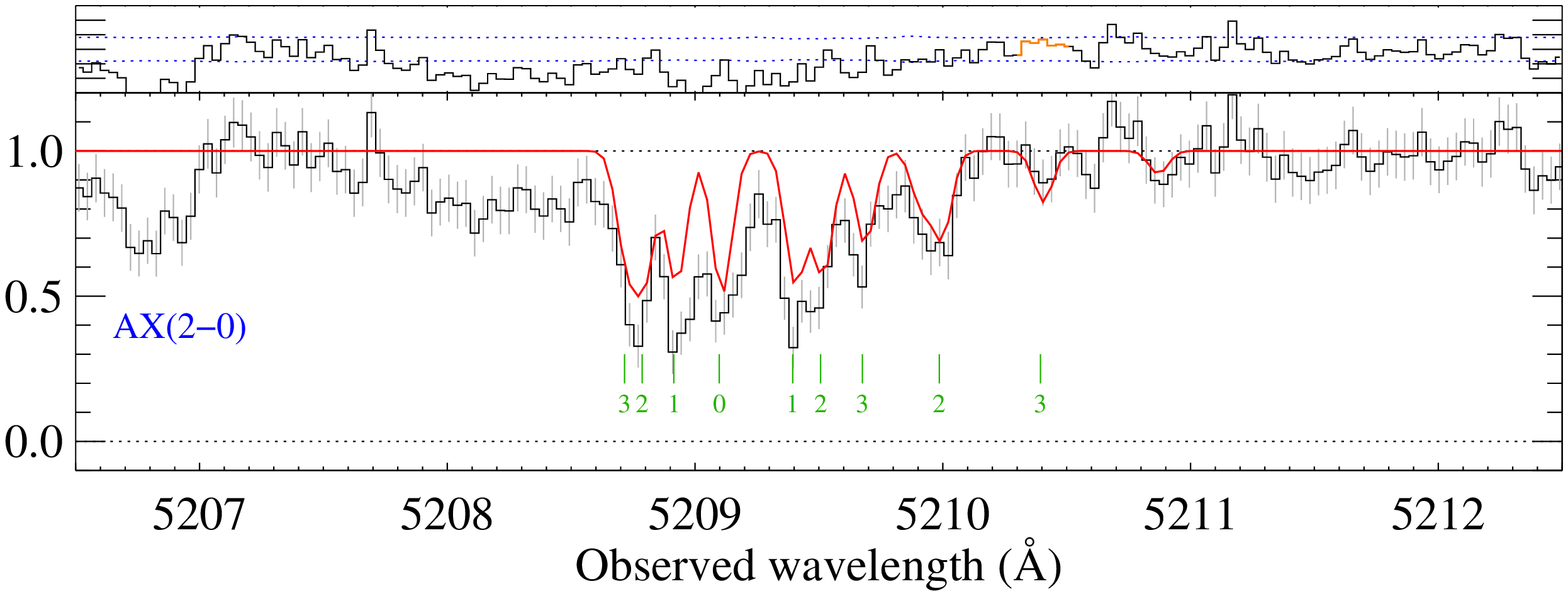} \\
    \includegraphics[bb=165 20 435 780, clip=,angle=90, width=0.48\hsize]{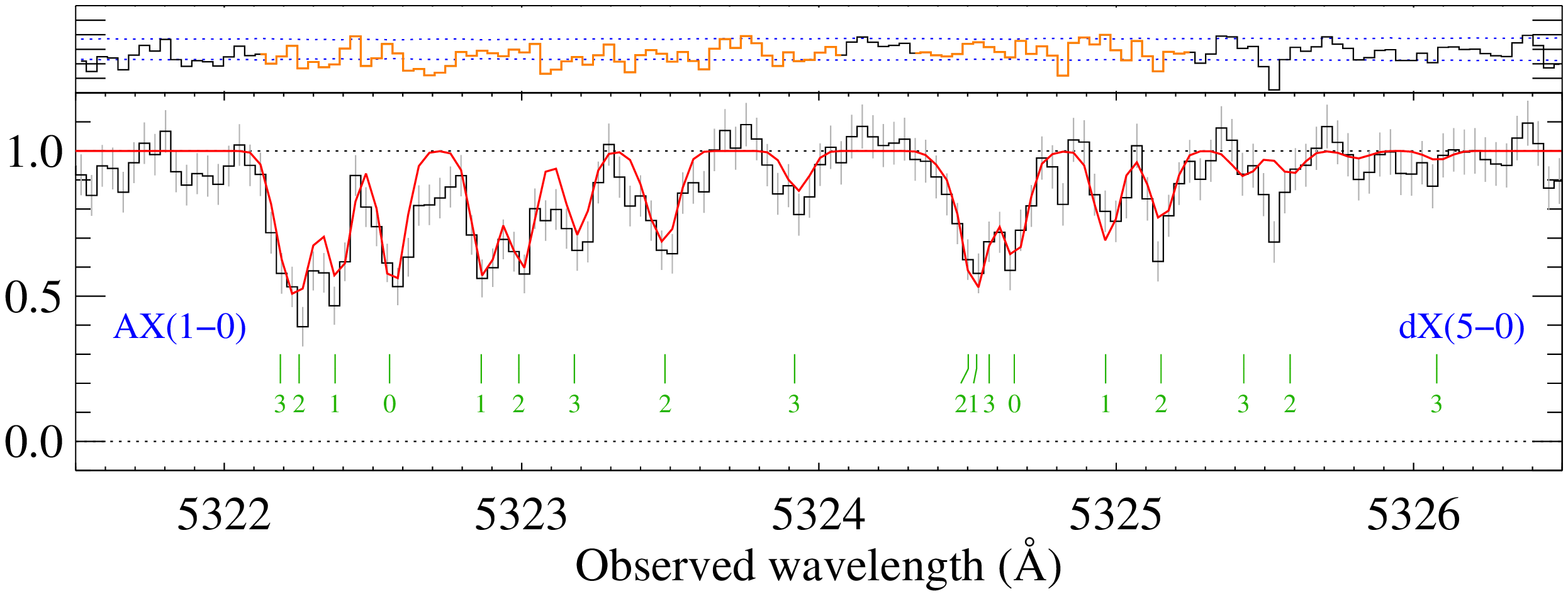}&
    \includegraphics[bb=165 20 435 780, clip=,angle=90, width=0.48\hsize]{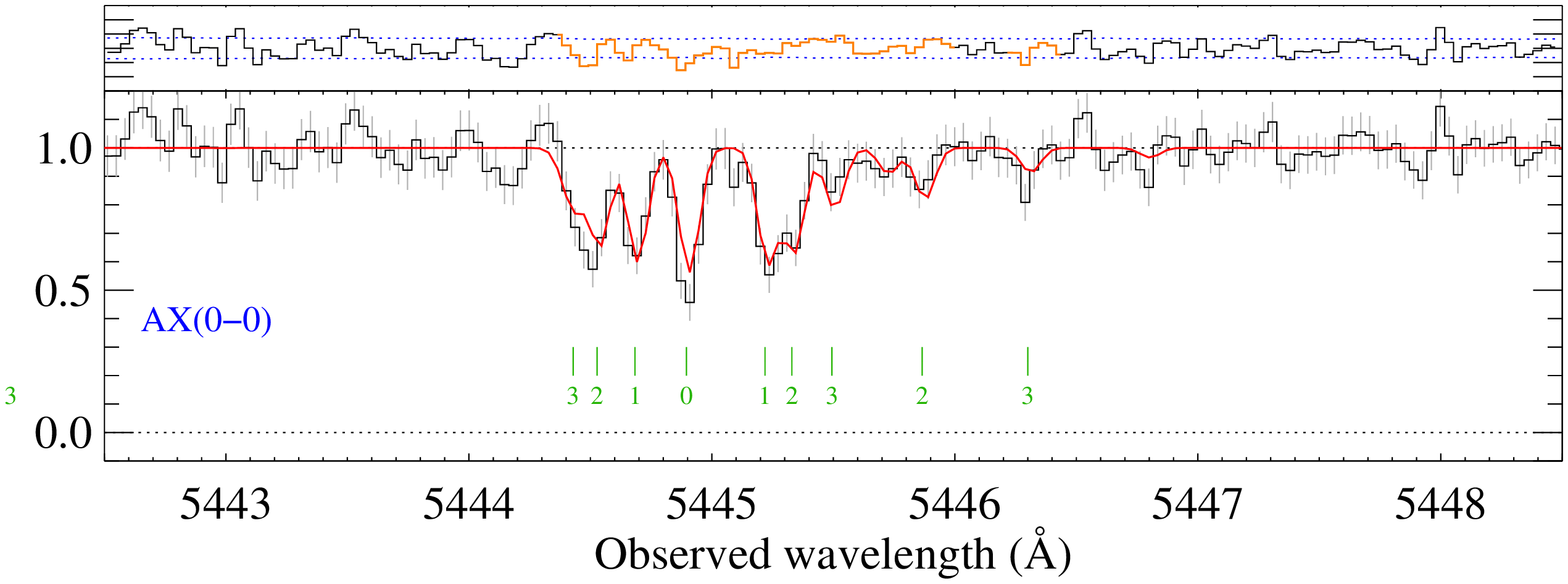}\\
\end{tabular}
\caption{Voigt-profile fit to the CO absorption bands labelled in blue in each panel (UVES data).
  Rotational levels from $J=0$ to $J=3$ are indicated as green tick marks. The panel above
  each region shows the residuals where the blue line indicates the $\pm$1\,$\sigma$ interval, and the orange regions
mark those used to constrain the fit. \label{f:CO}}
\end{figure*}

\begin{figure}%[!ht]
  \centering
  \includegraphics[bb=165 20 435 780, clip=,angle=90, width=\hsize]{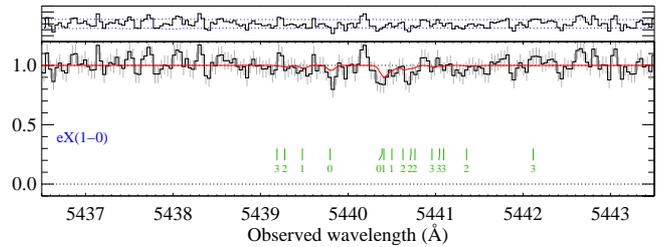}
  \caption{Tentative detection of $(1-0)$ band of the e$^3\Sigma^-$-X$^1\Sigma^+$ CO system (UVES data). The red profile correspond to the calculated
    synthetic spectrum, using the parameters obtained from fitting the AX and dX bands and molecular data from
    \citet{Eidelsberg03}.\label{f:COeX}}
\end{figure}

We detect CO absorption lines from ten bands, belonging to two systems: the A$^1\Pi(\nu')$-X$^1\Sigma^+(0)$ for $\nu'=0$ to 8
and the d$^3\Delta(5)$-X$^1\Sigma^+(0)$ inter-band system, see Fig.~\ref{f:CO}.  We also tentatively detect the e$^3\Sigma^-$-X$^1\Sigma^+$ system, although the lines remain
too weak to be significant (Fig.~\ref{f:COeX}). Rotational levels are unambiguously detected from J=0 to J=3. The J=4 lines are at
the noise level for most bands, but included in the fit. 
We used updated molecular data summarised in \citet{Dapra16}. Accurate wavelengths were obtained through calibration by laser
and VUV synchrotron studies \citep{Salumbides12, Niu13, Niu15}, while oscillator strengths and damping constants were carefully
re-evaluated taking into account an updated pertubation analysis.
We do not consider the $\nu'=5$ AX system, which is
completely blended with the extended \SiIV$\lambda$1393 absorption. Similarly, a large region of the (2-0) absorption band is contaminated
by unrelated absorption and ignored during the fitting. Finally, the (6-0) band is also partially blended with smooth absorption from the
\lya\ forest, but the latter is well modelled using a single component. We therefore included this band and left the intervening \lya\ parameters
free during the fitting process.
The other CO bands are apparently free from blending. 
We tied together the Doppler
parameter and redshift for the different rotational levels. 
We are therefore left with seven free parameters for CO: $b$, $z$ and the column densities for the 5 detected rotational levels.
We obtain a satisfactory fit with a global $\chi^2_\nu=1.1$, shown as the red profile in Fig.~\ref{f:CO}, with $z=2.525467$, $b=0.7$~\kms\ and obtain a total
CO column density of almost 10$^{15}$~\cmsq, which is the highest measured among high-$z$ quasar absorption systems to date. We further test the robustness of our measurement in the appendix.

From the non-detection of $^{13}$CO lines, we also constrain the isotopic ratio $^{12}$CO$/^{13}$CO to be higher than 40, assuming
the same Doppler parameter, redshift and excitation temperature for both molecules. This is comparable with values in the Solar
neighbourhood \citep[$^{12}$CO$/^{13}$CO$\sim 70$; ][]{Sheffer07}, meaning that a measurement of the CO isotopic ratio at high-$z$ should
be possible in the near future. This is particularly interesting since the isotopic ratio seems to be anticorrelated with $N$(CO) in
the Galaxy,
indicating $^{13}$CO enhancement through chemical fractionation in the denser and colder regions \citep[e.g.][]{Sonnentrucker07}.

%Automatically generated using results insynth79 on Wed Feb 10 15:04:54 2016
\begin{table}[!ht]
\centering
\caption{CO best-fit parameters}
\begin{tabular}{c c c }
\hline \hline
{\large \strut} Rot. level ($J$)  & $\log N($CO,$J)$      & $b$ (km\,s$^{-1}$)  \\
\hline
    0 & 14.43$\pm$0.12 & 0.71$\pm$0.03 \\
    1 & 14.52$\pm$0.08 &   ''          \\
    2 & 14.33$\pm$0.06 &   ''          \\
    3 & 13.73$\pm$0.05 &   ''          \\
    4 & 13.14$\pm$0.13 &   ''          \\
\hline
Total & 14.95$\pm$0.05 &   ''          \\
\hline
\end{tabular}
\end{table}

\begin{figure}
  \centering
  \includegraphics[bb=50 175 510 400, clip=,width=\hsize]{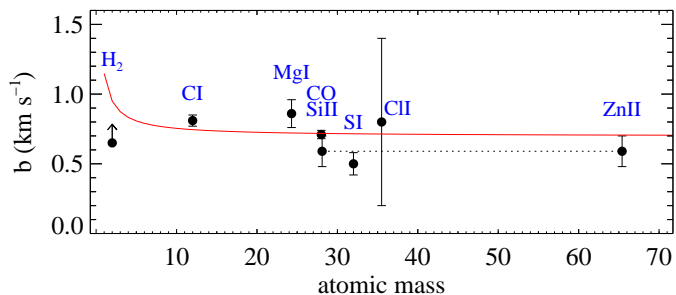}
  \caption{Doppler parameters and associated 1\,$\sigma$ uncertainties for different species in the cold component.
    The red curve represents the expected $b$-value for $T=50$~K and $b_{\rm turb}=0.7$~\kms.
    Note that the values for \SiII\ and \ZnII\ were tied together during the fit. The lower-limit
    to $b$(H$_2$) corresponds to thermal broadening only. \label{f:doppler}}
\end{figure}

\section{Metallicity and dust depletion \label{s:metallicity}}

\subsection{Metallicity in the atomic and molecular gas}

  In this section, we briefly discuss
  the metallicity in the different phases. We denote the abundance of a species M relative to hydrogen as
\begin{equation}
 [{\rm M/H}] \equiv \log(N({\rm M})/N({\rm H})) - \log(N({\rm M})/N({\rm H}))_{\odot}.
\end{equation}

\noindent where Solar abundances are taken from the photosperic values of \citet{Asplund09}.
H corresponds to the total hydrogen, i.e. including both neutral (\HI) and molecular (H$_2$) forms.
We measure $\log (N(\HI)+2N($H$_2))=21.07$. 
Using the undepleted zinc, we derive an overall super Solar metallicity, [Zn/H]~=~+0.46$\pm$0.45, where the large uncertainty is due to that on the
narrow component which contains most of the metals. Using phosphorus, we derive [P/H]$>$-0.04. This is a lower
  limit because only the strongest components are detectable and because of some phosphorus depletion in the ISM
  \citep[by about 0.2 to 0.5~dex, see e.g.][]{Lebouteiller05}. 
  While this is here again consistent with super-Solar metallicity, it
also indicates that the upper range from \ZnII\ is less likely. 

It is remarkable that if we assume a molecular fraction of one in the cold component and zero elsewhere,
then we get a lower limit to the metallicity of the ``warm'' gas to be [Zn/H]$_{\rm w} \ge -0.2\pm 0.1$, i.e., still consistent with Solar.
We also note that the bluest components ($v<-200$~\kms) of zinc could not be
constrained because of blending with unrelated lines, but these are expected to account for a marginal fraction of the
total metal column anyway. 
Using \SiII\ we obtain [Si/H]$_{\rm w} > -0.6$. This should be considered as a conservative lower limit 
since silicon depletion is expected to occur (see indeed Sect.~\ref{s:deple}).

While we cannot measure the \HI\ column density in individual components, we can expect that the metallicity in the cold component
is at least as high as in the rest of the profile and obtain a more realistic lower limit. We assume 

\begin{equation}
  {\rm [Zn/H]_c \ge [Zn/H]_{\rm overall}},
\end{equation}

\noindent where the index $c$ stands for ``cold'', i.e. associated to the molecule-bearing gas.
The molecular fraction in the cold component,
$f_{\rm c} = 2N($H$_2)/(2N($H$_2)+N(\HI)_{\rm c})$ can then be expressed as

\begin{equation}
f_{\rm c} \ga f {{N(\ZnII)_{\rm overall}} \over {N(\ZnII)_{\rm c}}} \approx 0.54.
\end{equation}

Conversely, if we assume that the cold component is fully molecular (i.e. $f_c=1$), we get an upper-limit to the metallicity
in that component, [Zn/H]$_{c} < 0.7 \pm 0.5$, while the lower-limit to the metallicity in the warm gas is [Zn/H]$_{\rm w} > -0.2$
(see Sect.~\ref{s:met}).

Because chlorine is associated to the H$_2$-bearing gas, its abundance can also be used to constrain the metallicity
of the latter using the relation from \citet{Balashev15}

\begin{equation}
{\rm [Cl/H]} = {\rm [Cl/H_2]} + \log{f}, 
\end{equation}

\noindent where

\begin{equation}
{\rm [Cl/H_2]} = \log\left( \frac{N({\mbox{Cl})}}{2N({\rm H}_2)} \right)  - \log\left(\frac{{\rm Cl}}{{\rm H}}\right)_{\odot}.
\label{ClH2}
\end{equation}

We measure [Cl/H$_2$]$\sim$0.4$\pm$0.3 using the fit with fixed Doppler parameter for the main component. The lower limit to the molecular fraction $f > 0.46$ (see Sect.~\ref{s:H2}) then
implies [Cl/H]~$>$~0.05$\pm$0.3 in the cold component. This is a strict limit on the metallicity since several studies
have argued for some depletion of chlorine, by about a factor of two  \citep[see e.g.][and references therein]{Moomey12}. The abundance
of chlorine is therefore consistent with the super-solar metallicity derived from zinc and phosphorus, assuming instrinsic solar ratio.
We also note that assuming uniform metallicity across the different components implies that about 95\% of H$_2$ resides in the main component.

\subsection{Dust abundance from depletion of refractory elements \label{s:deple}}

\citet{Ledoux03} have revealed the existence of a relation between the presence of molecular hydrogen
and both the overall metallicity \citep[see also][]{Petitjean06} and the observed depletion factors in
DLAs, i.e., the probability to detect H$_2$ is higher when the {\sl relative} abundances of metals (or dust)
are high. \citet{Noterdaeme08} further showed that the column density of H$_2$ is strongly related to
that of dust, quantified by the column density of iron missing from the gas phase \citep[$N({\rm Fe_{dust}})$,][]{Vladilo06}. 
However, the metallicity and depletion factors are only indicative of the average values over the whole absorption
path probed by the DLA while the $N({\rm Fe_{dust}})$ corresponds to an integrated value.
Indeed, metals probe gas over a wide range of physical conditions, making it generally difficult to associate a
given metal component to a molecular one.
Still, it has been possible to show that abundance ratios along the velocity profiles tend to show an enhanced
depletion factor at the velocity where H$_2$ is detected \citep{Rodriguez06}.

\begin{figure}
  \centering
  \begin{tabular}{c}
    \includegraphics[bb=70 220 530 398,clip=,width=\hsize]{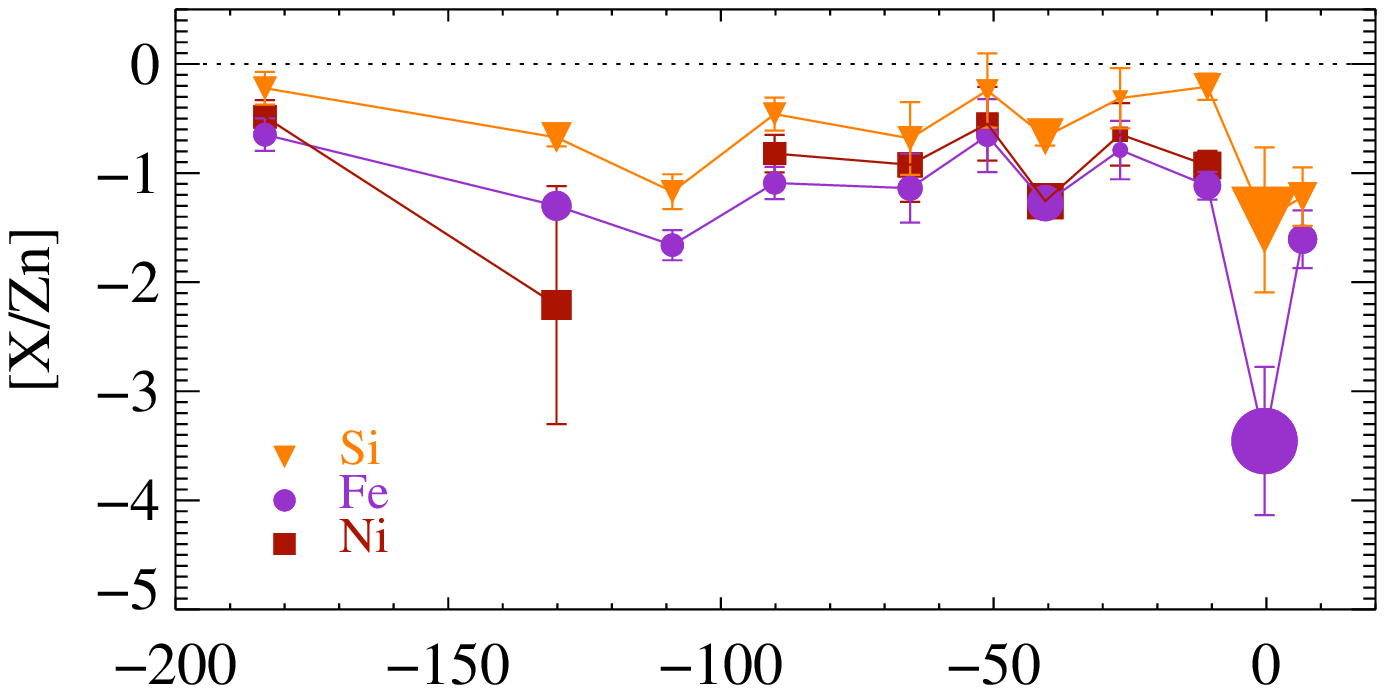}\\
    \includegraphics[bb=70 170 530 398,clip=,width=\hsize]{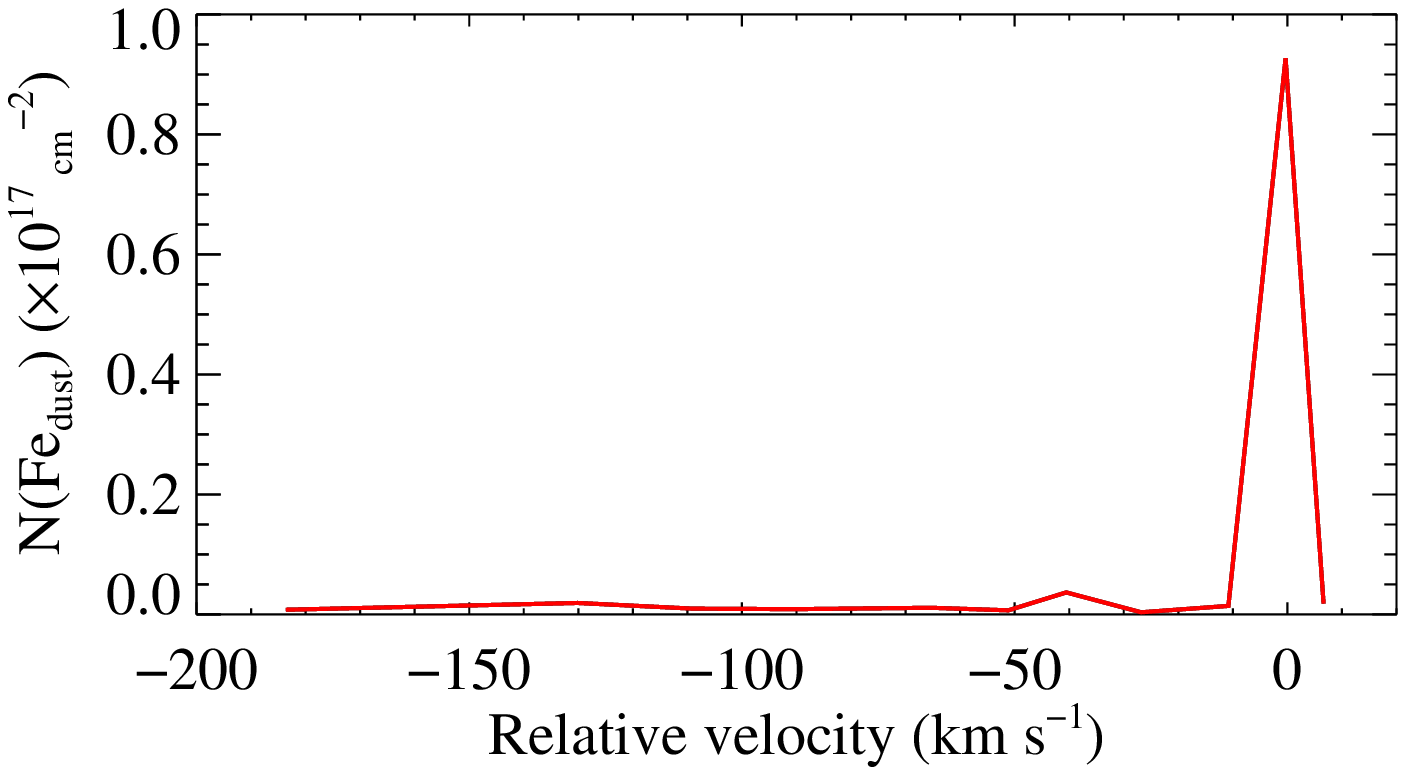}
    \end{tabular}
  \caption{Top: Depletion of silicon, iron and nickel relative to zinc in individual components (left axis). Note that
    several [Ni/Zn] are not available due to the column density of nickel being less than the detection limit. The size
    of each point is representative of the corresponding \ZnII\ column density. Bottom: Column density of
    iron locked into dust.\label{f:deple}}
\end{figure}

The system towards \jz\ presents an excellent opportunity to study this further, since the metal profile presents
a well defined narrow component corresponding to the molecular gas.
Fig.~\ref{f:deple} presents the observed depletion factors (Si, Ni and Fe relative to Zn) component
by component. The three patterns follow well each other, indicating that the abundances ratios are mainly dictated
by depletion onto dust grains, rather than differential nucleosynthesis. As expected, the cold narrow component
presents a high level of dust depletion, indicating that this component has a high {\sl relative} amount of dust.
However, a dusty component does not necessarily have a high integrated column density of dust, which will be more
naturally related to the column density (and hence detectability) of molecular species. We therefore compute the
column density of iron locked into dust grains component by component. The cold, molecule-rich component becomes
clearly visible and likely responsible for most of the extinction of the background quasar (see next section). We also notice
a secondary peak at $v\approx -40$~\kms. Interestingly, this corresponds to the location of a neutral chlorine component,
which also likely harbours H$_2$ molecules, though with a lower column density. This suggests that neutral chlorine could
be directly used as a ``high-resolution'' tracer of dust within a DLA.

\section{Extinction of the background quasar light \label{s:ext}}

The spectrum of the quasar \jz\ shows clear signs of being reddened, with a clear
2175~{\AA} absorption bump at the redshift of the DLA.
This, together with the detection of neutral and molecular species strongly suggests that the dust reddening is caused by the absorber. In the following analysis, we use the quasar template of \citet{Selsing16}.

Instead of assuming a fixed extinction curve parametrisation (e.g., Small Magellanic Cloud type as typically assumed in the litterature), we are able to constrain the extinction curve towards this quasar in high detail thanks to the long wavelength coverage of X-Shooter. For this purpose, we define continuum regions in the observed spectrum which are not strongly influenced by absorption (both telluric and from the DLA) or broad emission lines. In the near-infrared, we perform a 5-$\sigma$ clipping in order to discard outliers introduced by the removal of skylines in the data reduction.
We include photometry in the $K$-band from the UKIRT Infrared Deep Sky Survey (UKIDSS) and in band 1 from the
Wide-Field Infrared Survey Explorer (WISE). We are not able to include the redder bands from WISE, since the quasar template at this redshift only covers part of band 2 at 4.6 $\mu$m. However, we observe an offset between the UKIDSS photometry and the SDSS and WISE photometry. This is most plausibly due to variability of the quasar between the different epochs of observation. In order to correct for this offset, we scale the spectrum to the $z$-band of the SDSS photometry and subsequently scale the four UKIDSS bands to match the synthetic photometry calculated from the scaled spectrum.

The template of Selsing et al. is then smoothed with a Gaussian kernel ($\sigma$ = 7~pixels) to prevent the noise in the template to falsely fit noise peaks in the real data. In order to take into account the uncertainty in the template, we convolve the errors on the spectrum with the uncertainty estimate from the template.

We then fit the template to the data using 9 free parameters: 7 parameters to describe the extinction curve shape, a freely varying amount of dust, $A_V$, and an arbitrary scale since we do not know the intrinsic brightness prior to reddening.

\subsection{Parametrisation of the extinction law}
\defcitealias{FM2007}{FM2007}
We use a slightly modified version of the formalism from \citet[][hereafter, FM2007]{FM2007}:

\begin{equation}
  k(\lambda - V) = E(\lambda - V)/E(B - V) = c_1 + c_2 x + c_3 D ~,
\end{equation}

\noindent where
\begin{equation}
D = \frac{x^2} {(x^2 - x_0^2)^2 + x^2 \gamma^2}
\end{equation}

\noindent and $x = \lambda^{-1}$ refers to inverse wavelength in units of $\mu$m$^{-1}$ at the absorber rest-frame.
This corresponds to a linear component for the whole UV range (defined by $c_1$ and $c_2$) plus a 2175\~{\AA} bump, parametrised by $c_3$, $x_0$ and $\gamma$. FM2007 also consider a far-UV curvature component parametrised by $c_4$ at wavelengths shorter than $c_5$ (their Eq.~2). We do not
consider this component here (i.e. we set $c_4=0$) because we do not have enough data in the FUV and the quasar template is more uncertain
at very short wavelengths. In addition, preliminary fits to the data indicate that $c_4$ is little constrained and fully consistent with $0$. We therefore exclude it in the following analysis to simplify the fit without loss of generality.

In the infrared (IR), we use the power-law prescription of \citetalias{FM2007} assuming the correlation between $k_{\rm IR}$ and $R_{\rm V}$, thus yielding an extinction curve of the form (eq. 7 of \citetalias{FM2007}):

\begin{equation}
 k(\lambda - V) = (-0.83 + 0.63 \times R_V) \times x^{1.84} - R_V 
\end{equation}

Since this part of the extinction curve is beyond the spectral coverage we choose to reduce the two original IR parameters, by including the correlation between $k_{\rm IR}$ and $R_V$.
We use a spline interpolation as in \citetalias{FM2007}, in the optical range using one anchor point in the optical to ensure a correct normalisation in the $V$-band. In order to obtain a smooth and continuous transition between the various parts, we include two anchor points in the UV and two in the IR. We use the UV points $U_1$ and $U_2$ as defined in \citetalias{FM2007} at 2600 and 2700~{\AA}, respectively. In the IR, we anchor the spline at 0.75 and 1.0 ${\rm \mu m}^{-1}$ (similar to the $I_4$ and $I_5$ points of FM2007).

Finally, we convert the extinction curve from the original formulation in terms of $E(\lambda-V)$ to use $A_{\lambda}$:
\begin{equation}
A_{\lambda}/A_V = \frac{1}{R_V} E(\lambda-V)/E(B-V) + 1.
\end{equation}

\subsection{Fitting the extinction}

We fit the parameters using a Markov Chain Monte Carlo method as implemented in the python package {\sc emcee} \citep{Foreman-Mackey13}. This way we can include priors and parameter boundaries in a straightforward way. The shape parameters for the 2175~{\AA} bump, $x_0$ and $\gamma$, are given quite strong priors, since these parameters are observed to be very well behaved in many different environments \citep{FM1990, Gordon03}. As priors on the two parameters, we use the average values from \citet{Gordon03}, $x_0 = 4.57 \pm 0.01$ and $\gamma = 0.94 \pm 0.02$.

In order to give the photometry a more appropriate weight compared to the densely sampled spectral data, we calculate an effective number of pixels per filter. We calculate this quantity by integrating the filter transmission curves scaled to a maximum of $1$ and interpolated onto a grid with the same sampling as the spectral data. This way, pixels with high transmission are weighted more than `pixels’ with low transmission. The uncertainty for each filter is then divided by the square of this number.

The chain is initiated with 100 walkers located at initial locations around the best-fit from a quick $\chi^2$ minimisation. We then run the chain for 1200 iterations and discard the first 600 as burn-in. From the posterior distributions we obtain the best-fit parameters stated in Table~\ref{t:extlaw}. We furthermore provide the inferred $A_{\rm bump}$ which measures the strength of the 2175~{\AA} bump. This quantity is defined in the following way: $A_{\rm bump} = \pi c_3 /(2 \gamma R_V) \times A_V$.

\begin{figure*}
  \centering
  \includegraphics[width=\hsize]{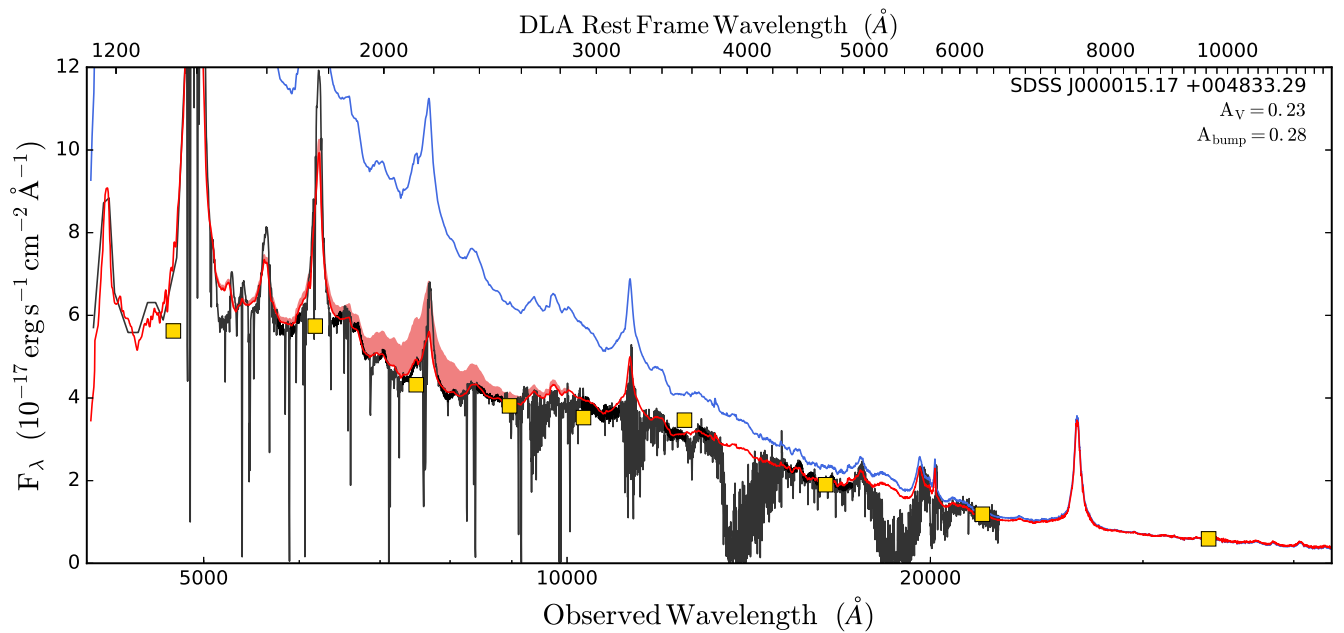}
  \caption{Combined 1D X-Shooter spectrum of \jz\ in black. The data has been replaced by the
    continuum over the \lya\ forest.
    The yellow squares indicate the photometry in $u$, $g$, $r$, $i$, $z$, $Y$, $J$, $H$, $K$, and $W1$ bands (left to right). The uncertainty on the photometry is smaller than the extent of the square marker. The blue line shows the unreddened, smoothed template by \citet{Selsing16}, and the red line indicates the same template reddened by the best-fit extinction curve by an amount of $A_V = 0.23$. The red shaded area marks the strength of the 2175~{\AA} bump. The upper edge of the shaded region corresponds to an extinction curve with no 2175~{\AA} bump. \label{f:sed}}
  \end{figure*}

\begin{figure}
  \centering
  \includegraphics[width=\hsize]{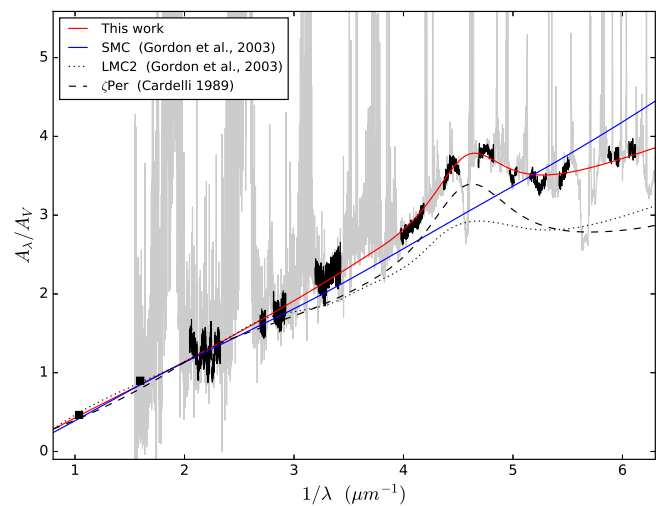}
  \caption{Extinction curve as function of inverse wavelength at the rest-frame of the DLA normalised to the $V$-band extinction $A_V$. The gray line shows the full X-Shooter spectrum divided by the template of \citet{Selsing16}. The black regions indicate the wavelength regions used in the fit, and the two black squares indicated the photometry in the $K$ and $W1$ bands. The red, solid line shows the best fit extinction curve, a clear 2175~{\AA} bump is observed in the data. For comparison, the extinction curves from Gordon et al. (2003) are shown as solid blue (SMC), and gray dotted (LMC2) lines. The dashed line corresponds to a Milky-Way extinction law with the single parameter, $R_V$, corresponding to $\zeta$Per \citep{Cardelli89}. A full derivation of the extinction curve towards this star is not available. \label{f:ext}}
  \end{figure}

The best solution is shown in Figure~\ref{f:sed} where the reddened template is plotted on top of the spectral and photometric data. In Figure~\ref{f:ext}, we show the inferred extinction curve. 
For comparison, we also show the average extinction curves towards the Small Magellanic Cloud (SMC) and Large Magellanic Cloud supershell (LMC2) from \citet{Gordon03} as well as a Milky-Way extinction curve for the $R_V$ measured towards the B star
$\zeta$Per by \citet{Cardelli89}.

\begin{table}
  \centering
  \caption{Derived extinction curve parameters and associated 1\,$\sigma$ uncertainties. \label{t:extlaw}}
\begin{tabular}{c c}
\hline
\hline
Parameter    &    Best fit value    \\
\hline
$c_1$          & -2.62  $\pm$ 0.18   \\
$c_2$          &  2.24  $\pm$ 0.13   \\
$c_3$          &  2.70  $\pm$ 0.18   \\
$x_0$          &  4.593 $\pm$ 0.005  \\
$\gamma$      &  0.85  $\pm$ 0.02    \\
$R_V$          &  4.13 $\pm$ 0.39    \\
$A_V$          &  0.23  $\pm$ 0.01   \\
$\rm{log}(s)$  &  0.043 $\pm$ 0.003  \\
\hline
\end{tabular}
\end{table}

The uncertainty quoted on the $A_V$ from the best fit only includes the formal
statistical error. This error is not fully representative as it does not take
into account the intrinsic variations of the UV slope of the quasar.
The slope of the quasar might vary with respect to the used template spectrum,
which would lead us to infer a wrong amount of extinction. We have estimated
this systematic effect on our best-fit $A_V$ by varying the slope of the
used quasar template, by multiplying the template with a power-law normalized
at 5500~{\AA}. We note that this approach is only a rough approximation since
the quasar shape is poorly described by a single power-law at all wavelengths.
However, from about 1200~\AA\ to 1~$\mu$m in the quasar rest-frame, this is
a reasonable approximation. For each variation in the intrinsic slope, we fit the
data again. In this fitting procedure, we keep the extinction curve parameters fixed,
since varying the intrinsic slope and $R_V$ simultaneously leads to a completely
degenerate fit with non-physical fit parameters.

For a shallower slope (by +0.2~dex), we obtain a best-fit $A_V$ of 0.12~mag.
Conversely, for a steeper slope (smaller by $-$0.2~dex), we recover a {\it larger} best-fit value of
$A_V$=0.34~mag. This change in slope is consistent with the average spread of
intrinsic slopes observed in the literature \citep{VandenBerk01,Krawczyk15}.
Although the different slopes provide acceptable fits to the data, the best
fit is obtained with the original quasar template.

As mentioned above, changing the slope of the template will inevitably change
the slope of the recovered extinction curve. Although we cannot fit these two
quantities together, we can require the fit to reproduce a value of $R_V$
consistent with an average Milky Way sight-line ($R_V=3.1$), which is obtained
for a change in slope of +0.04~dex, which in return yields a best-fit $A_V$ of 0.17~mag.

\section{Modelling of the physical conditions in the cold cloud \label{s:phys}}

In this section, we aim at understanding the structure of the cold gas by modelling the physical conditions using the version
c13 of the spectral simulation code Cloudy \citep[last described in][]{Ferland13}. This code performs a self-consistent
calculation of the thermal, ionisation and chemical balance of both the gas and dust exposed to a radiation source, with
a full treatment of H$_2$ introduced by \citet{Shaw05}. The cloud is assumed to be old enough for all physical processes to be
in steady state.

\subsection{Geometry and turbulence}
We consider a plane-parallel geometry with radiation illuminating both surfaces of the cloud. Such geometry has been succesfully used
to reproduce the physical conditions in typical interstellar clouds \citep[e.g.][]{vanDishoeck86}.
We consider constant density models and stop the calculation
when reaching the observed H$_2$ column density (instead of $N(\HI)$ whose measurement encompasses the whole profile).
While H$_2$ is also likely present in more than one component, most of it should be found in the main cold component in
which CO is also found. We consider a turbulent broadening of 0.7~\kms,
as derived from the Doppler parameter of heavy elements, see Fig.~\ref{f:doppler}. This is mostly important for its effect on the CO
self-shielding with a negligible effect on H$_2$ due to the strong damping wings.

\subsection{Incident radiation field and cosmic rays}
The Haardt-Madau ionising UV background from both galaxies and quasar \citep[see][]{Haardt12} is included
at the absorber's redshift, and so is the CMB radiation. We also consider the presence of a local
source of UV radiation by adding a blackbody radiation with a temperature of $T=40\,000$~K, to simulate the presence of hot stars.
We parametrise the intensity of this blackbody radiation by $\chi$, the ratio of the assumed incident blackbody radiation to
the \citet{Habing68} field (compared in the range 0.44 to 1 Ryd). Since we aim at understanding primarily the conditions in the cold
cloud, we take into account the attenuation of the incident radiation after it went through neutral gas, removing photons
between 1 to 4 Ryd.
Cosmic rays are also included as they play a major role inside molecular regions, becoming the main source of ionisation
and impacting the ion-molecule chemistry in the cold gas. Indeed, the cosmic ray ionisation rate, $\zeta_{\rm H}$, is generally deduced from the
abundance of chemical ions in our own Galaxy, where it is also found to vary by a large amount between different regions
\citep[e.g.][]{Federman96}. We note however that this remains an active area of research, with more recent studies pointing towards
an average Galactic value one order of magnitude higher than previously found \citep[see e.g.][]{Indriolo07}.

\subsection{Abundances}
We set the metal abundances to 2.5 times Solar, as derived from the abundance of undepleted zinc, and assume intrinsic Solar ratio
for all species. We apply the observed depletion factor for iron and silicon, which are observed in their dominant ionisation stages.
Since we have no measurement of the {\sl total} abundance of other species, we apply the default depletion values for the cold
medium in the Galactic disc as compiled in Table~7.7 of the Hazy1 documentation of Cloudy. 

\subsection{Model with standard Galactic grains}
As a first test, we start modelling the cloud using a canonical Milky-Way ISM dust grain mixture, with an abundance scaled to the metallicity,
i.e. we set the dust grain abundance 2.5 times the Galactic ISM value. Instead of running large grids of parameters, we
vary the main input parameters individually and study their effect on the predicted column densities. These parameters
are: total hydrogen volumic density ($n_H$), strength of UV field ($\chi$) and cosmic ray ionisation rate $\zeta_{\rm H}$.

The initial density is mostly determined by matching the observed relative population of the \CI\ fine structure
level with the computed ionisation, chemical and thermal balance. Note that the density of different colliders is
calculated self-consistently across the cloud. We find that densities in the range 40-100~cm$^{-3}$ predicts \CI\ ratios
consistent with the observed ones. In turn, the CO/\CI\ ratio is strongly underpredicted by a factor of about 30.
This issue was also raised by \citet{Sonnentrucker07}, who noted that most published models of translucent
clouds predict less CO than observed for a given $N($H$_2)$.
We
find that decreasing $\chi$ does not help (as similarly concluded by \citealt{Bensch06} when modelling the emission
from the dark cloud Barnard 5 in the Perseus complex). \citet{Shaw08} showed that the CO column density increases almost linearly with
$\zeta_{\rm H}$, but increasing $N($CO) this way leads to a strong overproduction of \CI. We therefore have to re-evaluate our initial
assumptions but find that varying other parameters such as the shape of the incident UV field or the geometry of the cloud
does not help either.

\subsection{Model with small dust grains}
Interestingly, the abundances of most molecular and neutral species observed at $\zabs=2.5$ towards \jz\ are very similar
to those observed in the Perseus cloud along $\zeta$Per, that was successfully modelled by \citet{Shaw08}. These authors
used a higher number of small grains compared to a standard mixture to approximate the observed $R_{\rm V}$ and $E(B-V)$.
The $\zeta$Per extinction curve is also quite similar to that observed towards \jz, although the latter has a
steeper UV slope, probably indicating a smaller average grain size.
Small grains seems indeed to be a key
ingredient to increase the column density of CO with respect to that of carbon and molecular hydrogen. In other words, a higher total
surface of grains favors CO without going too deep into the cloud. \citet{Shaw16} also highlighted the need of increased grain
surface area to reproduce the CO column density towards SDSS\,J1439$+$1117 \citep{Srianand08}.

We therefore move to a second series of models, using a dust grains size distribution containing more small grains. 
Typically, the distribution for each grain type (graphite and silicates) is parametrised as a single power law 
with the form $dn/da \propto a^{-q}$, where $n$ is the number of grains with radius in the range [$a$,$a+da$]. \citet{Nozawa13}
have shown that a graphite-silicate model can reproduce well the range of extinction curves seen
in the Milky-Way and Small Magellanic Cloud with a remarkably constant power law index $q\approx 3.5 \pm 0.2$, with a
cutoff at small ($a_{min}\sim 0.05$~$\mu$m) and large ($a_{max}\sim 0.2-0.3$~$\mu$m) grain radii. These authors note that they could
not determine well the small grain cutoff due to lack of data a short wavelengths, but that these have little effect on other 
parameters anyway. We fixed $q=3.5$ and find that a silicate-graphite mixture with a silicate-to-graphite ratio increased by 40\%
compared to canonical ISM mixture, $a_{min}=0.001$~$\mu$m and $a_{max}=0.15$~$\mu$m for both grain types reproduces well the extinction
curve derived in Sect.~\ref{s:ext}. In addition, we verified that the abundances of different metals
locked into the grains are consistent with the observed metallicity for the assumed depletion pattern to within 0.1~dex, or better.

Using the small-grains model, we are able to reproduce most of the column densities to within 0.4~dex
or better for most neutral, ionised, and molecular species, see Table~\ref{t:modout}, for
$\chi=0.5$ and $\zeta_{\rm H}=2.5\times10^{-15}$~s$^{-1}$ (Table~\ref{t:modin}). 
The column densities of \MgI\ remain however overpredicted by about 0.8 and the high rotational levels of H$_2$
are underpredicted.
It is likely that the actual depletion of magnesium is much higher
than assumed here \citep[see][]{DeCia16}. Indeed, for the high volumic density and molecular fraction estimated here, we can expect an order of
magnitude stronger depletion of magnesium \citep[see][]{Jensen07}, in which case the model perfectly matches the observations.
Finally, most of the column density in the high-$J$ levels of H$_2$ could arise mostly from an additional warmer component,
which is not modelled here. We also note that the predicted visual extinction ($A_V \sim 0.7$) is higher than what we derived through SED fitting or seen towards $\zeta$Per, but within a reasonable factor, given the uncertainties on the measurement and on 
the intrinsic quasar brightness. 

In conclusion, it is remarkable that our model reproduces fairly well the observed quantities with the small number
of parameters considered. Further fine-tuning of the parameters could still be performed, and other parameters previously
fixed could be varied. For example, a harder UV flux, with black-body
temperature of $T \sim 60\,000$~K increases the CO column density by about 0.07~dex (hence a better agreement) while changing
little the other predicted values.  However, such fine-tuning is not the purpose of this paper, given
the other uncertainties (in particular on the metallicity) and assumptions involved.
We also remind that the true geometry of the cloud and its density profile is likely more complex than assumed.
Lastly, we note that the chemistry of CO is an evolving research field. At the low densities of the ISM, CO can be formed through different
paths and sequences of reaction. For example, radiative association between C$^+$ and H$_2$ (or with a much lower efficiency
with H$^0$) leads 
to CH$^+$, which then reacts with O to produce CO. Several chemistry networds (such as hydrogen-oxygen)
are initiated by cosmic rays, hence their importance in controlling the production of CO.
However, as discussed by \citet{Goldsmith13}, other physical phenomena can also raise the temperature of the gas,
leading to an increase of CO abundance, such as shock heating \citep{Elitzur78} or Alfv\'en waves \citep[e.g.][]{Federman96,Visser09}. Finally, dissipation of energy from supersonic turbulence can deeply modify the chemistry
of the gas \citep{Godard09}.

\begin{table}
  \centering
  \caption{Cloudy input parameters  \label{t:modin}}
  \begin{tabular}{c c}
    \hline \hline
    Parameter & Value \\
    \hline
    $n_{\rm H}$        & 80~cm$^{-3}$ \\
    $\zeta_{\rm H}$    & 2.5$\times10^{-15}$~s$^{-1}$ \\
    $\chi$           & 0.5 \\
    turbulence      & 0.7~\kms \\
    Metallicity\tablefootmark{a}            & 2.5~$Z_{\rm \odot}$ \\
    $\log {\rm (C/H)}$    & -3.57 \\
    $\log {\rm (O/H)}$    & -3.13 \\
    $\log {\rm (Na/H)}$    & -6.06 \\
    $\log {\rm (Mg/H)}$    & -4.70 \\
    $\log {\rm (Si/H)}$    & -5.49 \\
    $\log {\rm (Cl/H)}$    & -6.50 \\
    $\log {\rm (Fe/H)}$    & -6.50 \\
    $\log {\rm (Ni/H)}$    & -7.38 \\
    \hline
  \end{tabular}
    \tablefoot{
      \tablefoottext{a} {Abundance} for all species except those listed below. The gas-phase abundance for
      the listed depleted species correspond to the same intrinsic metallicity after appyling the depletion
      factors described in the text. }
  \end{table}

\begin{table}
  \centering
      \caption{Comparison of observed and predicted values in the cold component \label{t:modout}}
  \begin{tabular}{ccc}
    \hline \hline
    Quantity\tablefootmark{a} & observed value & model prediction  \\
    \hline
    \HI                      & $<20.9$        & 20.7           \\  
    CO                       & 14.95$\pm$0.05 & 14.72          \\    
    \CI                      & 16.22$\pm$0.07 & 16.40          \\
    \CI,$J=0$                & 16.10$\pm$0.08 & 16.19          \\
    \CI,$J=1$                & 15.54$\pm$0.14 & 15.93          \\
    \CI,$J=2$                & 14.67$\pm$0.11 & 15.04          \\
    \SI                      & 14.85$\pm$0.18 & 14.53          \\
    \ClI                     & 14.63$\pm$0.29 & 14.50          \\
    \MgI\tablefootmark{b}    & 14.10$\pm$0.10 & 14.92          \\
    \NaI                     & 15.00$\pm$0.30 & 14.51          \\
    $T_{\rm 01}$ (K)           & 51$\pm$2       & 56        \\    
    \hline
  \end{tabular}
  \tablefoot{
    \tablefoottext{a} {Except} for $T_{\rm 01}$, all quantities correspond to column densities expressed
    in $\log{(\cmsq)}$.
    \tablefoottext{b} {See} text for the apparently strong mismatch between model and observations.}
  \end{table}

\subsection{Characteristics of the cloud}

\begin{figure}
  \centering
  \begin{tabular}{c}
    \includegraphics[bb=50 220 510 400, clip=,width=\hsize]{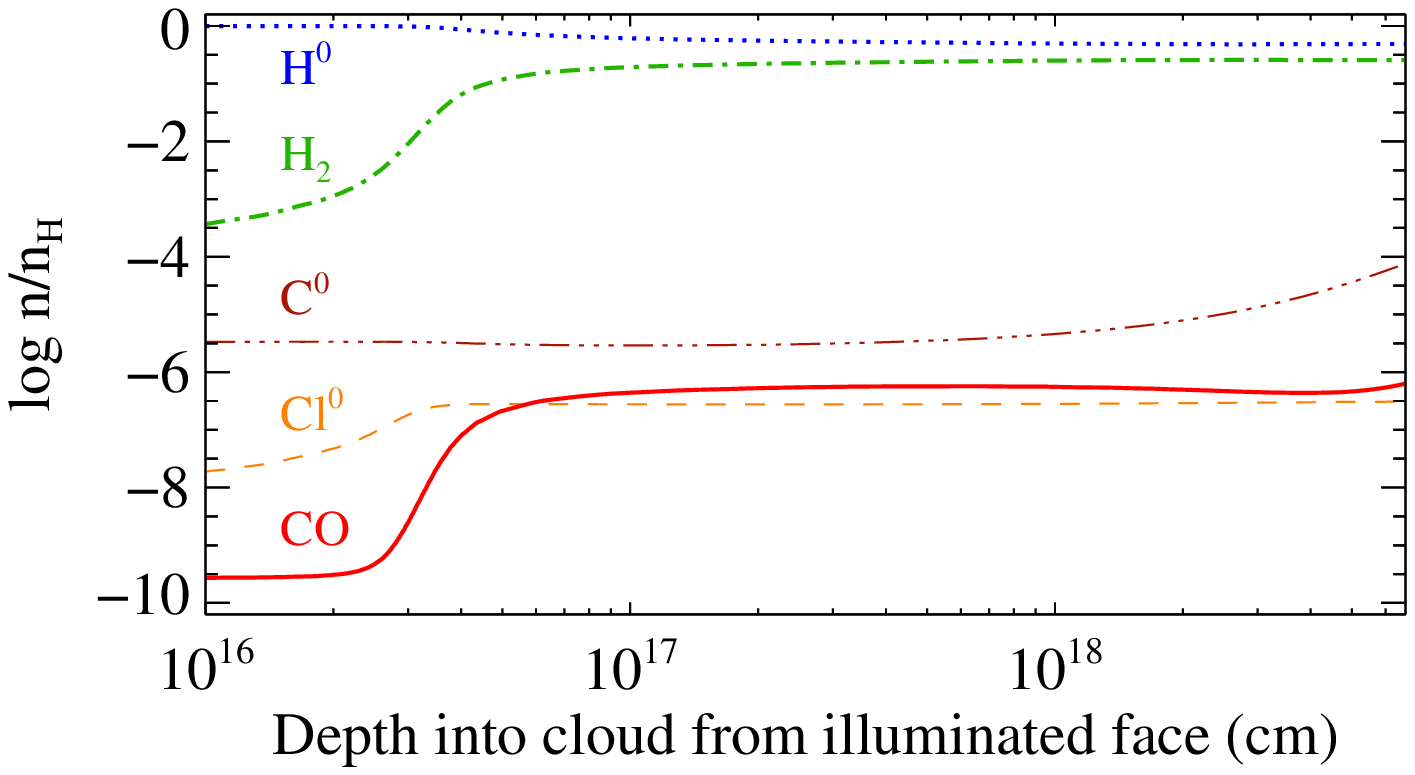}\\
     \includegraphics[bb=50 220 510 400, clip=,width=\hsize]{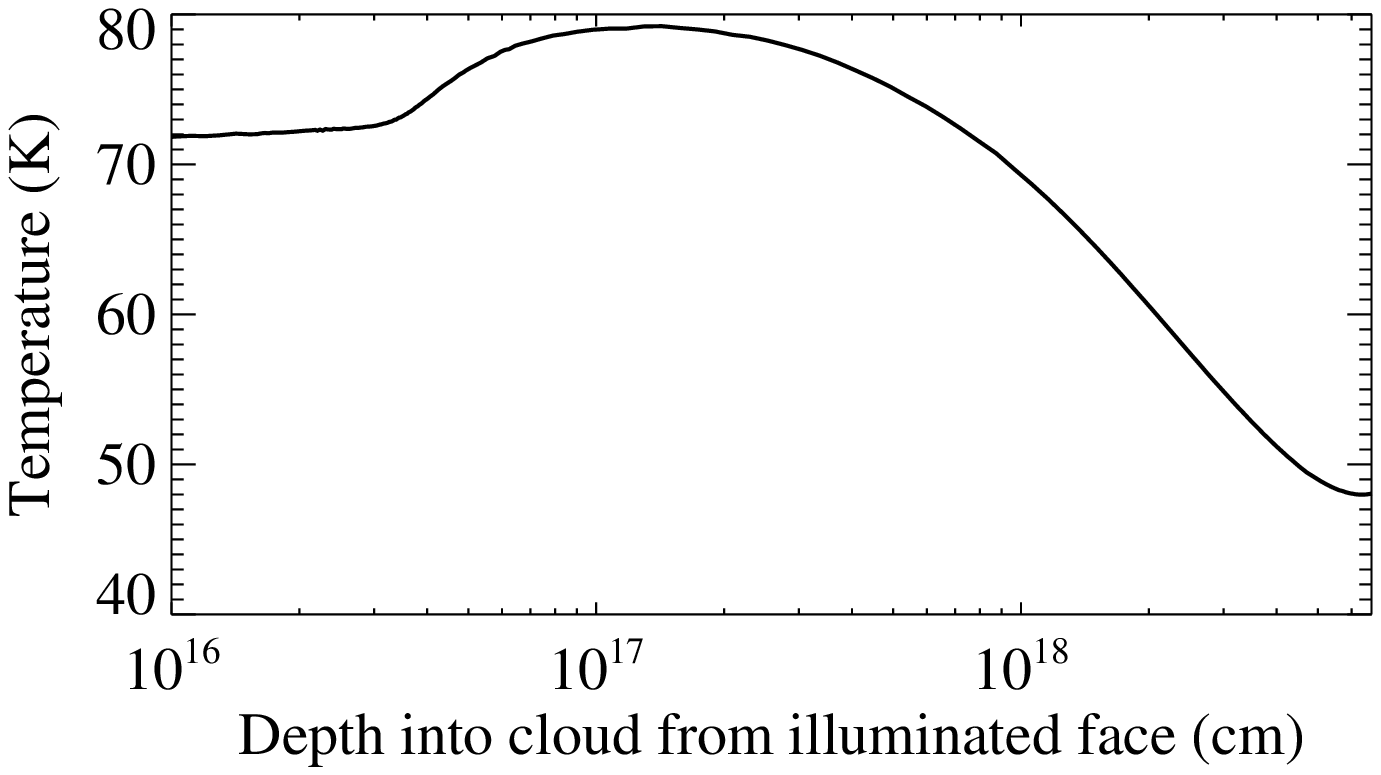}\\
    \includegraphics[bb=50 175 510 400, clip=,width=\hsize]{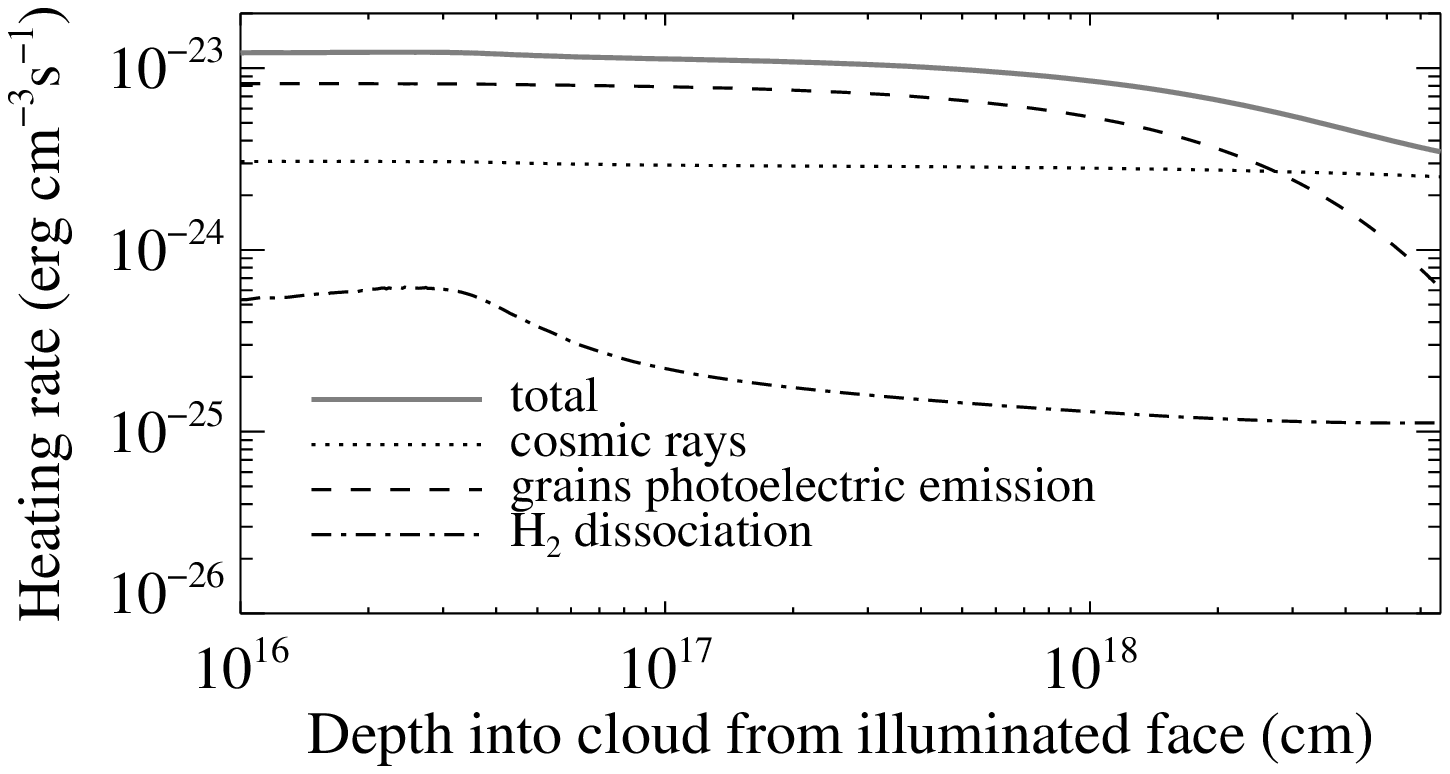}
  \end{tabular}
  \caption{Top panel: variation of the abundances of neutral atomic hydrogen (blue dotted), molecular hydrogen (green dashed-dotted), neutral chlorine (orange dashed) neutral carbon (brown, dashed-triple dotted) and
    carbon monoxide (red) as a function of depth into the cloud. Note that for H$_2$, the curve corresponds to half the molecular
    fraction (since $f = 2n_{\rm H2}/n_{\rm H}$).
    Middle panel: Variation of the temperature across the cloud. Bottom panel: Heating
    rates for the main sources. The cooling rate, which is largely dominated
    by [\CII]$157\mu$m emission, equals the total heating rate since the cloud is assumed in
    thermal equilibrium.  \label{f:cloudydepth}}
  \end{figure}

The top panel of Fig.~\ref{f:cloudydepth} shows the abundances of neutral atomic hydrogen, molecular hydrogen, neutral chlorine
and carbon monoxide
relative to the total amount of hydrogen as a function of the depth into the cloud. Only half of the profile
is shown, the other half being symmetric with the assumed geometry.
We find a total cloud size to be about 4~pc along the line of sight. The line of sight passes quickly through a sharp
\HI-to-H$_2$ transition, which is very well traced by neutral chlorine and after which the H$_2$ molecular fraction
stays constant at about 50\%. This incomplete conversion is due to destruction of H$_2$ in the cloud center, inside
the self-shielding layer \citep[see also][]{Liszt15}. 
The abundance of CO starts to be significant immediately after the \HI-to-H$_2$ transition. Indeed the chemical networks
leading to the formation of CO with H$_2$ as starting point are much more efficient than those starting with atomic
hydrogen \citep[see][]{Goldsmith13}. In addition, H$_2$ participates in the shielding of CO, through a few but important
overlaps with FUV CO electronic bands \citep[e.g.][]{Glassgold85,vanDishoeck88,Bensch06,Visser09}. The abundance of
  \CI\ does not vary much inside the molecular cloud, and do not directly follow H$_2$ nor CO. This is due to the abundance
  of \CI\ being mostly determined by the ionisation/recombination balance. In addition, neutral carbon does neither benefit
  from H$_2$ shielding nor participate directly in the molecules chemistry (which rather involve C$^+$). The fact that \CI\
  is considered a tracer for H$_2$ is therefore more indirect: \CI\ probes the conditions that favors the presence of H$_2$, without
  locally following the latter within layers of a given molecular cloud.

The temperature (central panel) varies slowly between $\sim$~70-80\,K and 48\,K towards the center of the cloud, in agreement
with the average temperature measured from the column densities in the first H$_2$ rotational levels. Indeed,
  we predict T$_{\rm 01} \sim 56$~K, very close to the measured value.
Photoelectric effect on dust grains is the main heating
source in the
external layers on the cloud while cosmic rays become the main heating source in the cloud center. Heating by
H$_2$ photo-dissociation contributes for a small fraction to the total heating and naturally decreases after the
\HI-to-H$_2$ transition. 
Cooling is in turn largely dominated by [\CII]$157\mu$m emission. The model predicts $\log N(\CII^*)=15.64$. Unfortunately, we cannot test this prediction since the measurement is 
 affected by a large uncertainty.

\section{Temperature of the Cosmic Microwave Background at $z=2.53$ \label{s:cmb}}

\begin{figure}
  \centering
\includegraphics[bb=70 170 500 580,clip=,width=0.95\hsize]{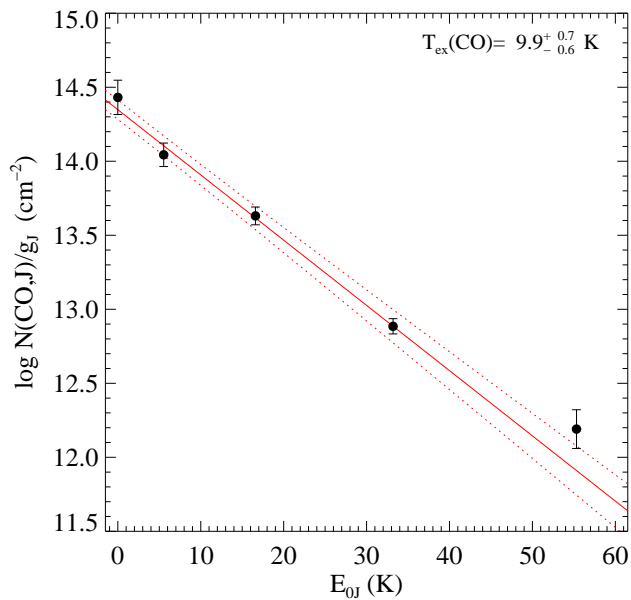}
\caption{CO excitation diagram. The solid line gives the best fit to the rotational
  population using a Boltzmann distribution, with the associated uncertainty shown as dashed
  line. \label{f:coext}}
\end{figure}

\citet{Srianand08} and \citet{Noterdaeme11} have shown that the excitation
of CO at high redshift is largely dominated by the CMB radiation, i.e.,
that other excitation processes such as collision are negligible, in which
case $T_{\rm ex}($CO$) \approx T_{\rm CMB}$.
More recently, \citet{Sobolev15} estimated a correction to apply if we do take into account
collisional excitation with hydrogen atoms and H$_2$ molecules. The correction
depends almost linearly on the total volumic density and on the kinetic temprature.
In addition,
since the main collision partner is H$_2$, the correction increases with the molecular fraction.
Using $f=0.5$, $T=50$~K and $n_{\rm H}=80$~\cmsq, their Eq.~7 implies an excess temperature of
$\Delta T=$~0.3\,K. Applying this correction to the observed CO excitation temperature towards
\jz\ ($T_{\rm ex}($CO$)=9.9^{+0.7}_{-0.6}$~K, see Fig.~\ref{f:coext}) we obtain $T_{\rm CMB}(z=2.53)=9.6^{+0.7}_{-0.6}$~K in excellent
agreement with the expected CMB temperature in the standard hot Big-Bang theory ($T_{\rm CMB}=T_{0}(1+z)=9.61$~K).
Using the statistical equilibrium radiative transfer code RADEX \citep{vanderTak07}, that takes as input
the density of various colliders, the kinetic temperature, CO Doppler parameter and column density, we
also obtain very similar results, within 0.1~K. We finally remind that since the expected
  excess temperature is almost linearly dependent on the both kinetic temperature and the density, 
  good constraints on these quantities are highly valuable.

Our Cloudy model predicts $T_{\rm ex}($CO$)=10$~K when the expected
CMB radiation field at $z=2.53$ is included, in agreement with the observed
temperature. To test further the effect of the CMB radiation, we run a model
setting the CMB background temperature to the value measured at $z=0$ instead \citep[$T_{0}=2.726$~K, ][]{Fixsen09},
while keeping all other parameters the same as previously. This model predicts a CO excitation temperature
of 3.8~K, that we derive from the predicted population of the first three rotational levels since higher rotational
levels do not follow the Boltzmann distribution anymore.
The predicted temperature would then be similar to that seen in diffuse Galactic environments \citep{Burgh07} but
about 10$\sigma$ away from the measured value towards \jz.
We also note that the difference between
CMB temperature and CO excitation temperature tends to vanish at high redshift, where
excitation by CMB becomes strongly dominant over collisional excitation.

Our result suggests ways to improve the constraints on the evolution of the CMB
temperature at high redshift. From the distribution of the CO rotational levels
alone, any departure from the Boltzmann distribution will be due to non-CMB (presumably
collisionnal) processes. This means that a measurement remains possible using
the low rotational levels, as in \citet{Noterdaeme10b}. However, a good understanding
of the physical conditions in the cloud, and in particular the temperature and density
-- that can be very effectively constrained by the observations of low-$J$ H$_2$ lines
and \CI\ fine-structure levels-- allows one to correct for collisional excitation. The
excess temperature for diffuse molecular cloud will typically be of the order of 0.3~K,
i.e. a few times larger than the expected measurement uncertainty achievable with
future high resolution spectographs on extremely large telescopes. This will
therefore need to be carefully evaluated. 

Cyanogen (CN) is known to be an even better thermometer of the CMB temperature, with negligible
contribution from collisions. Very accurate measurements have been performed using
absorption spectroscopy towards nearby stars, including observations towards $\zeta$Per
\citep[e.g.][]{Roth95,Ritchey11}. The similarity with our line of sight provides a good hope
that CN lines will also be detectable in systems like \jz. Indeed, a strong correlation between $N($CO)
and $N($CN) is observed in our Galaxy \citep[e.g. Fig.~18 of][]{Sonnentrucker07}, with
$N$(CN) of a few times 10$^{12}$~\cmsq\ at the CO column density of our system. 
However, obtaining high S/N and
resolution in the NIR where the CN$\lambda$3875 lines are redshifted is technically challenging.
In the case of \jz, the lines unfortunately fall in a region of low transmission due to water
vapour in the atmosphere. This means that the search for similar molecular systems at higher redshift
(e.g. $z\sim 2.9-3.5$ for the CN lines to be redshifted in the H-band) needs to be continued. The CMB contribution
for such systems will also be higher and their constraint on the evolution of CMB temperature stronger.

\section{Direct search for star-formation activity \label{s:sfr}}

Motivated by the similarity between this absorbing cloud and the local ISM, we search for the emission
associated with active star formation. The main indicator of star formation, available to us in the
data at hand, is the nebular emission lines from the photo-ionised gas in the vicinity of young stars.
The most prominent of these lines in the rest-frame UV and optical are the two principal Balmer lines
(H$\alpha$ and H$\beta$) and the two doublets from singly and doubly ionised oxygen
([\ion{O}{ii}]$\lambda\lambda$3726,3728 and [\ion{O}{iii}] $\lambda\lambda$ 4959, 5007), which at the
redshift of the DLA fall in the NIR spectrum from X-Shooter.
Unfortunately the H$\alpha$ line falls in the last order of the
NIR spectrum, which is corrupted by strong sky background. 
We therefore search for emission from the DLA counterpart using the following lines: H$\beta$,
[\ion{O}{iii}], and [\ion{O}{ii}]. However, we do not detect emission from any of the lines. 

For each PA, we first subtract the quasar trace in the 2D spectrum using a similar
approach as described in \citet{Krogager13} in order to search for the faint emission lines. In the
following, we have combined the observations taken with a same position angle.

We use an elliptical extraction aperture to search for the lines. In the spatial direction, we use a semi-minor axis extent of one full width at half maximum of the spectral trace (FWHM = 5 pixels). In the spectral direction, we use a semi-major axis of $400$~km~s$^{-1}$ around the expected location given from the redshift of the DLA. However, we observe no flux in any of the PAs. We derived the detection limit for the [\ion{O}{ii}] line by modelling 10\,000 emission line profiles with various line fluxes (drawn randomly in log-space between $1 \times 10^{-18}$ and $1\times10^{-16}~{\rm erg}~{\rm s}^{-1}~{\rm cm}^{-2}$), line widths (drawn randomly between 50 and 350~km~s$^{-1}$), and line strength ratio for the two transitions of the doublet \citep[randomly drawn between 0.35 and 1.5; see][]{Seaton57}. Each model line profile (assumed to be Gaussian) is convolved with the spatial broadening function, determined from the spatial profile of the quasar trace, to create a 2D line model. We subsequently add noise to the 2D line model according to the noise model from the 2D spectrum. For each realisation, we extract the flux and noise within the elliptical aperture mentioned above. This results in a robust limit on the detectable flux within the aperture, given the fact that we do not know the intrinsic line width of the [\ion{O}{ii}] line.
We quote all the non-detections for the individual PAs as 3$\sigma$ upper limits in Table~\ref{t:oii}.

\begin{figure}
  \centering
  \includegraphics[width=0.9\hsize]{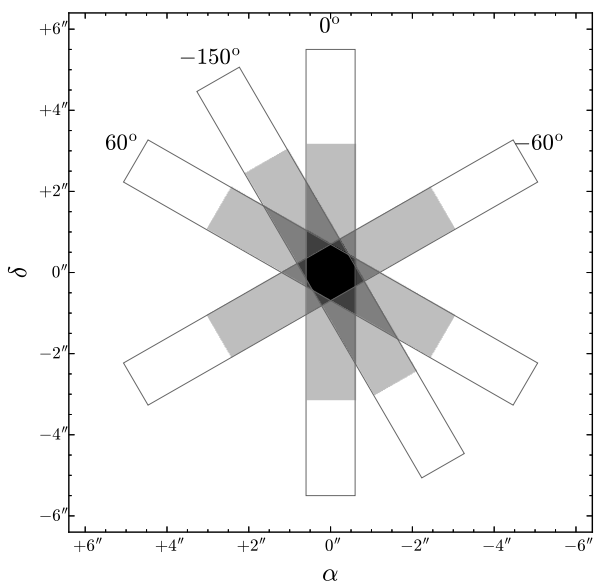}
  \caption{
The positions of the X-Shooter slits on the sky relative to the quasar position. The thin gray lines indicate the full X-Shooter slit length (11 arcsec). The colour coding indicates effective exposure time in the part of the slit available to our analysis (see text): darker colour corresponds to higher exposure time. For each slit position, the position angle is given at the end of the slit. \label{f:em}}
\end{figure}

Due to the nodding strategy used for the X-Shooter observations, only the central $6.4\arcsec$ (out of the total $11\arcsec$ slit length) are suitable for the analysis. However, in the overlap of all the slits, we are able to combine the individual limits to improve the constraint on the line flux. We combine the individual 2D spectra to look at emission but recover no detection. Using the same elliptical aperture as mentioned above, we measure a 3$\sigma$ limit consistent with the value obtained from the combination of the 4 individual limits. The combined limit measured in the combined stack of all spectra is given in the last row of Table~\ref{t:oii}.
This central region roughly corresponds to a circular aperture with a 1.2 arcsec diameter, which at the redshift of the DLA translates to 5.0~kpc. The configuration of slits is shown in the schematic map of exposure time in Figure~\ref{f:em}.

We can use the constraint on the line flux of [\ion{O}{ii}]$\lambda$3727 to put a limit on the star
formation rate. Instead of using the relation from \citet{Kennicutt98b}, which is calibrated using local
galaxies, we use the empirical dust-uncorrected $L[\OII]-$SFR relation from \citet{Kruehler15}. This relation
was constructed using high redshift GRB-DLA host galaxies and implicitely accounts for dependence on galactic properties
\citep[e.g.][]{Kewley04}.
The inferred star formation rates are summarised along with the flux limits in Table~\ref{t:oii}.

\begin{table}
  \caption{Limits on star-formation rate from the non-detection of [\OII]. 
    \label{t:oii}}
  \centering
  \begin{tabular}{c c c}
    \hline \hline
   {\large \strut} PA    & $F([\OII])$  & SFR \\
   {\large \strut} (deg) & ($10^{17}$~erg\,s$^{-1}$\,cm$^{-2}$) & (M$_{\rm \odot}$\,yr$^{-1}$) \\
    \hline
    0     & $<$ 4.0                             & $<$ 22 \\ 
 -150     & $<$ 3.0                             & $<$ 16 \\ 
 -60      & $<$ 3.8                             & $<$ 21 \\ 
 +60      & $<$ 4.3                             & $<$ 24 \\ 
 combined  & $<$ 1.9                            & $<$ 9.9 \\
 \hline
    \end{tabular}
  \end{table}

\section{Summary \label{s:concl}}

In this work, we present the detection and detailed analysis of an exceptional molecular absorber
at $\zabs=2.53$ towards the quasar \jz, being one of the very few absorption systems featuring
CO absorption lines known to date (see \citealt{Noterdaeme11} and references therein, \citealt{Ma15}).
Using both high resolution and multiwavelength spectroscopic observations, we derive the chemical
composition, dust depletion and extinction as well as the molecular content of this cloud. Our
main findings are the following:

(1) The absorption system is characterised by super-Solar metallicity (about 2.5~$Z_{\rm \odot}$). Although the uncertainty
on this value is large (by a factor of about 2-3), it is much higher than the metallicities seen in
the overall population of \HI-selected DLAs at this redshift \citep[see e.g.][]{Rafelski12}. 
The observed depletion of refractory elements is typical of cold gas in the Galactic disc and peaks at the location
where molecular gas is found. Higher spectral resolution data would be desirable to constrain better the amount of metals in this
very narrow component ($b\sim$0.7~\kms).

(2) The DLA has a molecular fraction reaching almost 50\% overall, being the largest value measured till
now in a high-$z$ absorption system, for a total neutral hydrogen column density of $\log N({\rm H}) = 20.8$. This
corresponds to a neutral gas surface density of $\Sigma_{\rm HI} \approx 5$~M$_{\rm \odot}$\,pc$^{-2}$ and
a molecular hydrogen surface density $\Sigma_{\rm HI} \approx 4.4$~M$_{\rm \odot}$\,pc$^{-2}$. 
These surface densities are very similar to what is seen across the Perseus molecular cloud from high spatial
resolution emission observations \citep{Lee12}, despite the \HI\ and H$_2$ surface densities being
measured directly along a pencil beam line of sight in our study while derived from dust maps in the latter.
This also shows that \HI-to-H$_2$ transition in our high-$z$ system is likely following the same physical processes as in our Galaxy.

(3) We further use the different molecular and atomic species to constrain the actual physical conditions in the cold gas with
the help of the spectral synthesis code Cloudy. We find that the column densities can be well reproduced by a
cloud with density around $n_{\rm H} \sim 80$~cm$^{-3}$ inmersed into
a moderate UV field, similar to the local interstellar radiation field. We show that a high cosmic ray ionisation rate together with the presence of small
dust grains -- consistent with the depletion pattern, the steep extinction curve, and presence of a 2175~{\AA} bump -- can explain the high
CO fractional abundance. The model also predicts a kinetic
temperature around 50~K, in perfect agreement with that derived from the excitation of H$_2$.
The CO abundance rises immediately after the \HI-to-H$_2$ transition, thanks
to the efficient chemistry paths involving H$_2$, together with pre-shielding of CO. About half of the hydrogen
remains in atomic form in the cloud interior, well inside the \HI-to-H$_2$ transition layer, due to H$_2$ destruction by cosmic rays. This can also explain the high HD/2H$_2$ ratio through chemical enhancement of HD compared to H$_2$.
We must however keep in mind that, in addition to cosmic rays, several other processes heating the gas can also enhance the production
of CO \citep[][]{Goldsmith13} and that our model suffers from uncertainties in several imput parameters, such as the metallicity.
The detection of more molecular species in this system, together with comparison of different codes
\citep[e.g. the Meudon PDR code, ][]{LePetit06} should break degeneracies between parameters such as dust abundance and cosmic ray rate
and lead to a better understanding of the physical processes at play.
Our study suggests that the presence of strong C\,{\sc i} lines, detectable at low spectral resolution \citep[as in][]{Ledoux15},
is a good indicator for high molecular hydrogen column density (in self-shielded regime), but this (or directly detecting
damped H$_2$ lines as in \citealt{Balashev14}) is not a sufficient condition to get CO in detectable amounts. Since small grains
seem to play a crucial role in the production of CO, 
selecting systems that, in addition to \CI, also present a steep extinction curve and/or the presence of a 2175~{\AA} bump should significantly
increase the probability to detect CO.

(4) While our study shows that the line of sight towards \jz\ has chemical and physical characteristics
similar to those found in diffuse molecular regions of the Perseus cloud, we show that the former absorbing cloud is immersed into a
warmer cosmic microwave background radiation. We derive the temperature of the CMB radiation at the absorber's redshift from the excitation of CO lines,
correcting for a small temperature excess due to collisional excitation.
The temperature we obtain is in perfect agreement with the adiabatic cooling expected
in the standard hot Big-Bang theory, $T_{\rm CMB}(z)=T_0 \times (1+z)\approx 9.6$~K at $z=2.53$.

Final remarks:
The discovery presented here was facilitated by the steadily increasing number of available spectra of faint quasars, together
with sensitive instruments on large telescopes. The involved amounts of carbon monoxide remain however far too low to be
detectable in emission (even if it were at $z=0$), making the cloud being an example of CO-dark molecular gas
\citep{Wolfire10}. Interestingly, such furtive phase may actually contain a large fraction
of the total molecular gas in galaxies \citep[e.g.][]{Smith14}. Efforts should therefore be pursued to detect more absorption
systems like the one presented here at high-redshift. 
These will provide excellent targets for detailled high-resolution studies (including CMB temperature) using the next generation of
extremely large telescopes.

\begin{acknowledgements}
  We warmly thank the anonymous referee for detailed comments and suggestions.
  We are very grateful to ESO for support during the preparation of OBs, for carrying out the observations in
  service mode as well as allowing us to accomodate X-Shooter observations in an originally UVES progamme
  (096.A-0924(B)). 
  We thank Robert Carswell for modifying for us his Voigt-profile fitting code {\sc vpfit}, providing us an 
  updated version (v10.3) ahead of public release. 
  We also thank Patrick Boiss\'e, Steve Federman and Jacqueline Bergeron for useful
  discussions during the analysis. We thank Gargi Shaw for help with running Cloudy.
  PN and PPJ acknowledge support from the Agence Nationale de la Recherche
  under grant ANR-12-BS05-0015. PN, NG, PPJ and RS acknowledge support from the Indo-French Centre for the Promotion
  of Advanced Research under project No.~5504-2.
  JK acknowledges support from the European Union's Seventh Framework Programme for research and innovation under
  the Marie-Curie grant agreement no. 600207 with reference DFF-MOBILEX--5051-00115. SB acknowledges support from the Russian Science Foundation under grant 14-12-00955.
  TK acknowledges support through the Sofja Kovalevskaja Award to P. Schady from the Alexander von Humboldt Foundation of Germany.  
  MTM thanks the Australian Research Council for \textsl{Discovery Project} grant DP130100568 which supported this work.
  This work has made use of the SDSS-III/BOSS
  database. Funding for SDSS-III has been provided by the Alfred P. Sloan Foundation, the Participating Institutions, the National Science Foundation, and the U.S. Department of Energy Office of Science. The SDSS web site is \url{http://www.sdss.org/}.
SDSS-III is managed by the Astrophysical Research Consortium for the Participating Institutions of the SDSS-III Collaboration including the University of Arizona, the Brazilian Participation Group, Brookhaven National Laboratory, Carnegie Mellon University, University of Florida, the French Participation Group, the German Participation Group, Harvard University, the Instituto de Astrofisica de Canarias, the Michigan State/Notre Dame/JINA Participation Group, Johns Hopkins University, Lawrence Berkeley National Laboratory, Max Planck Institute for Astrophysics, Max Planck Institute for Extraterrestrial Physics, New Mexico State University, New York University, Ohio State University, Pennsylvania State University, University of Portsmouth, Princeton University, the Spanish Participation Group, University of Tokyo, University of Utah, Vanderbilt University, University of Virginia, University of Washington, and Yale University. 
 
\end{acknowledgements}

\bibliographystyle{/scisoft/share/texmf/aa/aa-package/bibtex/aa}
\bibliography{J0000}

\FloatBarrier
\begin{appendix}
  \section{Robustness of the CO measurement \label{ACO}}

When estimating best-fit parameters, {\sc vpfit} takes as input the normalised spectrum
and the resolution provided by the user. This means that continuum placement uncertainties
and errors due to the knowledge of the spectral point spread function (SPSF) are not reflected
in the error estimates.

\begin{figure*}
\centering
\begin{tabular}{ccc}
  \includegraphics[bb=192 20 435 780, clip=,angle=90, width=0.45\hsize]{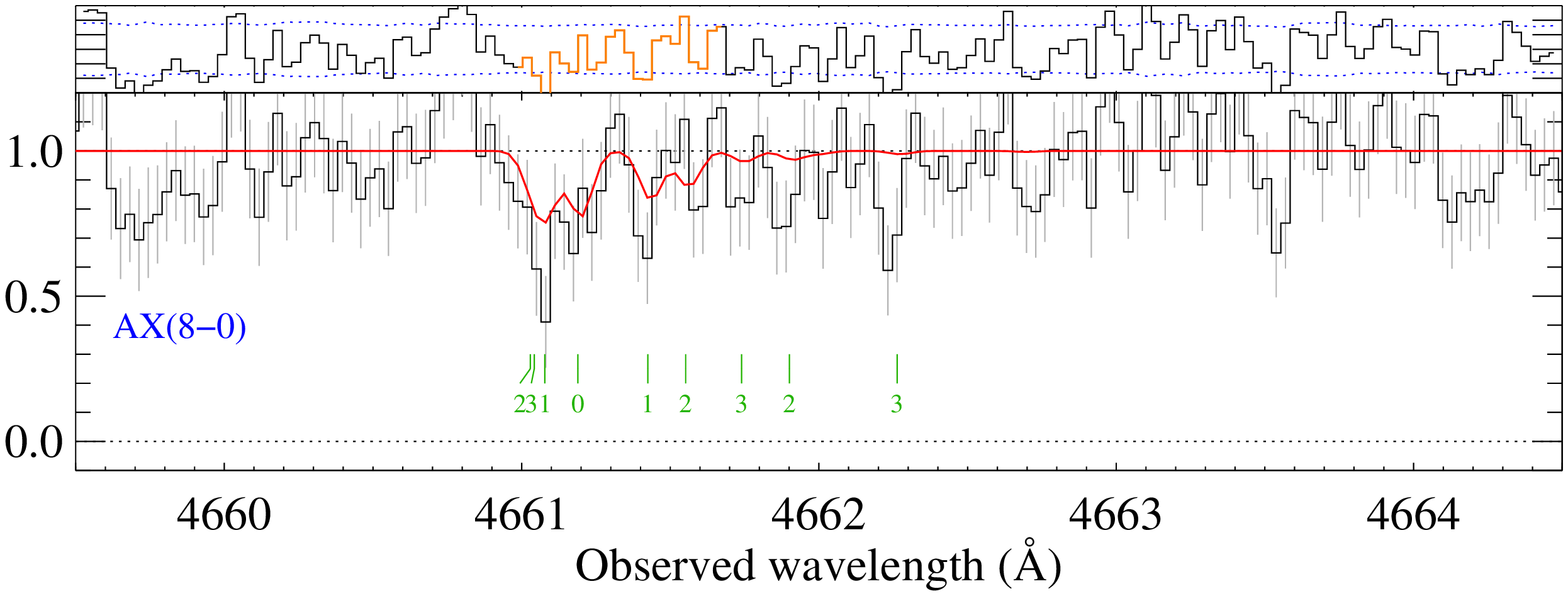}&
  \includegraphics[bb=192 20 435 780, clip=,angle=90, width=0.45\hsize]{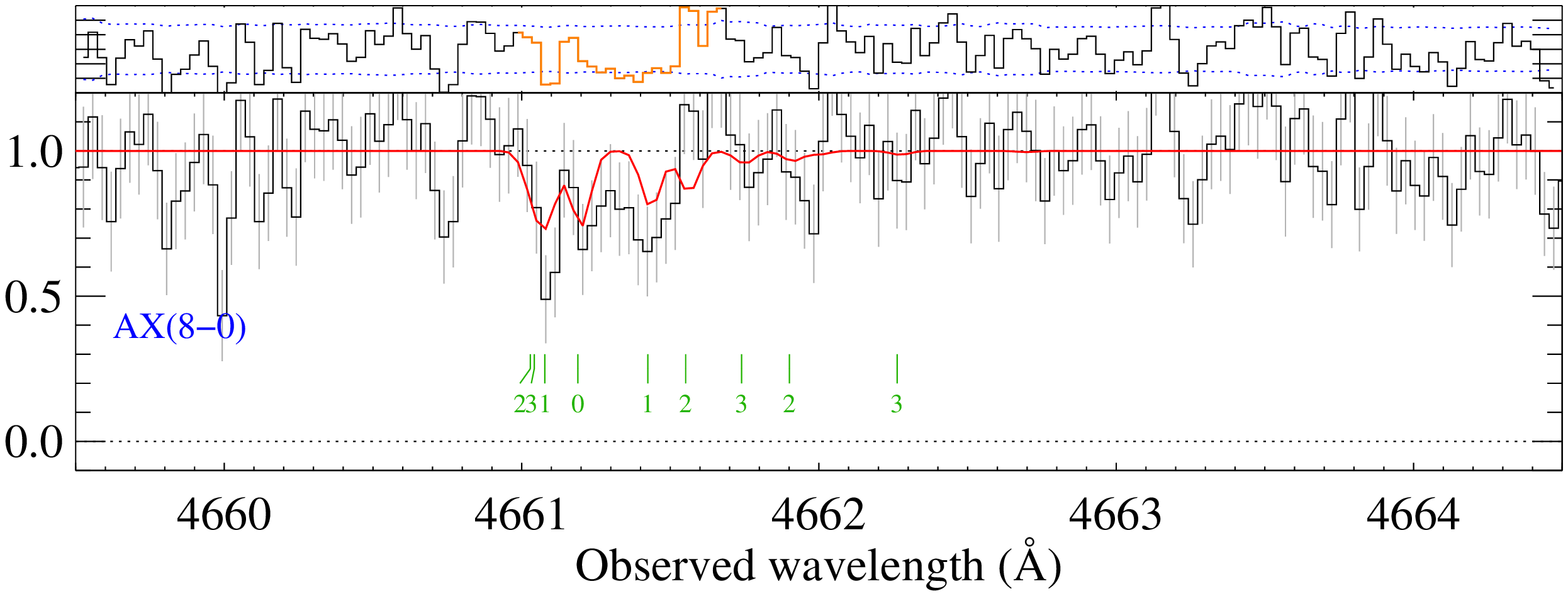}\\
  
  \includegraphics[bb=192 20 435 780, clip=,angle=90, width=0.45\hsize]{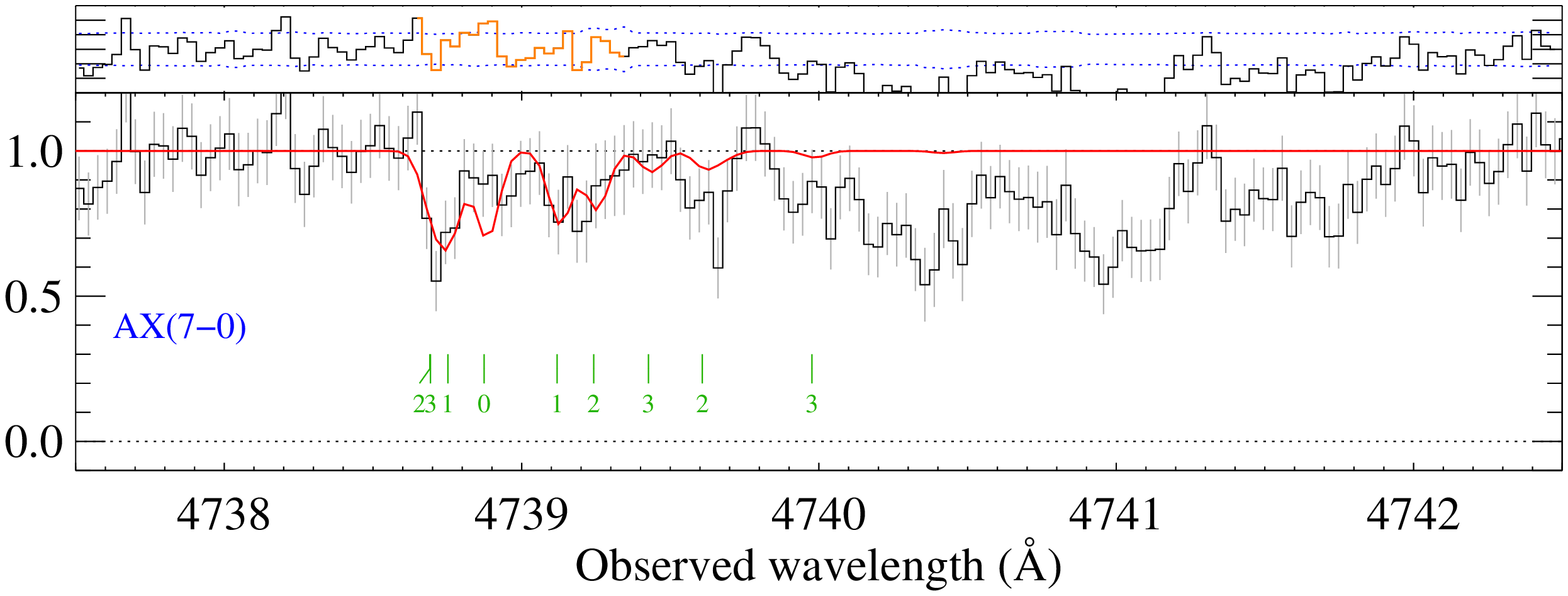}&
  \includegraphics[bb=192 20 435 780, clip=,angle=90, width=0.45\hsize]{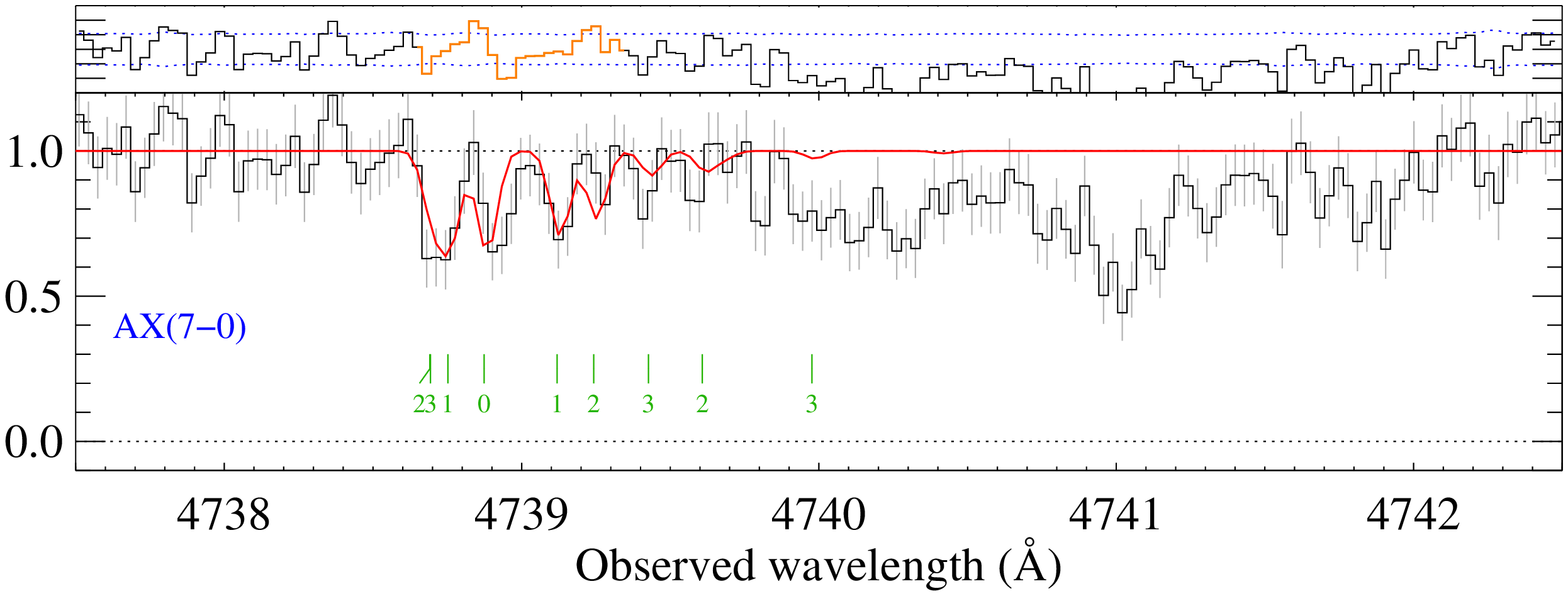}\\
  
  \includegraphics[bb=192 20 435 780, clip=,angle=90, width=0.45\hsize]{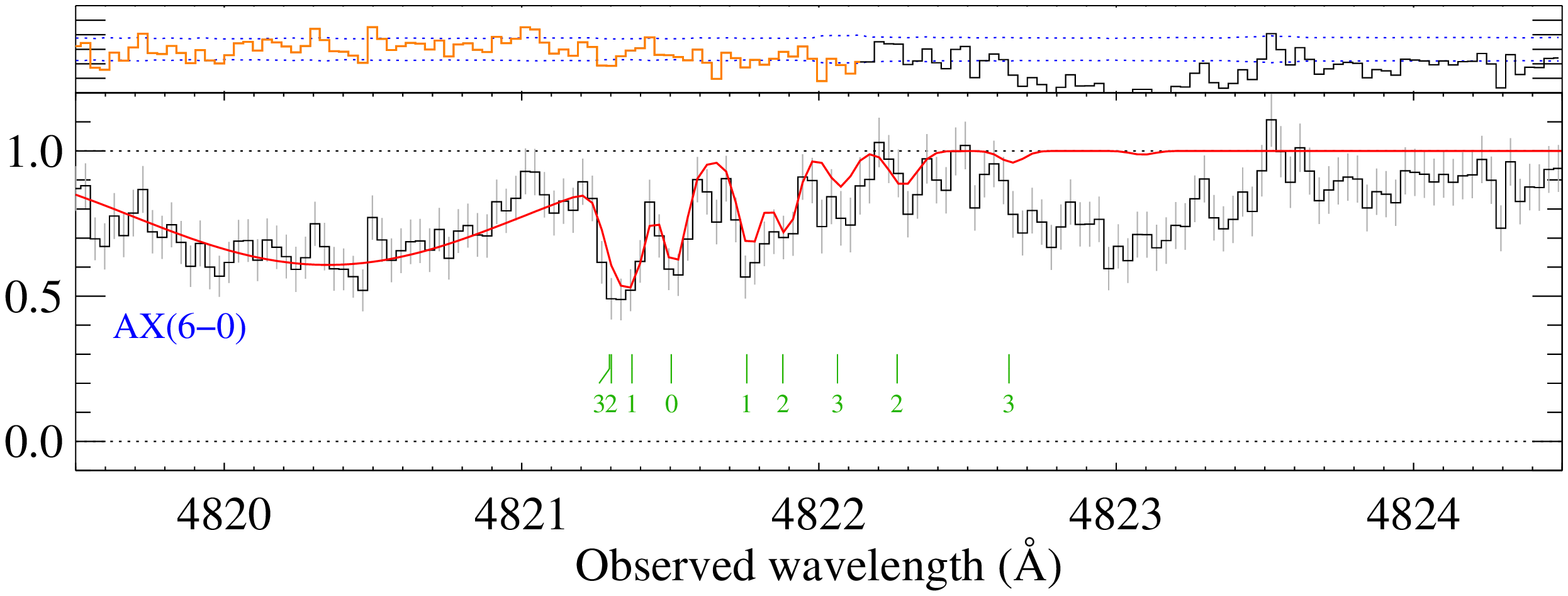}&
  \includegraphics[bb=192 20 435 780, clip=,angle=90, width=0.45\hsize]{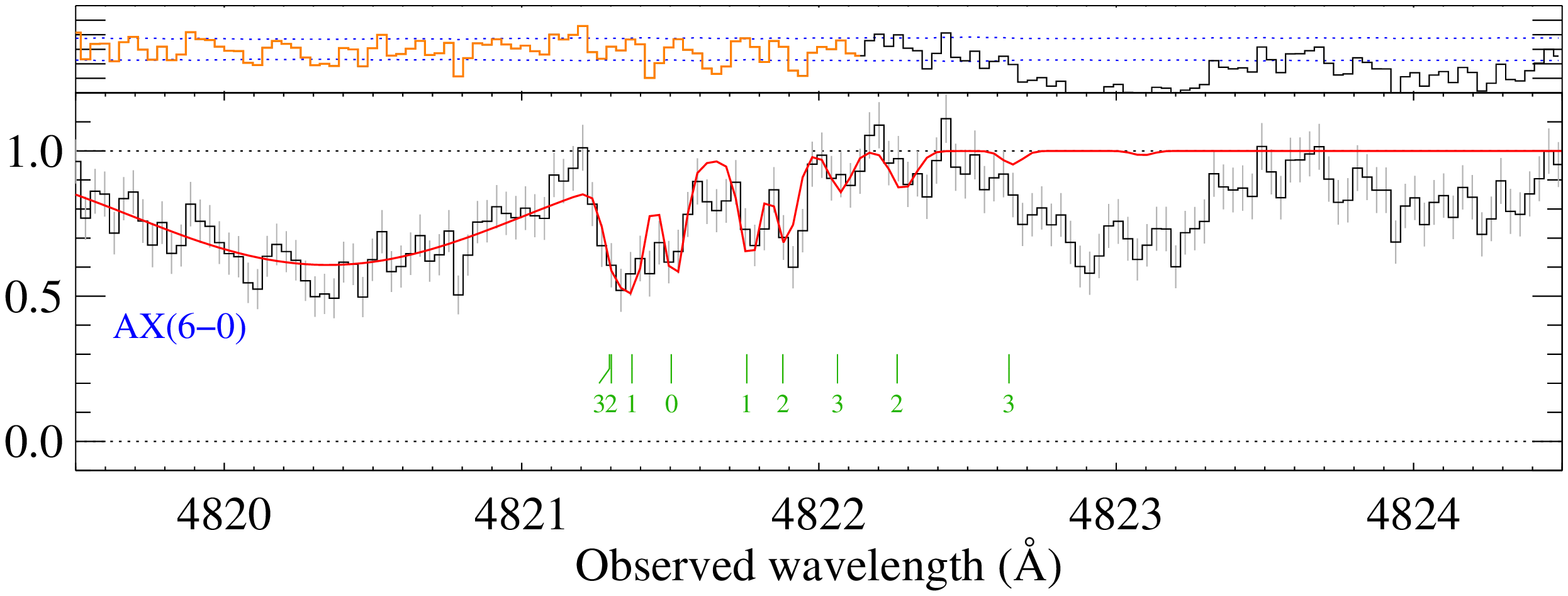} \\

    \includegraphics[bb=192 20 435 780, clip=,angle=90, width=0.45\hsize]{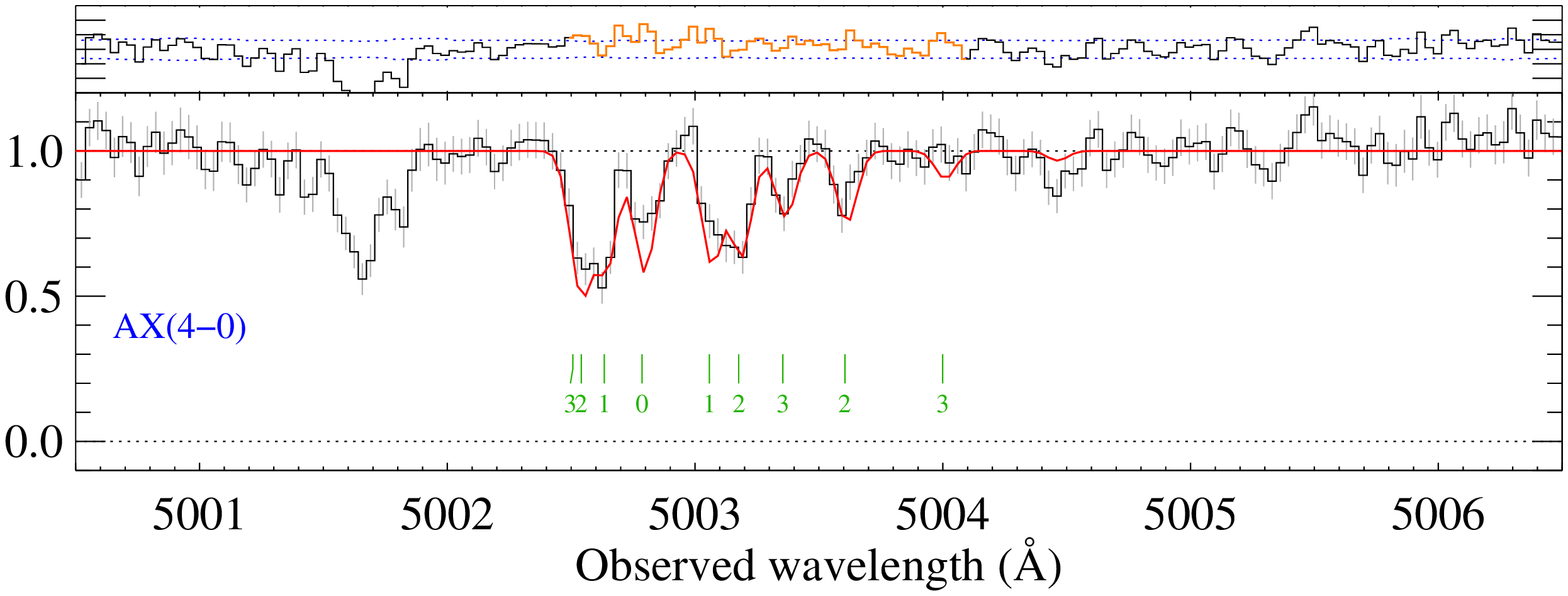}&
    \includegraphics[bb=192 20 435 780, clip=,angle=90, width=0.45\hsize]{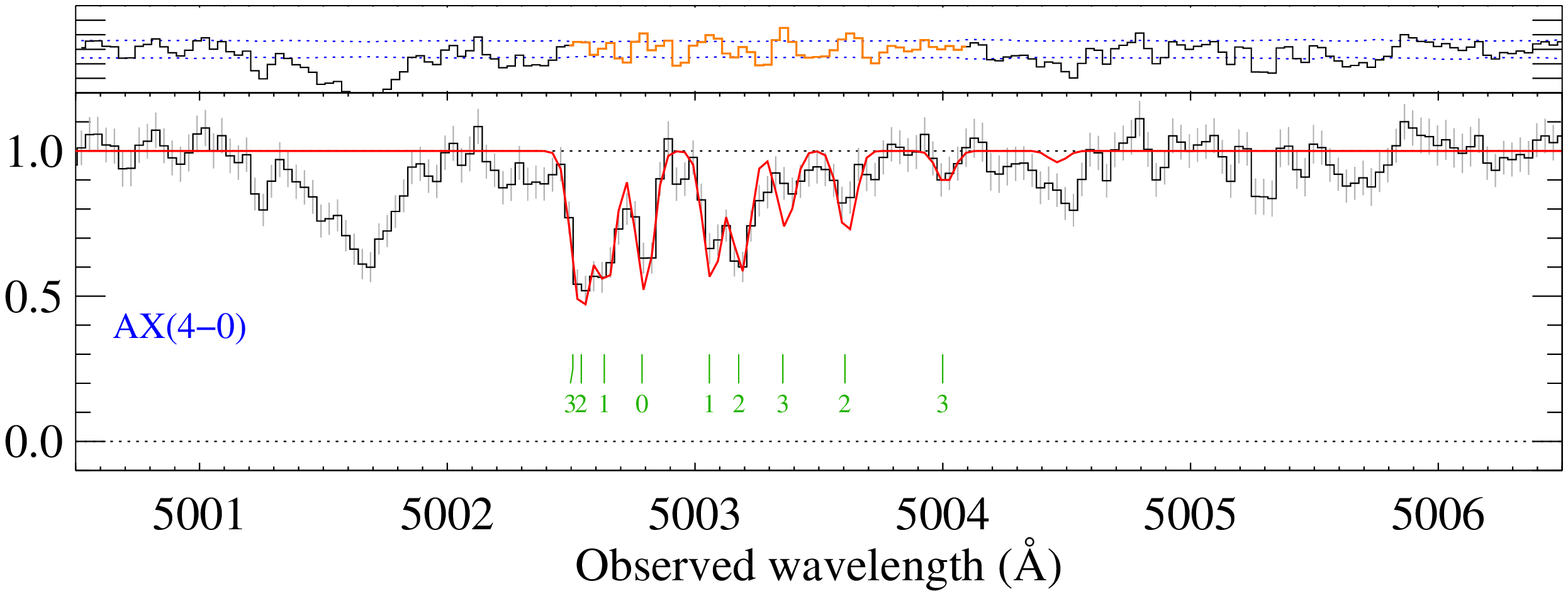} \\

    \includegraphics[bb=192 20 435 780, clip=,angle=90, width=0.45\hsize]{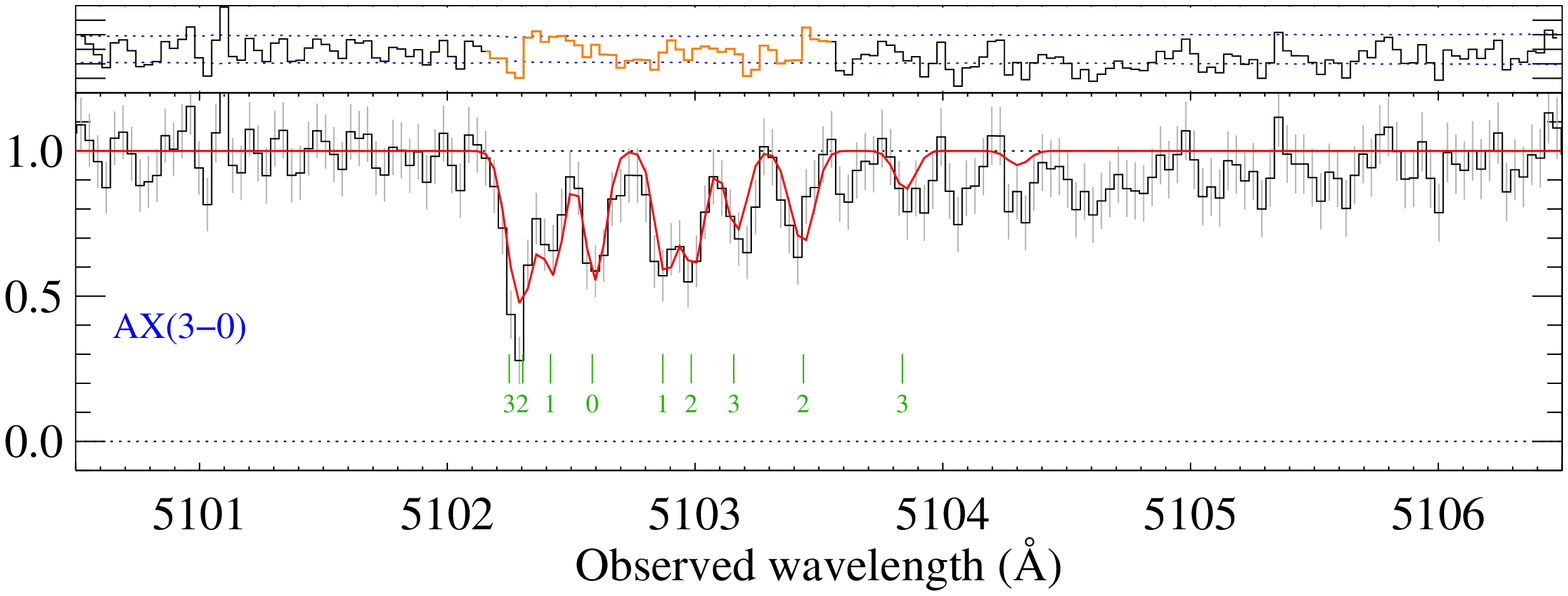}&
    \includegraphics[bb=192 20 435 780, clip=,angle=90, width=0.45\hsize]{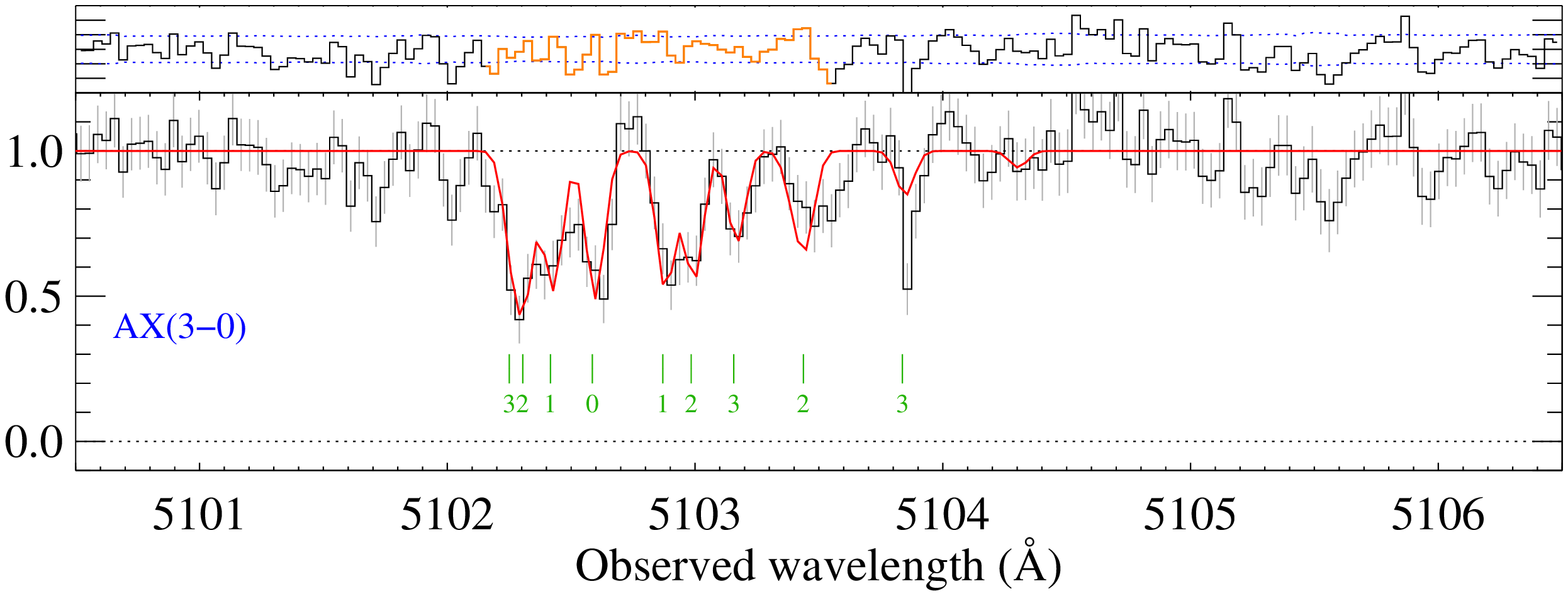} \\

    \includegraphics[bb=192 20 435 780, clip=,angle=90, width=0.45\hsize]{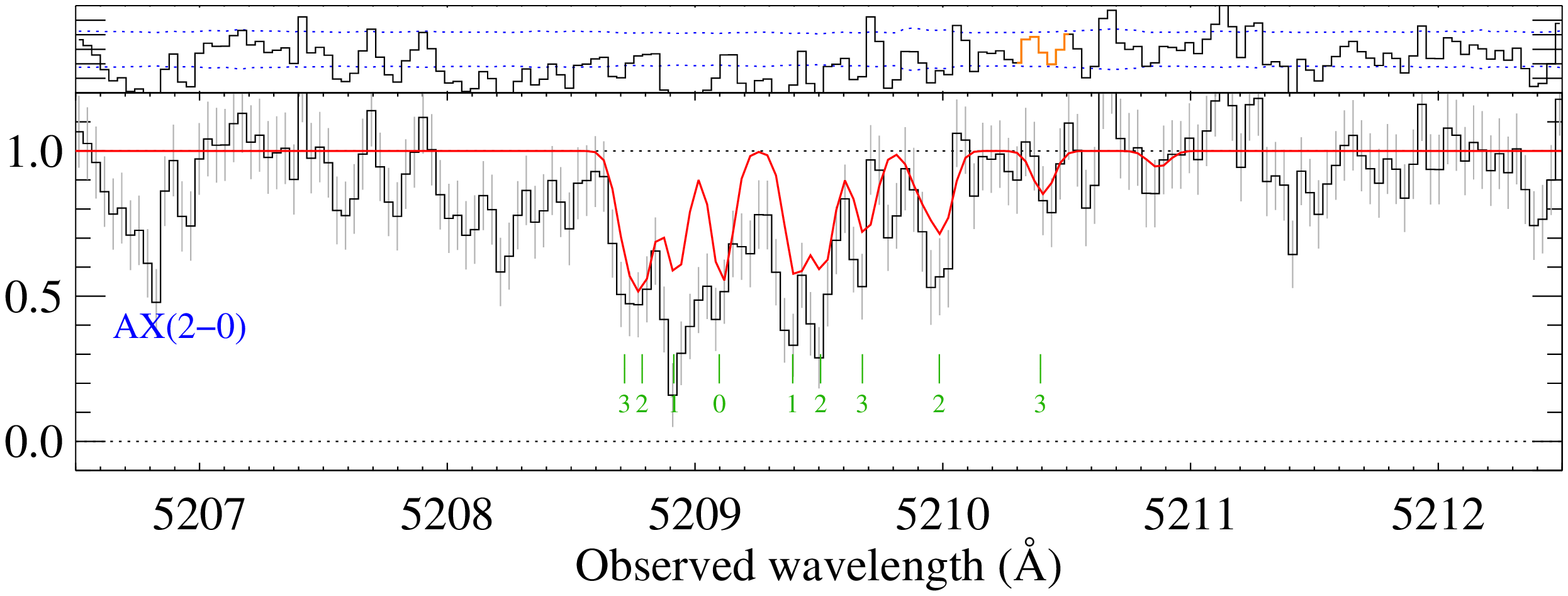}&
    \includegraphics[bb=192 20 435 780, clip=,angle=90, width=0.45\hsize]{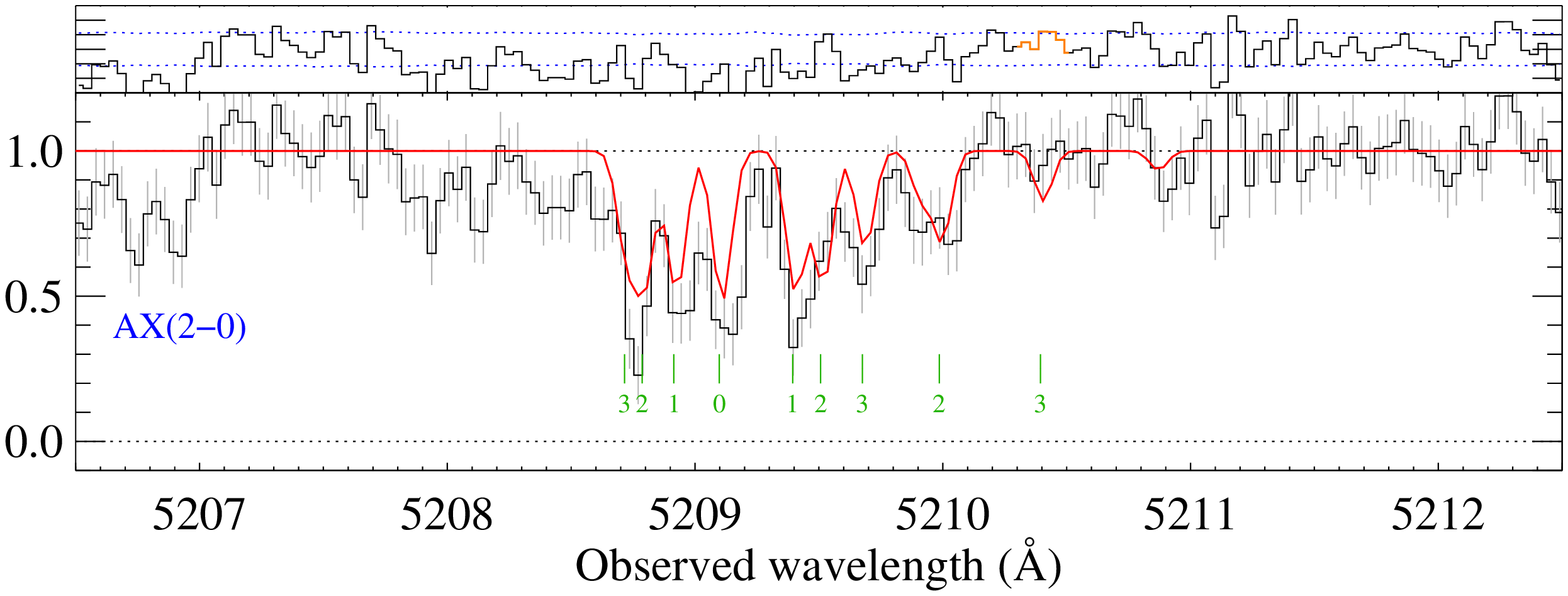} \\

    \includegraphics[bb=192 20 435 780, clip=,angle=90, width=0.45\hsize]{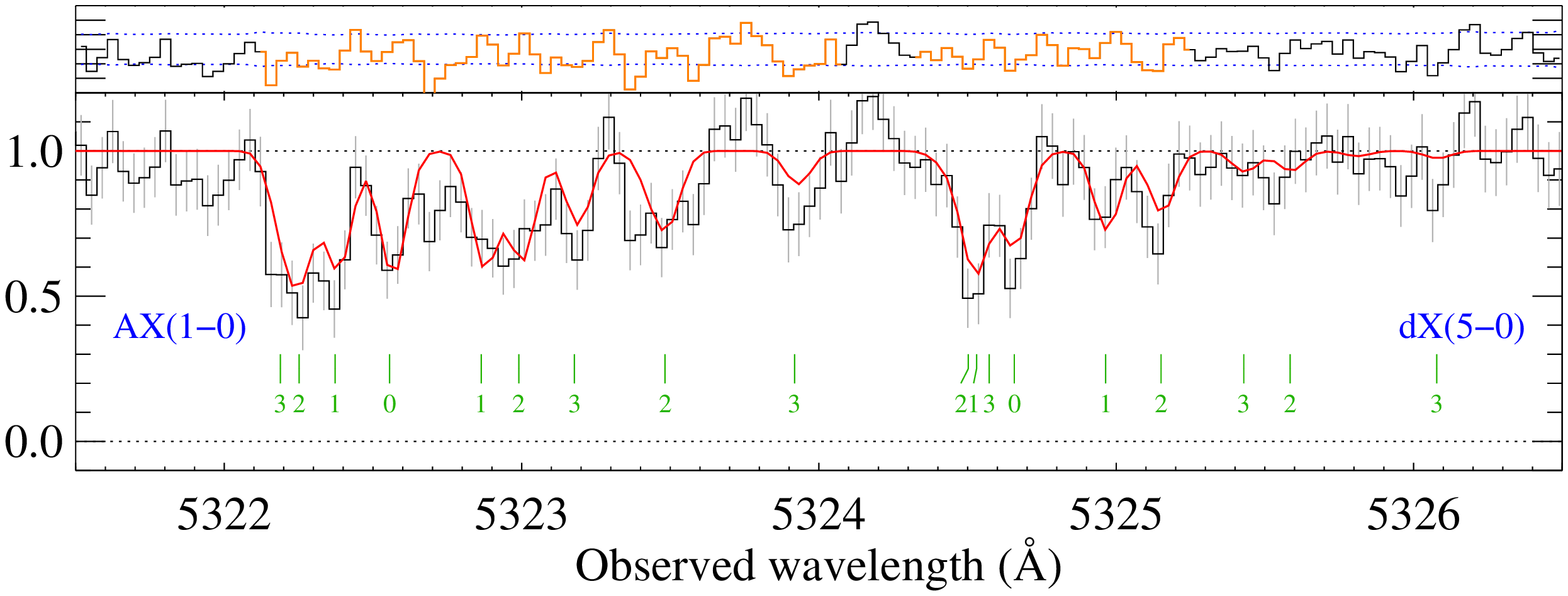}&
    \includegraphics[bb=192 20 435 780, clip=,angle=90, width=0.45\hsize]{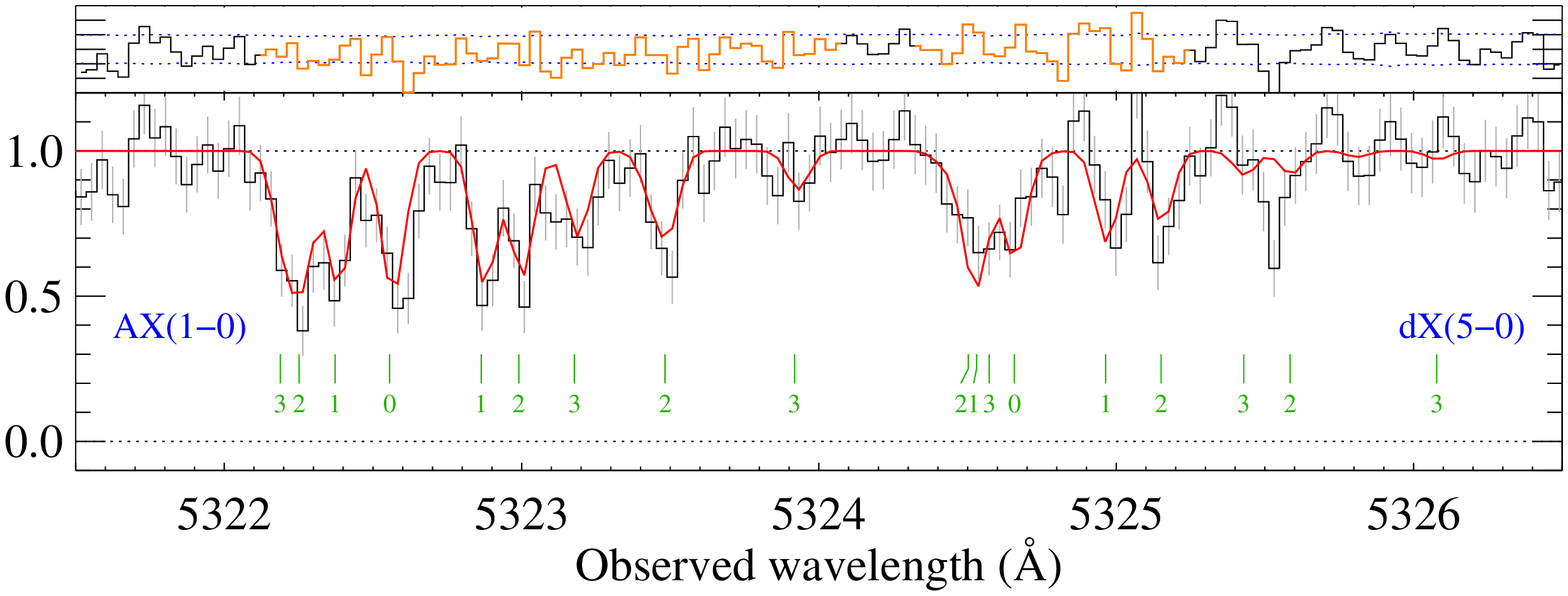} \\

   \includegraphics[bb=192 20 435 780, clip=,angle=90, width=0.45\hsize]{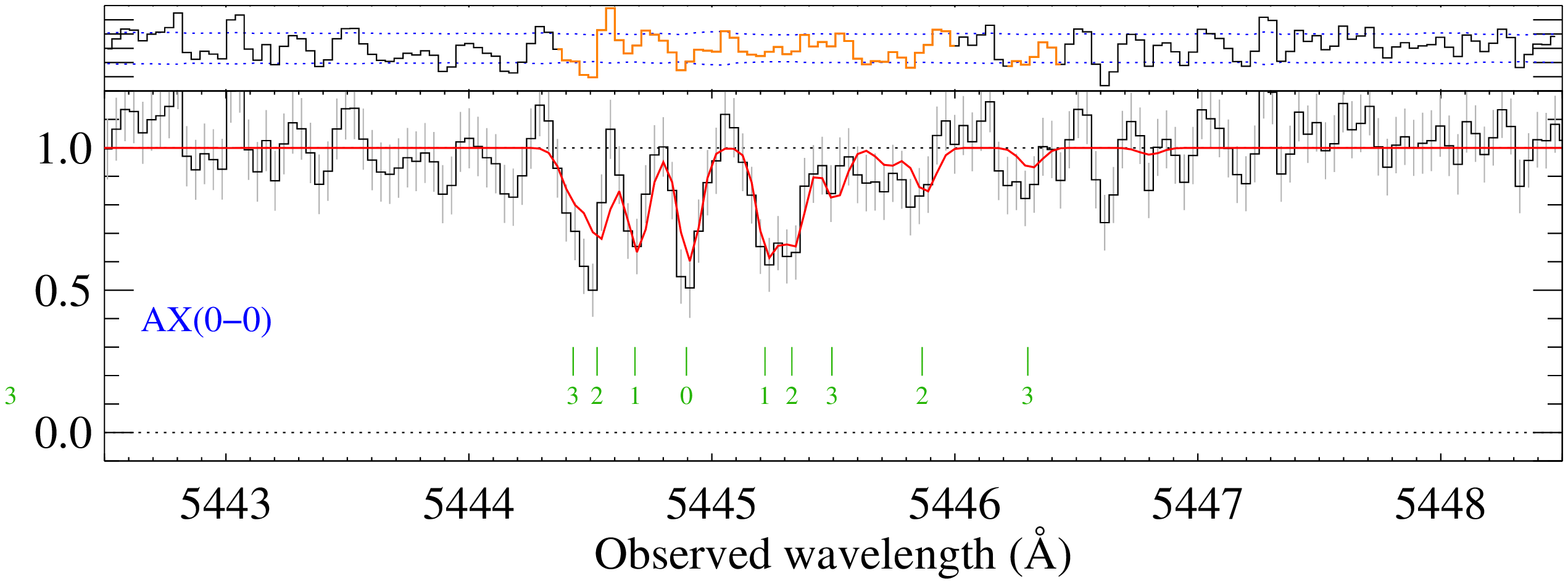}&
   \includegraphics[bb=192 20 435 780, clip=,angle=90, width=0.45\hsize]{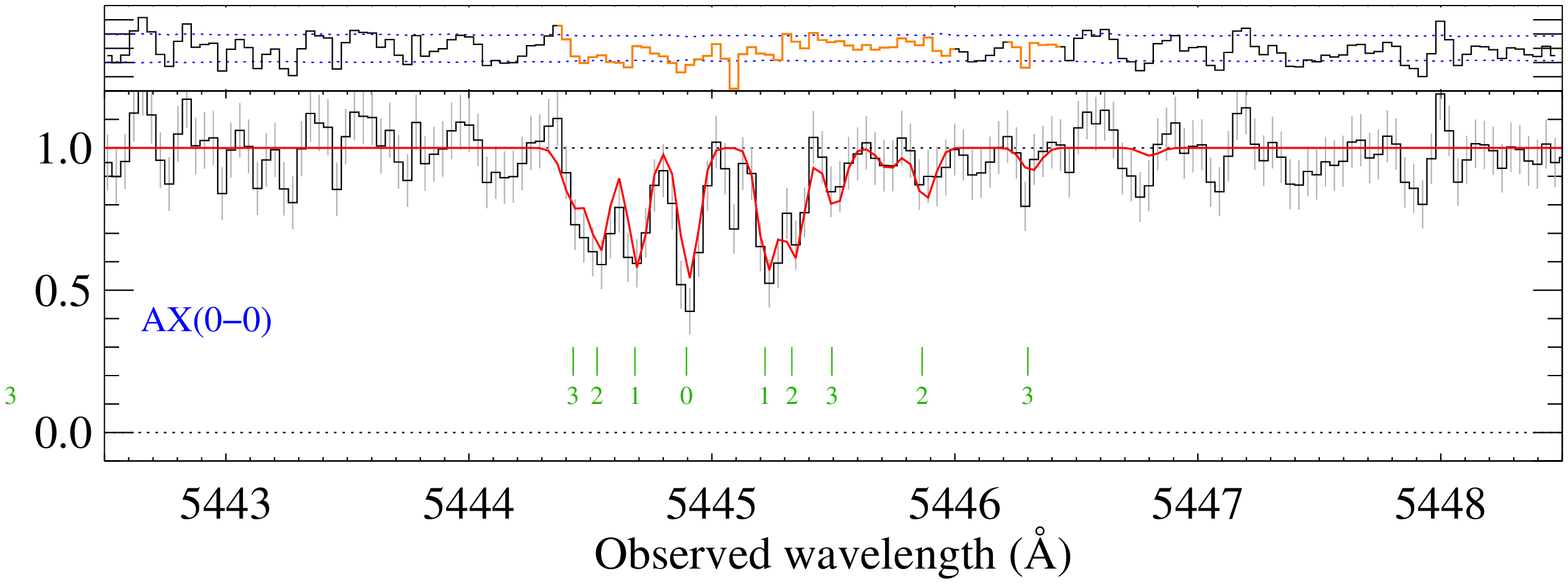} \\

\end{tabular}
\caption{Voigt-profile fit to the CO absorption bands, labelled in blue in each panel.
  The left (resp. right) panels correspond to data taken with the 0.9'' (resp. 0.7'') slit. Rotational levels from $J=0$ to $J=3$ are indicated as green tick marks. The panel above
  each region shows the residual, with the blue line indicating the $\pm$1\,$\sigma$ interval, and the orange regions
that used to constrain the fit. \label{f:COi}}
\end{figure*}

\begin{figure}%[!ht]
\centering
\begin{tabular}{c}
\includegraphics[bb=50 222 570 660,clip=, width=\hsize]{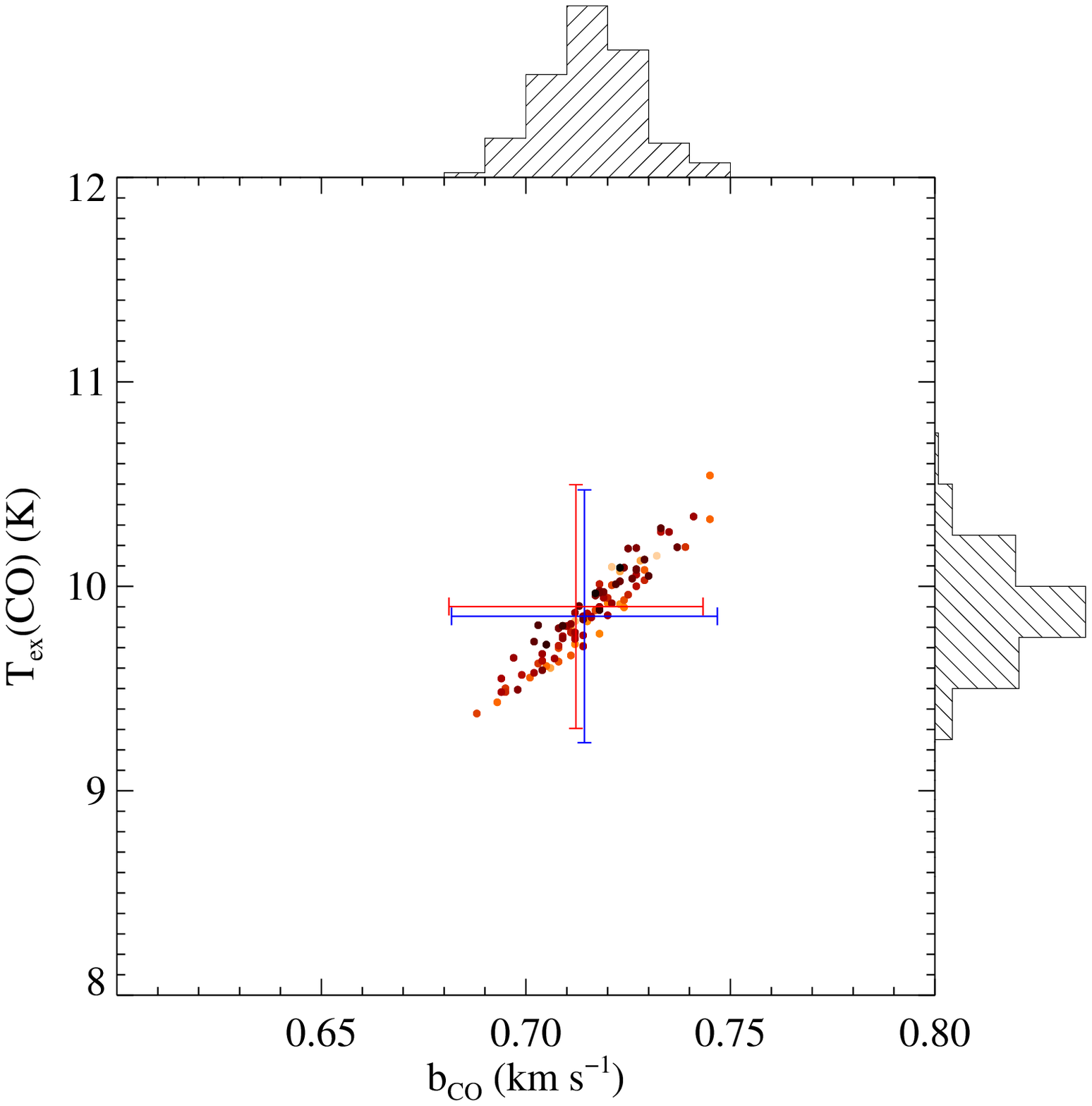}\\
\includegraphics[bb=50 170 570 580,clip=, width=\hsize]{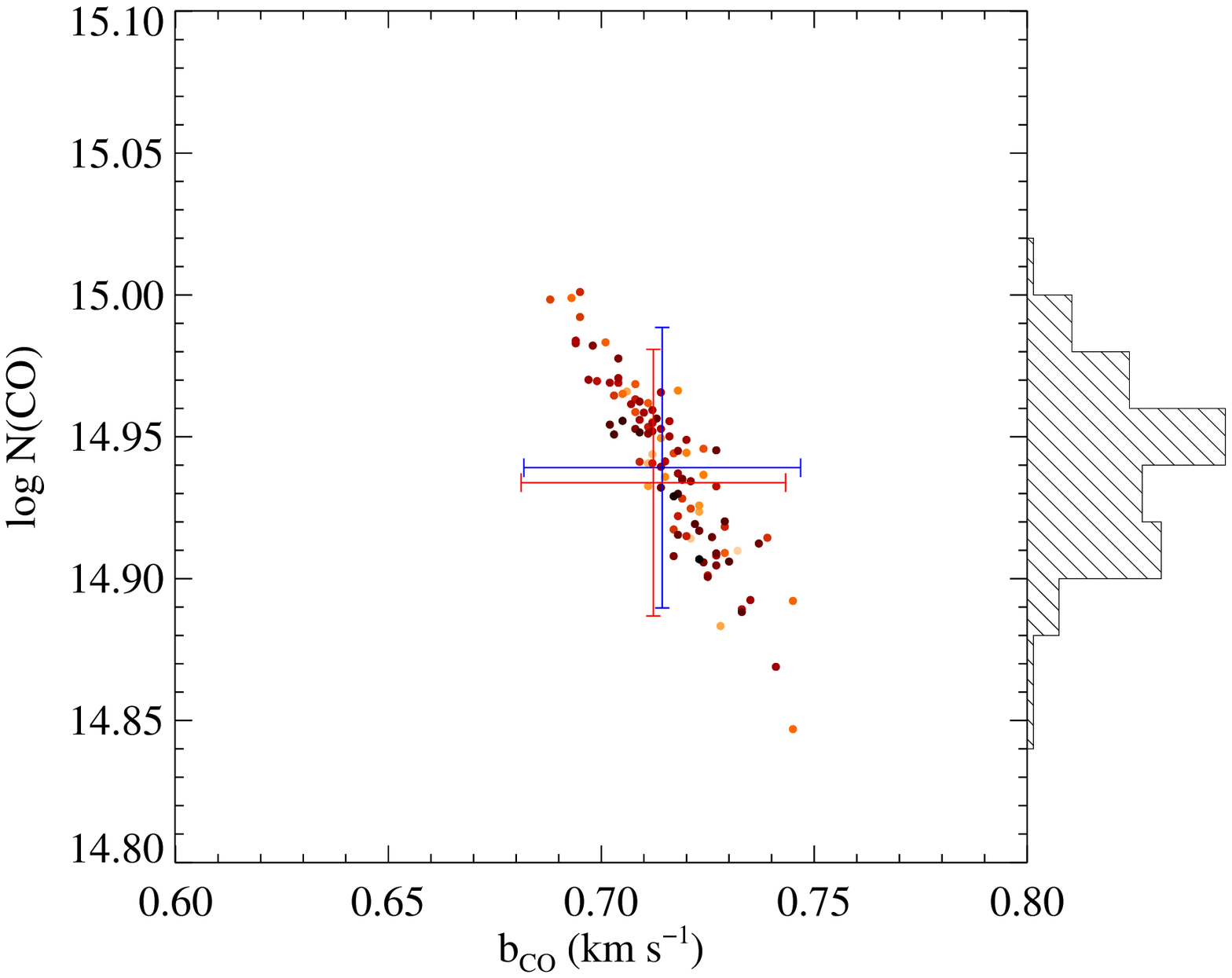}\\
\end{tabular}
\caption{Effect of continuum placement uncertainties on the total 
CO column density, excitation temperature and Doppler parameter.
The histograms represent the distributions of $\log N($CO),  
$T_{\rm ex}$(CO) and $b$ for 100 realisations where the continuum for each fitting 
region has been modified independently and randomly.
The colour of each point
indicates the reduced $\chi^2$ that varies from 1.08 (darkest) to 1.23 (lightest).
The blue (resp. red) error bar corresponds to the best fit value using the combined UVES spectrum (resp.
both UVES spectra fitted simultaneously).\label{f:shaky}
}
\end{figure}

We test the effect of SPSF uncertainties by refitting the data, using a range of
spectral resolution from 4.75 to 5.25~\kms. 
The resulting parameters remain well within their
associated fitting uncertainty, with total column density varying by only $\pm$0.05~dex,
excitation temperature (see Fig.~\ref{f:coext}) varying by $\Delta T_{\rm ex} \sim 0.1$~K and Doppler parameter basically
unchanged ($<$2\%).
We also fitted the two red spectra simultaneously, instead of combining them (Fig.~\ref{f:COi}). The results are also very close
to what we get from the combined spectrum, as can be seen from the blue and red error bars on Fig.~\ref{f:shaky},
with a slight improvement in the $\chi^2_{\nu}=1.03$ probably thanks to the better evaluated SPSF.

To estimate the effect of continuum placement uncertainties, we independently renormalised each region
around CO absorption bands by randomly modifying the local continuum slope and intercept, following their
respective normal error distribution.  The distribution of CO column
densities, Doppler parameter and excitation temperature obtained for hundred realisations of this ``shaky'' continuum
procedure is shown on Fig.~\ref{f:shaky}. Clearly, the effect of continuum placement uncertainty is well within the error estimates
from fitting the lines but highlights the correlation between $b$, $N$ and $T_{\rm ex}$. For example, a smaller
Doppler parameter will result in a higher column density and a lower excitation temperature. It can also be
seen from this figure that the most deviant points generally also have the highest $\chi^2_{\nu}$-values.

\end{appendix}

\end{document}